\documentclass{article}

\usepackage{arxiv}

\usepackage{amsmath}
\usepackage{graphicx}
\usepackage{enumerate}

\usepackage{url} % not crucial - just used below for the URL 
\usepackage[utf8]{inputenc}
\usepackage{amsmath, amsfonts, amssymb, bm}
\usepackage[makeroom]{cancel}
\usepackage{mathtools}
\usepackage{bbm}
\usepackage{comment}

\usepackage{xcolor}
\usepackage{subcaption}
\usepackage{enumitem}
\usepackage[hang,flushmargin]{footmisc} 

\usepackage{natbib}
\usepackage[colorlinks = true,
linkcolor = darkerteal,
urlcolor  = black,
citecolor = darkteal,
anchorcolor = darkteal]{hyperref}

\usepackage{tikz, circuitikz, setspace}
\usepackage[ruled]{algorithm2e}

\newcommand{\N}{\mathrm{N}}

\definecolor{darkteal}{RGB}{0,102,102}
\definecolor{darkerteal}{RGB}{0,76,76}

\title{Decoding Neuronal Ensembles from Spatially-Referenced Calcium Traces: \\ A Bayesian Semiparametric Approach}

    \author{ \href{https://orcid.org/0000-0001-5034-7414}{\includegraphics[scale=0.06]{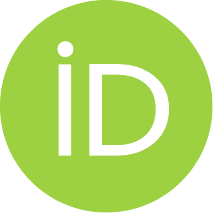}\hspace{1mm}Laura D'Angelo} \\
	Department of Economics,\\ Management and Statistics   \\
        University of Milano-Bicocca\\
	{laura.dangelo@unimib.it} \\
	\And
	\href{https://orcid.org/0000-0003-2978-4702}{\includegraphics[scale=0.06]{orcid.pdf}\hspace{1mm}Francesco Denti} \\
	Department of Statistical Sciences\\ 
        University of Padova\\
	{francesco.denti@unipd.it} \\
	\AND
	\href{https://orcid.org/0000-0002-5403-0040}{\includegraphics[scale=0.06]{orcid.pdf}\hspace{1mm}Antonio Canale} \\
	Department of Statistical Sciences\\ 
        University of Padova\\
	{antonio.canale@unipd.it} \\
    \And
	\href{https://orcid.org/0000-0002-6363-9907}{\includegraphics[scale=0.06]{orcid.pdf}\hspace{1mm}Michele Guindani} \\
	Department of  Biostatistics\\ 
    University of California, Los Angeles\\
	{mguindani@g.ucla.edu} \\
}

\date{}

\renewcommand{\headeright}{}
\renewcommand{\undertitle}{}
\renewcommand{\shorttitle}{Decoding neuronal ensembles from spatially-referenced calcium traces%:\\ a Bayesian semiparametric approach
}

\hypersetup{
pdftitle={Decoding neuronal ensembles from spatially-referenced calcium traces: a Bayesian semiparametric approach},
pdfsubject={},
pdfauthor={D'Angelo, Denti, Canale, Guindani},
pdfkeywords={Calcium imaging, Neuroimaging,  Dependent Dirichlet process, Gaussian process, Mixture model, Spatial clustering},
}

\begin{document}
\maketitle

\begin{abstract}
Understanding how neurons coordinate their activity is a fundamental question in neuroscience, with implications for learning, memory, and neurological disorders. Calcium imaging has emerged as a powerful method to observe large-scale neuronal activity in freely moving animals, providing time-resolved recordings of hundreds of neurons. 
However, fluorescence signals are noisy and only indirectly reflect underlying spikes of neuronal activity, complicating the extraction of reliable patterns of neuronal coordination. We introduce a fully Bayesian, semiparametric model that jointly infers spiking activity and identifies functionally coherent neuronal ensembles from calcium traces.  Our approach models each neuron's spiking probability through a latent Gaussian process and encourages anatomically coherent clustering using a location-dependent stick-breaking prior. A spike-and-slab Dirichlet process captures heterogeneity in spike amplitudes while filtering out negligible events. We consider calcium imaging data from the hippocampal CA1 region of a mouse as it navigates a circular arena, a setting critical for understanding spatial memory and neuronal representation of environments. Our model uncovers spatially structured co-activation patterns among neurons and can be employed to reveal how ensemble structures vary with the animal's position. 
\end{abstract}
\vspace{0.1cm}

% keywords can be removed

\clearpage

\section{Introduction}
Neurons are fundamental cells of the nervous system, responsible for receiving sensory input from the external world and transmitting information throughout the body. Understanding how neurons communicate and coordinate their activity is central to studying the mechanisms underlying brain function and dysfunction. In recent years, remarkable technological advances have enabled scientists to visualize and measure neuronal activity in freely moving animals. Among these innovations, calcium imaging via miniaturized, head-mounted microscopes has become central to basic neuroscience research \citep{Aharoni2019}. This method exploits fluctuations in intracellular calcium ions that occur during neuronal firing, using fluorescent indicators to visualize these changes. 
When a neuron activates, calcium ions rapidly enter the cell, producing a transient increase in fluorescence; levels return to baseline once activity ceases. The resulting fluorescence trace provides a proxy for neuronal firing, with peaks corresponding to active periods \citep{calcium_review2}.
A key feature of calcium imaging is that it enables the simultaneous recording from a large number of neurons \emph{in vivo}, providing insight into how cells in different brain regions work together to encode and process information during specific tasks.  This technique has opened unprecedented opportunities to study neural population dynamics in realistic behavioral settings, with broad applications spanning addiction \citep{Siciliano2019, Beacher2021}, learning and memory \citep{Grienberger2022}, neuropsychiatric disorders \citep{Chen2021, Terashima2022}, and many other areas.

While the resulting data are rich in information, they also pose substantial statistical challenges. The raw fluorescence traces are noisy, indirect proxies for spiking activity, and their temporal resolution is limited by the relatively slow decay kinetics of calcium indicators compared to the rapid underlying firing activity. When multiple spikes occur in rapid succession, this may cause these individual firing events to blend together in the recorded movie, making their identification more challenging \citep{Shibue2020}. 
The observed fluorescence traces are affected by measurement noise and post-processing variability, which can obscure true firing events and complicate inference \citep{Gauthier2022}. Moreover, neural activity often exhibits weakly synchronized and spatially structured patterns, modulated by behavioral context such as the animal’s position and trajectory.
For example, hippocampal neurons known as \emph{place cells} activate selectively at specific spatial locations and form ensembles encoding an internal representation of the environment \citep{Krupic2018, OKeefe1971hippomap, moser2014grid}. Identifying such a structure requires methods that cluster neurons based on shared temporal dynamics while allowing comparison across spatial contexts.

\subsection{In vivo imaging of hippocampal  CA1 neuronal ensembles}
\label{sec:data}
We are motivated by the data collected by \citet{chen2023anatomical}, who performed calcium imaging of hippocampal CA1 neurons in mice freely exploring arenas of different shapes (circular, square, and triangular).
The hippocampal CA1 region is essential for memory formation, spatial navigation, and cognitive processing. 
Each  CA1 place cell becomes active when the animal occupies a specific location within its environment, its \emph{place field}. This spatial coding is allocentric, meaning it is defined relative to external environmental cues rather than the animal's own body. The place fields of CA1 neurons are not static; they can shift or ``remap'' in response to changes in the environment or context, allowing the hippocampus to flexibly encode multiple environments \citep{Robinson2020, ChiuENEURO2023}.  Thus, the specific trajectories the animal takes contribute to the formation of highly organized spike-timing relationships among interconnected hippocampal neurons \citep{OKeefe1971hippomap, bookHippoc, bird2008hippocampus, moser2014grid}. These synchronized activation patterns unfold over several seconds and can be systematically matched with the animal’s specific trajectories through the environment.

\begin{figure}[t]
    \centering
    \includegraphics[width=\linewidth]{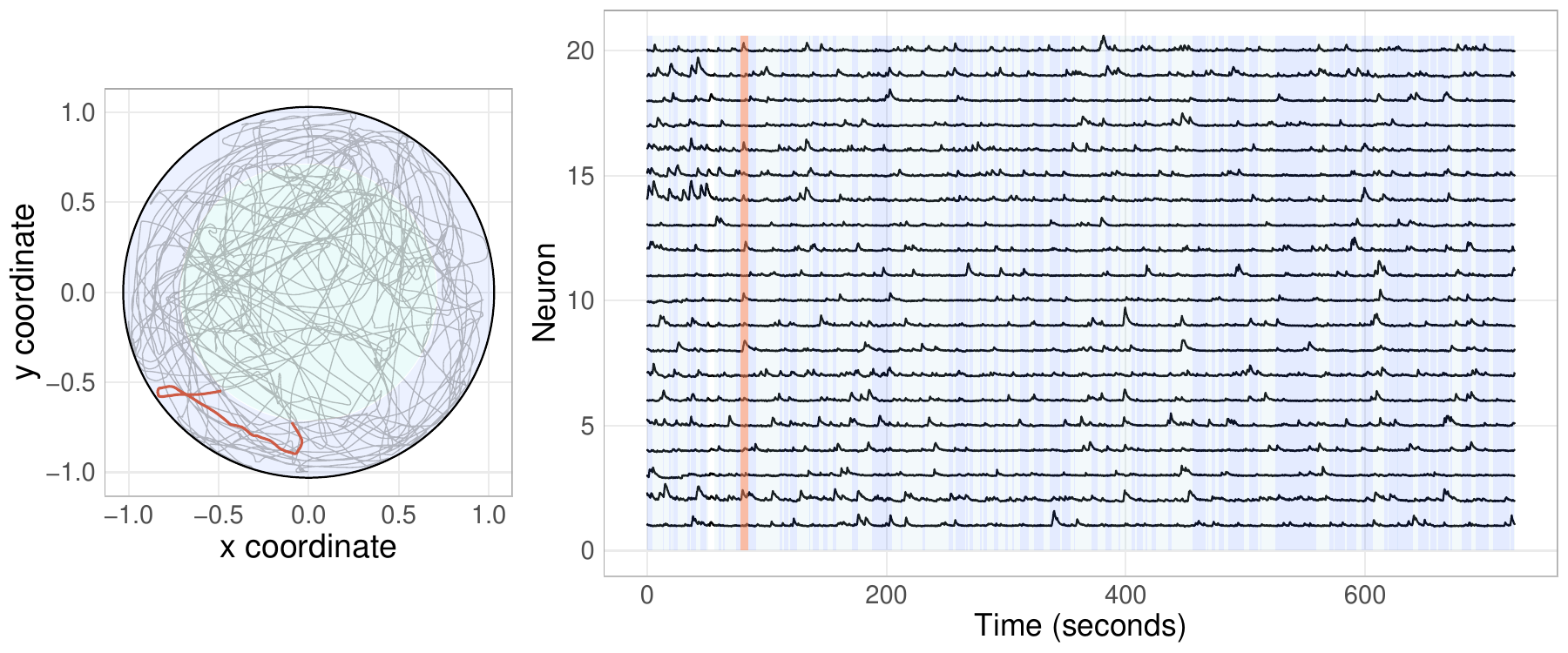}
    \caption{Left panel: mouse movements within the environment (continuous line). The background color corresponds to the two regions of the arena. 
    Right panel: example fluorescence traces of 20 neurons. The background colors correspond to the mouse position in the two regions over time. The orange highlights correspond to a representative window. }
    \label{fig:mouse_position}
\end{figure}

In this paper, we focus on a dataset of calcium imaging recordings from hippocampal CA1 neurons in a mouse exploring a circular arena. Mice were prepared via viral transfection to express a calcium indicator in excitatory CA1 neurons and implanted with a GRIN lens for \textit{in vivo} imaging. Neuronal activity was recorded using head-mounted miniscopes as the animals engaged in various behaviors. Here, we focus on the data collected in a circular environment: the mouse's movements are displayed in the left panel of Figure~\ref{fig:mouse_position} with a continuous line. The arena is divided into two spatial regions of approximately equal area: the center region (the inner circle) and the outer ring (the annular area between the center and the arena boundary). By focusing on these two regions, one can compare the temporal and anatomical organization of co-active neuronal clusters as the mouse moves between the center and the periphery of the arena, thereby revealing how spatial context may shape hippocampal coding. The right panel of Figure~\ref{fig:mouse_position} shows the fluorescence traces of a few representative neurons, with the background color corresponding to the mouse position in the two regions over time. 

\citet{chen2023anatomical} analyzed these data using a two-stage pipeline, where fluorescence traces are first deconvolved to estimate spike trains and then clustered to identify groups of co-active neurons. While this approach provides useful insights, it does not fully account for uncertainty in spike inference nor for the spatial organization of neurons.
To address these challenges, we propose a Bayesian framework that jointly models spike inference and neuronal clustering, while incorporating spatial information and propagating uncertainty throughout the analysis. This enables the identification of functionally coherent neuronal ensembles and their evolution across spatial contexts. To better position our contribution, we next review existing approaches and their limitations.

\subsection{Review of existing approaches and novel contributions}
\label{sec:related_methods}

The standard approach to analyzing calcium imaging data typically involves a two-step pipeline: first, deconvolve the fluorescence traces to estimate the underlying neuronal activity, commonly referred to as the \emph{spike train}; then use these estimated signals for downstream analysis \citep{shen2022}. A number of methods have been developed to efficiently and accurately extract spike trains from calcium imaging data, including those by \citet{vogelstein2010fast, Mishchenko2011,pnevmatikakis2016, deneux2016, jewell2018, jewell2019}. 
The resulting deconvolved traces are then used to address a range of research questions, including estimation of firing rates, tuning curves, receptive fields, and clustering or visualization of neuronal activity \citep{adler2012temporal, humphries2011spike, diana2019bayesian, paiva2010inner}.

Recent studies suggest that neuronal activity is driven by low-dimensional, time-evolving latent dynamics, motivating approaches that focus on population-level structure rather than individual neurons, whose inference can be unreliable due to noise in calcium recordings. Classical methods for such analyses include principal component analysis \citep[PCA; ][]{churchland2012neural, Levi9807} and locally linear embedding \citep{STOPFER2003991,BROOME2006467}. 
More recently, latent variable models have provided more flexible frameworks for characterizing these dynamics. In particular, Gaussian process factor analysis \citep[GPFA;][]{Yu2008, Duncker2018, Jensen2021} represents neural activity through low-dimensional latent factors with Gaussian process priors to induce temporal smoothness. Deep learning approaches have also been adapted to calcium imaging data; for example, \cite{Radical} introduce a Recurrent Autoencoder for Discovering Imaged Calcium Latents (RADICaL) extending \cite{LFADS} via a recurrent autoencoder that models deconvolved signals and infers smooth, low-dimensional neural trajectories.

While intuitive and modular, the two-step pipeline has important limitations and, in the presence of noise, can be severely biased \citep{ganmor2016}. 
It treats deconvolution as a preprocessing step, ignoring uncertainty in the inferred spike trains, and often relies on generic analysis techniques that do not incorporate external covariates, such as the spatial organization of neurons, or the noisy, probabilistic nature of firing. 

More recently, integrated methods that perform both deconvolution and downstream analysis within a unified framework have been proposed to improve accuracy and reduce error propagation. 
\citet{shen2024time} applied such an approach to estimate firing rates in longitudinal studies, while \citet{dangeloetal2023biometrics} analyzed the heterogeneity of neuronal responses to varying experimental conditions. 
However, these methods analyze only a single neuron at a time, making them unable to investigate the spatial organization of neurons in the brain and precluding effective discovery of potential neuronal cluster activity. 

%these methods analyze only a single neuron at a time, making them unable to account for the spatial organization of neurons in the brain and precluding effective analysis of neuronal cluster activity. 
 
Unified, multivariate analyses have been proposed by \cite{Rupasinghe} to estimate signal and noise correlations from two-photon fluorescence observations, and by \cite{Prince2021} to infer a hierarchical structure of latent dynamics. Finally, \cite{Keeley2026} extended hidden Markov models, GPFA, and dynamical systems models originally devised for spike trains to calcium imaging data, by integrating a likelihood that accounts for the spiking dynamics and autoregressive behavior of fluorescence traces. However, none of these approaches allows external information (e.g., the anatomical organization of cells) to be introduced in the learning process. 

In this paper, we propose a fully Bayesian, semiparametric model that performs both tasks, deconvolution and clustering, within a unified framework. Our model builds on the biophysical calcium dynamics model of \citet{vogelstein2010fast}, but extends it to incorporate a novel latent-variable framework for clustering neurons based on their spatiotemporal activation patterns. 
Our approach simultaneously deconvolves calcium traces and identifies spatially organized neuronal ensembles, even when firing is asynchronous or partially masked by noise. We achieve this by modeling neuron-specific spike probability trajectories using latent Gaussian processes (GPs) and incorporating anatomical information via a spatially informed nonparametric mixture prior that encourages anatomically nearby neurons to be grouped together. 
While the proposed formulation falls within the class of latent variable models, it differs from existing approaches in two key aspects. First, it provides a representation that is directly interpretable in biological terms: the inferred partition corresponds to functionally organized neuronal ensembles, while the cluster-specific latent trajectories describe their temporal firing dynamics. In contrast, methods based on PCA or generic latent factors rely on geometric projections that are often difficult to interpret, and deep learning approaches typically offer limited insight into the inferred structure. Second, our framework explicitly addresses the limitations of existing approaches by jointly modeling deconvolution, clustering, and spatial structure, while propagating uncertainty throughout the analysis in a coherent way. The fully Bayesian formulation provides a principled theoretical framework for constructing flexible and comprehensive probabilistic representations of the phenomenon under study. Nonparametric priors are used to avoid rigid distributional assumptions and to integrate the functional and anatomical structure of the analyzed neural population.
Overall, the proposed approach provides a unified and interpretable framework for studying neuronal population dynamics, enabling the identification of functionally coherent ensembles and the study of
their temporal and anatomical organization from calcium imaging data.
%Overall, the proposed approach provides a unified and interpretable framework for studying neuronal population dynamics, enabling the identification of spatially structured ensembles and their temporal evolution from calcium imaging data. 

The remainder of this paper is organized as follows. Section~\ref{sec:model} introduces the Bayesian semiparametric model for joint deconvolution and clustering and discusses posterior inference. Section~\ref{sec:application} details results from our hippocampal data analysis, highlighting the method's ability to compare neuronal ensemble structures across distinct mouse positions.  Additional analyses of the hippocampal data and simulation studies are reported in the Supplementary Materials. Finally, some concluding remarks are provided in Section~\ref{sec:Discussion}.

\section{Model for deconvolution and clustering}\label{sec:model}

\begin{figure}[t]
    \centering
    \includegraphics[width=\linewidth]{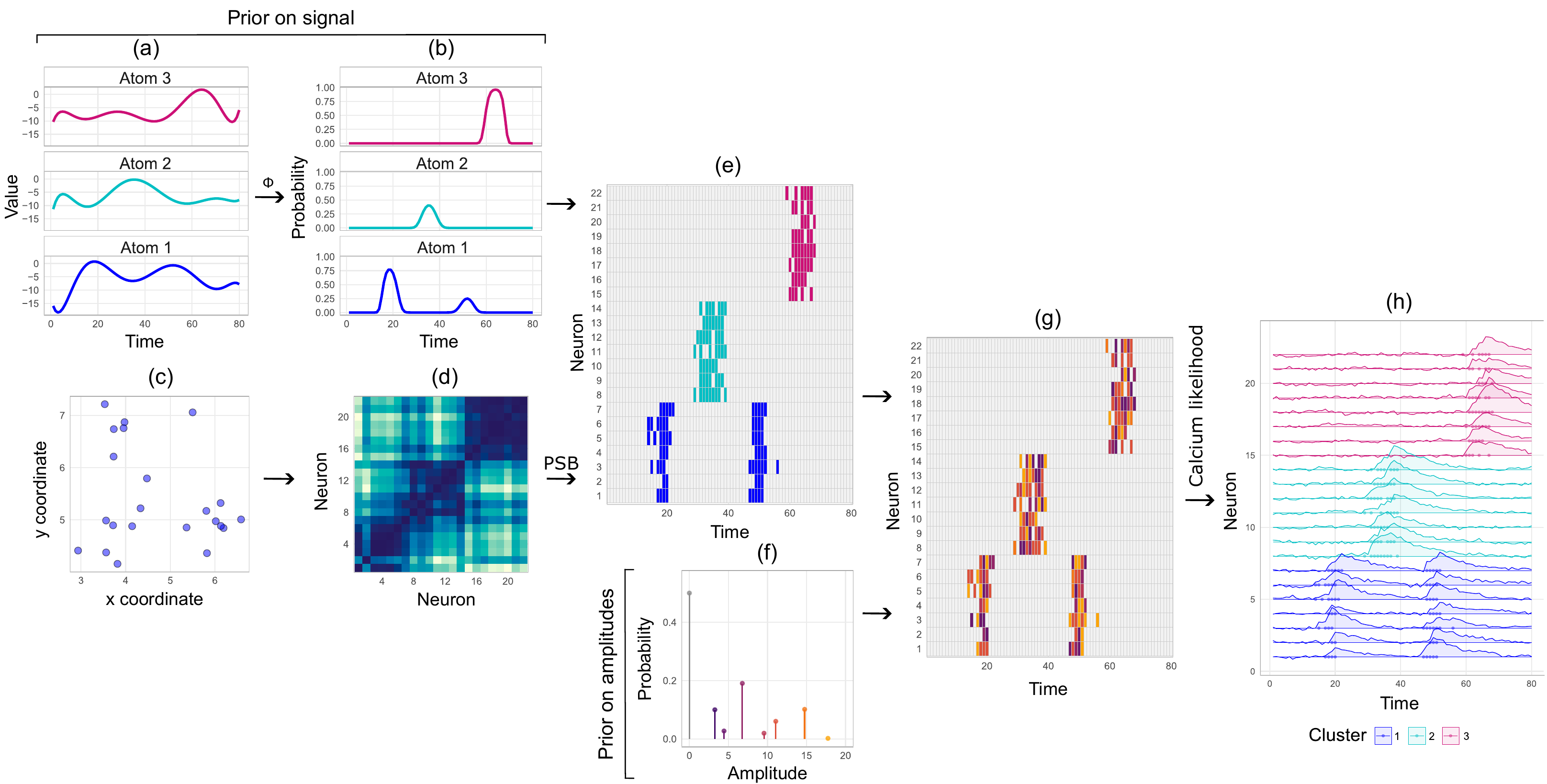}
    \caption{Representation of the model. The top-left panels (a) show the mixture atoms of the nonparametric prior on the signal. These are realizations of GPs, which are transformed via the probit function ($\Phi$) into probability trajectories that model the temporal dynamics underlying the activations (b). The second row shows how the mixture weights are constructed via the PSBP, starting from the information about the neurons' locations (c), summarized by the proximity matrix (d). These two components are then combined to allocate each neuron to an activation cluster, which underlies the observed synchronized firing patterns (e). Panel (f) instead shows the nonparametric prior on the amplitudes, a mixture of a DP and a Dirac mass at zero. The atoms of this distribution model the height of each spike, accounting for their heterogeneity (g). Finally, the obtained signal is convolved with the noisy autoregressive calcium likelihood to obtain the observed fluorescence traces (h). }
    \label{fig:graphical_rep}
\end{figure}

In this section, we formulate a Bayesian semiparametric model to simultaneously deconvolve calcium traces and estimate clusters of neurons with similar temporal activity patterns.  Figure~\ref{fig:graphical_rep} illustrates the model, providing an intuitive view of the prior distribution and how the various components, introduced in the following, interact.

\subsection{Calcium concentration model}
To model the fluorescence traces and infer the underlying spiking activity, we employ a popular biophysical parametric model originally proposed by \citet{vogelstein2010fast}.  This model captures the physiological mechanisms underlying the imaging technology through a system of two equations. The first relates the observed fluorescence to the calcium concentration, the second describes the calcium dynamics and how it responds to firing events.
Specifically, let $y_{i,t}$ denote the observed calcium level of neuron $i$ at time $t$, for $i=1,\dots,n$ and $t=1,\dots,T$, and let $c_{i,t}$ be the true calcium concentration, cleaned of the measurement error. \citet{vogelstein2010fast} model assumes 
\begin{equation}
y_{i,t} = b_i + c_{i,t} + \epsilon_{i,t}\,,\qquad
c_{i,t} = \gamma\, c_{i,t-1} + s_{i,t}\cdot a_{i,t} + \eta_{i,t}
\label{eq:calcium_dyn}
\end{equation}
where $b_i$ is a baseline term that centers the resting calcium concentration at zero, while 
$\epsilon_{i,t} \sim \N(0,\sigma^2)$ and $\eta_{i,t} \sim \N(0, \tau^2)$ are random Gaussian noises. The second equation describes how firing activity influences calcium levels. The binary variable $s_{i,t}\in\{0,1\}$ indicates the presence or absence of a spike for neuron $i$ at time $t$, while  $a_{i,t}\in(0,\infty)$ describes its amplitude, when $s_{i,t} = 1.$
In other words, when $s_{i,t} = 0$, the calcium dynamics follow an autoregressive process of order one, where the decay is governed by the parameter $\gamma$. If neuron $i$ fires at time $t$, $s_{i,t} = 1$, and the calcium concentration increases instantaneously with a spike amplitude $a_{i,t}>0$. 

We assume independent conjugate priors for the parameters that control the calcium dynamics: $c_{i,0} \sim \N(0,C_0)$, $b_i\sim \N(b_0,B_0)$, $1/\sigma^2 \sim \mathrm{Gamma}(\alpha_{\sigma},\beta_{\sigma})$, and $1/\tau^2 \sim \mathrm{Gamma}(\alpha_{\tau},\beta_{\tau})$. 
Under the assumption that the process is stationary with a positive correlation between the calcium level at consecutive times, we constrain $\gamma \in (0,1)$  assuming $\gamma\sim \mathrm{Beta}(\alpha_{\gamma}, \beta_{\gamma})$.

\subsection{Calcium activation model and clustering}

The binary activation state indicators, $s_{i,t}$, serve as the key parameters for characterizing neuronal activity over time. 
For each neuron $i=1,\dots,n$, the sequence $\bm{s}_i = (s_{i,1},\dots,s_{i,T})$ of these binary variables represents the spike train, which encodes discrete, event-based activity.
Crucially, identifying shared patterns among the activation time series $(\bm{s}_1,\dots,\bm{s}_n)$ allows us to detect functionally coupled neuron clusters, i.e., groups of neurons that exhibit synchronized firing. 
A major challenge in identifying co-activating neurons, however, is that firing events are often not strictly synchronous or temporally overlapping~\citep{chen2023anatomical}. Additionally, some observed spikes may result from measurement errors or occur randomly, making them irrelevant for cluster identification. 
Consistently, we would like these clusters to comprise all neurons with a similar activation pattern, even if the series might not be perfectly aligned or might differ for some occasional or isolated spikes.

To this end, we introduce an underlying neuron-specific process that describes the temporal trajectory of spike probability, and we perform clustering at this latent level. The key idea is that a similar evolution of the spike probability over time will produce similar spike trains.
We assume that for neuron $i$, for $t=1,\dots,T$, each $s_{i,t}$ is a realization of a conditionally independent Bernoulli random variable, whose probability depends on an underlying real-valued process $\tilde{\bm{s}}_i=(\tilde{s}_{i,1},\dots,\tilde{s}_{i,T})$, i.e.,
\begin{equation*}
s_{i,t}\mid \tilde{s}_{i,t} \overset{ind.}{\sim} \mathrm{Bernoulli}\left(\Phi(\tilde{s}_{i,t} )\right),
\label{eq::s_tildes}
\end{equation*}
where $\Phi(\cdot)$ is the cumulative distribution function of a standard Gaussian distribution.
Hence, the transformed process $\left(\Phi(\tilde{s}_{i,1}),\dots,\Phi(\tilde{s}_{i,T})\right)$ describes the temporal evolution of the spike probability of neuron $i$, for $i=1,\dots,n$. 

%We model the latent processes $\tilde{\bm{s}}_i$ using a discrete prior distribution to induce the desired partition of coactive neurons. 
We place a discrete prior on the latent processes $\tilde{\bm{s}}_i$, thereby grouping neurons that share a common latent spike-probability trajectory. Local anatomical organization in CA1 has been reported in some experimental settings \citep{chen2023anatomical,dombeck2010,Modi2014,Jimenez2020, WirtshafterDisterhoft2023}, whereas other studies have found no comparable pattern \citep{Taxidis2020,Slettmoen2026}. These differences may reflect the behavioral setting and the feature used to define neuronal similarity. We therefore allow neuronal location to inform the mixture weights without deterministically defining the partition.
Specifically, for $i=1,\dots,n$, we define
\begin{equation}
\tilde{\bm{s}}_i\mid \bm{\ell}_i,G_{\bm{\ell}_i}  \sim G_{\bm{\ell}_i},\quad\quad
G_{\bm{\ell}_i} = \sum_{k=1}^{\infty} \pi_k(\bm{\ell}_i)\cdot \delta_{\tilde{\bm{s}}^*_k },\quad
\quad\tilde{\bm{s}}^*_k \overset{iid}{\sim} \mathrm{GP}({\mu}_{\tilde{s}},\Omega).
\label{eq:mixture_psbp}
\end{equation}
The atoms $\tilde{\bm{s}}^*_k$ in Equation~\eqref{eq:mixture_psbp} are the unique values each latent process $\tilde{\bm{s}}_i$ can assume and thus the clusters' representative values.  We assume they are independent draws from a GP with mean $\mu_{\tilde{s}}$ and covariance function $\Omega(t,t')$, for $t,t'=1,\dots,T$, modeling the temporal dependence among spikes. 
As noted by several authors \citep[see, e.g.,][]{dombeck2010,dangeloetal2023biometrics}, neuronal spikes are not uniformly distributed over time. The prolonged duration of a calcium transient frequently reflects the summation of multiple spikes occurring in rapid succession. However, the slow decay of calcium fluorescence can hide the individual firing events. Explicitly modeling this temporal dependence can enhance spike detection and improve the identification of periods of neural activity.
The mixture weights defining $G_{\bm{\ell}_i}$ in Equation~\eqref{eq:mixture_psbp} are constructed using the probit stick-breaking process (PSBP) of \citet{rodriguez2011}. This instance of dependent Dirichlet process~\citep{Quintana2022} defines a collection of random distributions that become increasingly similar for spatially closer neurons. As a result, \textit{a priori}, neighboring neurons are more likely to be assigned to the same atoms. 
Specifically, we define $\Sigma$ as the $n\times n$ positive-definite matrix with entries $\Sigma(\bm{\ell}_i,\bm{\ell}_{i'}) = \exp\{-\theta \lVert \bm{\ell}_{i}-\bm{\ell}_{i'} \rVert^2 \}$, for $\theta>0$ and $i, i' =1,\dots,n$. Hence, $\Sigma$ measures the proximity between pairs of neurons. 
The probability of assigning neuron $i$ to cluster $k$, $\pi_k(\bm{\ell}_i)$, is defined using the location-dependent PSBP as 
\begin{equation*}
\pi_k(\bm{\ell}_i) = \Phi\big(\alpha_k(\bm{\ell}_i)\big) \prod_{r<k} \big\{ 1- \Phi\big(\alpha_r(\bm{\ell}_i)\big)\big\}
\end{equation*}
for $i=1,\dots,n$ and $k\geq 1$.
Here, $(\alpha_k(\bm{\ell}_1), \dots, \alpha_k(\bm{\ell}_n) )^T \sim \N_n \left(\bm{0},\, \Sigma \right)$, for $k\geq 1$, is a latent process modeling spatial dependence: for two neurons located in the same hippocampal region, $\Sigma(\bm{\ell}_i,\bm{\ell}_{i'})$ will be large, inducing a positive correlation in the corresponding components $\alpha_k(\bm{\ell}_i)$ and $\alpha_k(\bm{\ell}_i')$, and, consequently, on the mixing probabilities.

An intuitive formulation is obtained by introducing the cluster allocation variables $\zeta_i \in\{1,2,\dots\}$, such that $\zeta_i=k$ if neuron $i$ is assigned to the $k$-th activation cluster, with prior distribution $\mathbb{P}(\zeta_i=k\mid G_{\bm{\ell}_i}) = \pi_k(\bm{\ell}_i)$, for $k\geq 1$ and $i=1,\dots,n$. Therefore, all neurons characterized by $\zeta_i = k$ will be associated with the mixture atom $\tilde{\bm{s}}^*_{k}$, and thus with the same spike probabilities. 

To summarize, the discrete random measure in Equation~\eqref{eq:mixture_psbp} allows detecting neuronal ensembles characterized by similar, albeit not identical, spike trains, all driven by the same spike probability function $\Phi(\tilde{\bm{s}}^*_k)$, for each cluster $k\geq 1$. Additionally, the GP prior on the atoms of the discrete probability measures allows taking into account the temporal dependence among firing events. At the same time, the PSBP introduces a prior preference for anatomically proximal neurons to co-cluster.

\subsection{Spike amplitude model}

Once the prior for spatiotemporal clustering of the spike trains is defined, we are left only with the
specification of the distribution on the spike amplitudes $a_{i,t}$, for $i=1,\dots,n$ and $t=1,\dots,T$. 
We assume an outer spike-and-slab prior \citep{canale2017pitman}: when $s_{i,t}=0$, $a_{i,t}$ has a point mass at zero; when $s_{i,t}=1$, the distribution of $a_{i,t}$ is a discrete distribution $P$ following a Dirichlet process (DP) prior, i.e. 
\begin{equation}
    \{a_{i,t}\mid s_{i,t},P  \}\sim (1-s_{i,t})\delta_0+s_{i,t}P, \quad P\sim \mathrm{DP}(\alpha, P_0)
    \label{eq:prior_a}
\end{equation}
with $\alpha>0$ the concentration parameter, and $P_0$ the base measure. We assume for $P_0$ a shifted-gamma distribution, defined as the distribution of the random variable $a^*$ such that $a^*- \bar{a} \sim\mathrm{Gamma}(\alpha_{a}, \beta_{a})$. The presence of the parameter $\bar{a}>0$ allows discarding negligible amplitudes that would likely introduce noise in the signal.

The distribution of $a_{i,t}$ depends on the binary activation status $s_{i,t}$ and hence, indirectly, on the clustering of neurons and on the latent signal.
To clarify this dependence, it is useful to introduce the cluster allocation variables $\xi_{i,t}$, for $i=1,\dots,n$ and $t=1,\dots,T$, with values in $\{0,1,2,\dots\}$, where 0 represents the absence of a spike. These variables are defined so that $\{a_{i,t}\mid \xi_{i,t}=h\} = a^*_h$, with $a^*_0 = 0$ and $\{a^*_h\}_{h\geq 1}$ the atoms of the DP $P$ in Equation~\eqref{eq:prior_a}. The prior distribution of these allocation variables is categorical and depends on the cluster allocation $\zeta_i$ of neuron $i$, and on the corresponding latent series $\tilde{s}^*_{\zeta_i}$. Specifically, the distribution is $\mathbb{P}(\xi_{i,t} = h \mid \tilde{s}^*_{k,t}, P) = \omega^*_{k,t,h}$
with
\begin{gather*}
 \omega^*_{k,t,0} = \Phi\big( -\tilde{s}^*_{\zeta_i,t} \big),\quad\quad
 \omega^*_{k,t,h} = \Phi\big( \tilde{s}^*_{\zeta_i,t} \big) \,v_h \prod_{r<h} (1-v_r), \quad h\geq 1,
\end{gather*}
where the $\{v_h\}_{h\geq 1}$ are $\text{Beta}(1,\alpha)$ random variables, with $\alpha>0$, involved in the stick-breaking construction of the weights of the DP $P$; and $\tilde{s}^*_{\zeta_i,t}$ is the value of the realization at time $t$ of the latent GP associated with neuron $i$.

\subsection{Posterior inference}\label{sec::posterior_inference}

We perform posterior inference via Markov Chain Monte Carlo (MCMC), implementing a Gibbs sampler that leverages closed-form full conditional distributions for most model parameters.
The formulation in Equation~\eqref{eq:calcium_dyn} implies that the observations $y_{i,t}$ are conditionally independent given $c_{i,t}$, for $t=1,\dots,T$ and $i=1,\dots,n$. Hence, the distribution of an individual fluorescence trace $\bm{y}_i$ can be written as the product of $T$ Gaussian densities:
\begin{equation*}
f(\bm{y}_i\mid b_i, \bm{c}_i,\gamma,\bm{s}_i,\bm{a}_i,\sigma^2,\tau^2) = \prod_{t=1}^T \phi(y_{i,t}\mid\, b_i + \gamma\, c_{i,t-1} + s_{i,t}\cdot a_{i,t}, \: \sigma^2+\tau^2).
\label{eq:lik}
\end{equation*}
The steps for sampling $b_i$, $c_{i,t}$, $\gamma$, $\sigma^2$, and $\tau^2$ can be easily adapted from the updates outlined in Section~2.2 of \citet{dangeloetal2023biometrics}. 

Regarding the clustering of neurons, the update of the variables $\zeta_i$ is performed using the data augmentation strategy outlined in the original paper of \citet{rodriguez2011}. For simplicity, we used a truncated PSBP with a large number of components, as it constitutes a fair approximation of the original process \citep{rodriguez2011, ishwaran2001}. 
The full conditional distribution of the latent GP $\tilde{\bm{s}}^*_k$, for $k\geq 1$, depends on all the spike trains of neurons allocated to the $k$-th cluster, i.e., $\{ \bm{s}_i:\zeta_i=k, i=1,\dots,n \}$. We made use of the data augmentation scheme of \citet{albertchib} to relate the latent $\tilde{\bm{s}}^*_k$ with the Bernoulli variables.

Finally, to sample from the posterior distribution of the outer spike-and-slab DP prior on the amplitudes in Equation~\eqref{eq:prior_a}, we used a blocked Gibbs sampler. Because of the lack of conjugacy of the shifted-gamma prior on the amplitude values, we resorted to a Metropolis-within-Gibbs step. 
Details are reported in Section~\ref{Asec:posterior_inference} of the Supplementary Material. Code to fully replicate the analyses is available in a dedicated 
GitHub repository\footnote{\scriptsize{\url{https://github.com/laura-dangelo/Bayesian_Clustering_of_Spatially-Referenced_Calcium_Traces}}}.

Once the Markov chains for all model parameters and augmented variables have been obtained, to derive relevant inferences about the clustering structure, targeted post-processing procedures are needed. In particular, it is often of interest to identify a unique representative point estimate of the neurons' partition. This can be obtained from sampled cluster membership values $\zeta_i$, for $i=1,\dots,n$, across the iterations of the algorithm. In our analysis, activation clusters were defined as the optimal partition obtained by minimizing the variation of information loss, following the approach of \cite{wade2018} as implemented in the \texttt{salso} package \citep{salso, Dahl2022}. This approach also allows us to infer the number of clusters, as the cardinality of the estimated partition. Moreover, because the partition is sampled at each iteration of the algorithm, inference is not limited to a point estimate of the number of clusters; we can also compute its variance, thereby providing a measure of uncertainty.

\section{Hippocampal CA1 neuronal ensemble data analysis}\label{sec:application}

We applied our method to the calcium imaging dataset from \citet{chen2023anatomical}, which recorded neuronal activity in a mouse freely exploring a circular arena. The proposed approach is employed to investigate the spatial organization of hippocampal ensembles under real-world conditions. The activity of 325 neurons was recorded simultaneously over a 12-minute period.
The frame rate of the acquired calcium traces is 15 frames per second, resulting in $10{,}869$ time points. 
Consistent with the theory of position-dependent neuronal ensemble activation, we segmented the experiment into time windows based on the animal’s spatial coordinates. Previous works \citep{oneill,Ainge} have demonstrated that neuronal ensemble membership and activity patterns systematically vary with the mouse’s trajectory. We therefore divided the full experiment duration into brief intervals where the animal’s position remained approximately constant—a critical step for identifying meaningful activation clusters.
To this end, we first identified two distinct regions of approximately equal area: the center and the outer ring of the environment, highlighted with distinct colors in Figure~\ref{fig:mouse_position}. Then, we divided the experiment's total duration into time windows based on the mouse's position. This way, we expect firing patterns to be approximately stable within each window and to differ across windows. The duration of these windows ranges from six to 256 time points. 
Since the objective of the analysis is to identify co-activating neurons, we focused on windows with a duration of at least three seconds \cite[in line with the indications of ][]{chen2023anatomical}, corresponding to 45 time points.
As an illustration, in what follows, we analyze one of these windows, namely the one highlighted in orange in Figure~\ref{fig:mouse_position}. The analyses of additional windows are reported in Section~\ref{Asec:additional_windows} of the Supplementary Material.

We ran the algorithm outlined in Section~\ref{sec::posterior_inference} for $30{,}000$ iterations, discarding the first $25{,}000$ as burn-in. 
In the absence of precise knowledge about the spikes' amplitudes, we considered three different values for the shift parameter $\bar{a}$, namely $\{0, 0.5, 1\}$. 
To evaluate the robustness of our estimates, we performed a sensitivity analysis. The results showed minimal variation across parameter choices, so we focus here on those obtained with $\bar{a} = 1$. The sensitivity analysis is reported in Section~\ref{Asec:read_sensitivity} of the Supplementary Material.

\subsection{Analysis of activation clusters in a representative window}\label{sec:representative_win}
The considered window spans approximately seven seconds (102 time points) and comprises neuronal activity recorded while the mouse is near the boundary of the arena (see Figure~\ref{fig:mouse_position}). 
Our fully Bayesian analysis provides posterior point estimates of neuronal partitions along with their uncertainties.
Activation clusters were identified based on the value of cluster allocation $\zeta_i$, for $i=1,\dots,n$, throughout the MCMC iterations by minimizing the variation of information. 
This procedure yields nine activation clusters and nine singletons (neurons with no synchronized activity, i.e., outliers). Remarkably, one cluster comprises all neurons with no relevant activity (Cluster 1, 167 neurons). The remaining neurons were grouped into five large clusters, each containing between 18 and 32 neurons, and four smaller clusters, each comprising between three and 12 neurons.

\begin{figure}[t]
    \centering    \includegraphics[width=.8\linewidth]{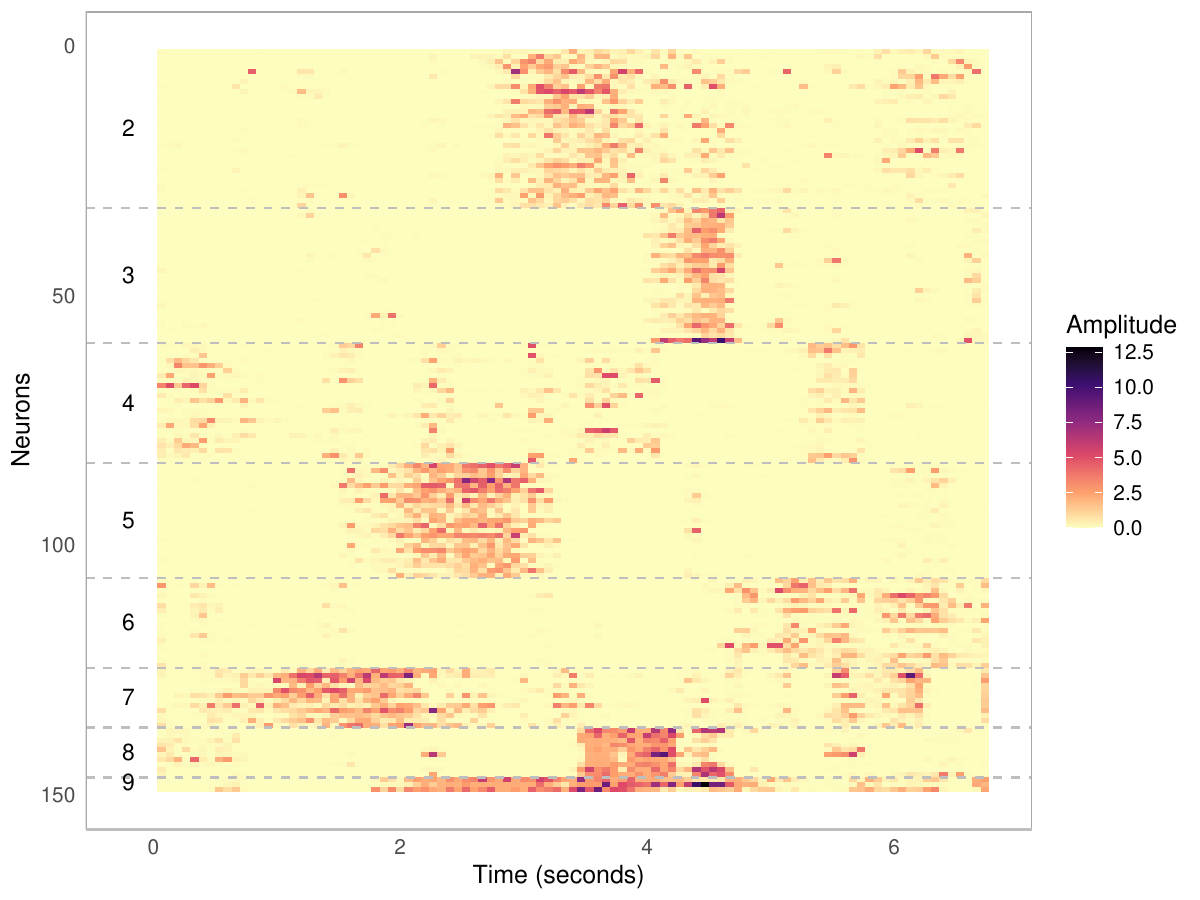}
    \caption{Estimated spike amplitudes over time for active neurons. The traces are sorted according to the neurons' cluster allocation (Clusters 2 to 9, left labels).}
    \label{fig:AA_by_cluster}
\end{figure}

Figure~\ref{fig:AA_by_cluster} shows the estimated spike amplitudes during the considered time window for the active synchronized neurons (i.e., Clusters 2 to 9), sorted by their cluster assignment. The plot clearly displays resting periods, characterized by $a_{i,t} = 0$, and periods of activity. 
Notably, the identified activation clusters consistently group neurons with synchronized firing patterns, even when their spike amplitudes exhibit substantial variability.
\begin{figure}
    \centering
    \includegraphics[width=\linewidth]{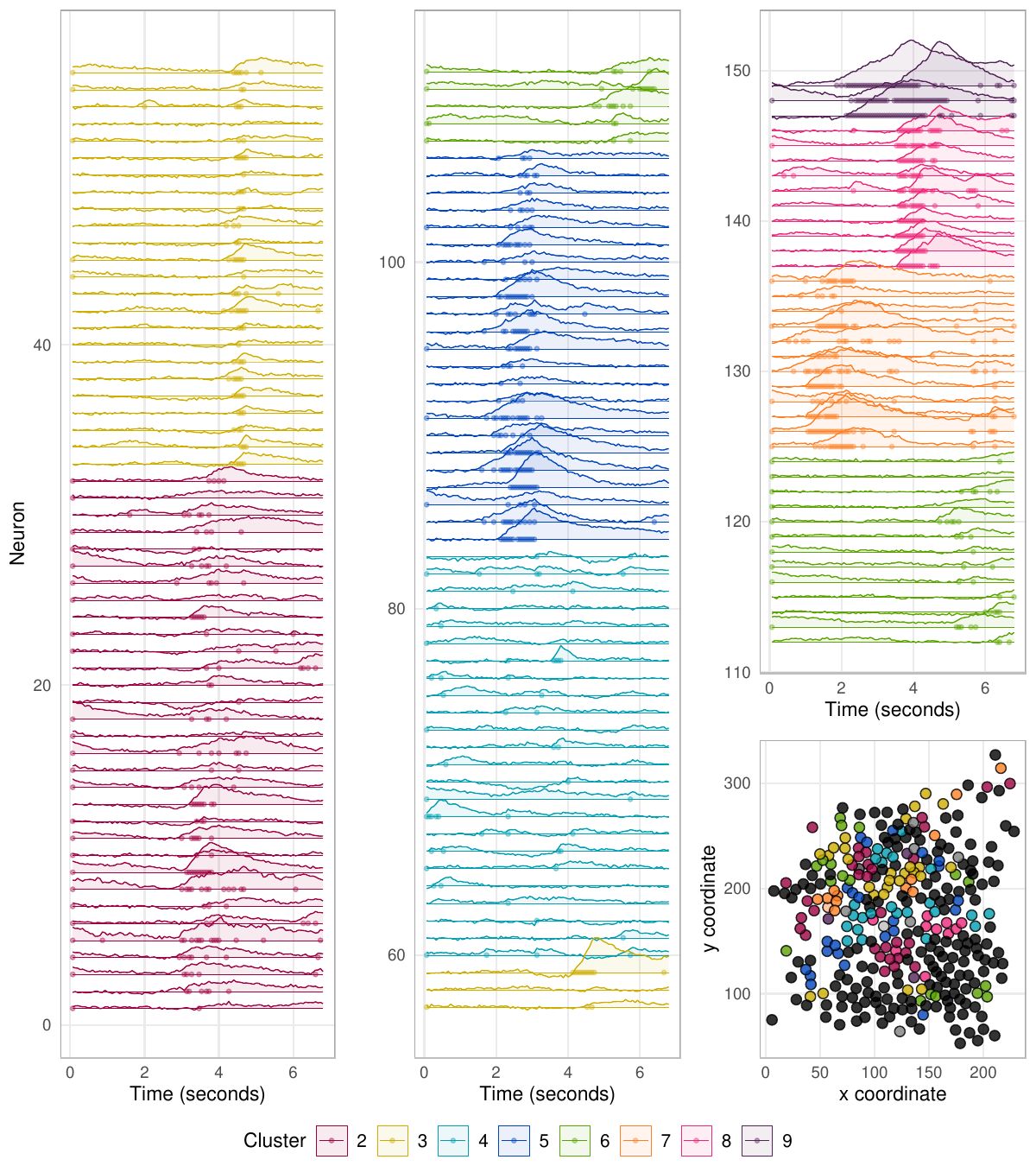}
    \caption{Left, center, and top-right panels: observed calcium traces sorted (and colored) by estimated cluster allocation (only active neurons). Points correspond to the times of the detected firing events. Bottom-right panel: neurons' location in the hippocampus, colored according to the estimated cluster allocation.}
    \label{fig:clusters_and_neurons}
\end{figure}
Figure~\ref{fig:clusters_and_neurons} shows the observed fluorescence traces for the same subset of active neurons, again sorted by their cluster allocation. Here, the noise in the series complicates the identification of spikes; however, it is still possible to observe a common trend across series within the same activation cluster. The bottom-right panel shows the anatomical distribution of these clusters. Co-activating neurons appear to form irregularly shaped anatomical patches, most clearly for Clusters~2 and~3. This pattern is qualitatively consistent with \citet{chen2023anatomical}. More broadly, the identification of co-activating groups is in accord with prior work suggesting that the animal's position is encoded not only by individual place cells, but also through coordinated activity across neuronal populations \citep{OKeefe1971hippomap,WilsonMcNaughton1993,HarrisEtAl2003}. In this sense, the localized clusters in Figure~\ref{fig:clusters_and_neurons} can be naturally interpreted as local functional ensembles within CA1.

\begin{figure}
    \centering
    \includegraphics[width=\linewidth]{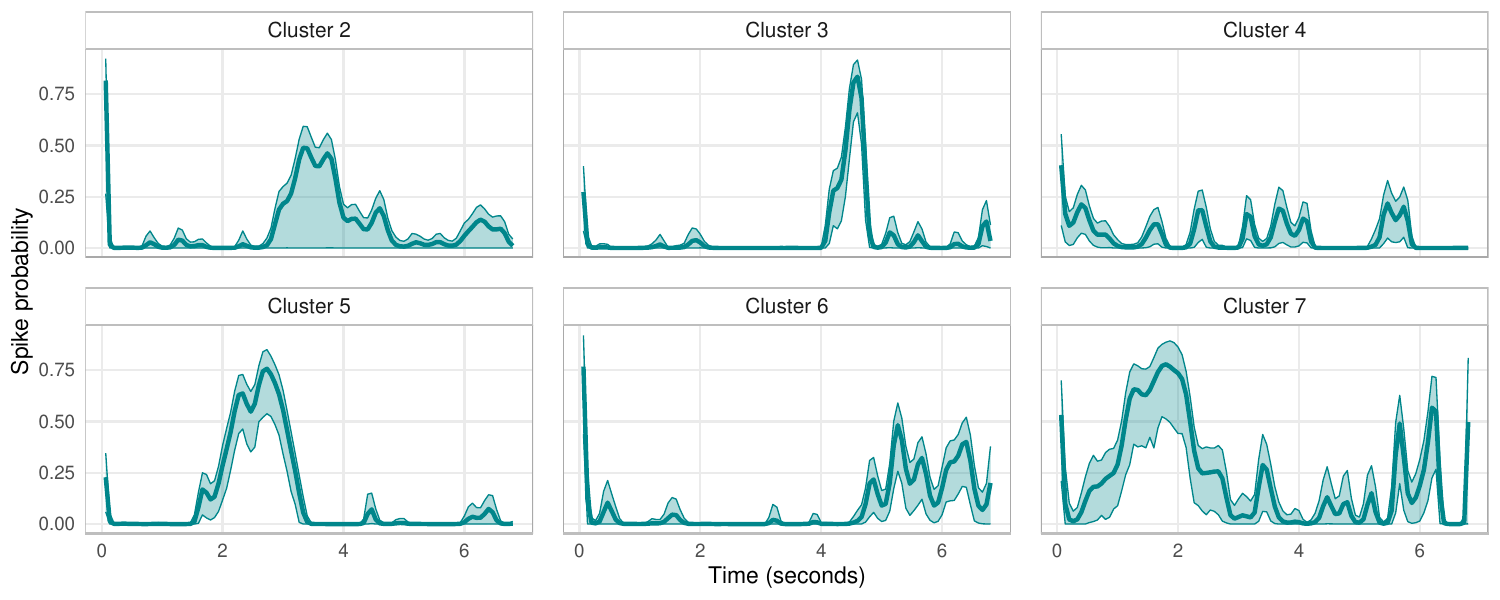}
    \caption{Posterior estimates of the underlying spike probability trajectories for the activation clusters with at least ten neurons. Shaded bands correspond to the point-wise 95\% HPD credible intervals.}
    \label{fig:GP_win24}
\end{figure}

In our model, neurons within the same activation cluster share a common spike-probability trajectory, making these latent functions of direct interest. Figure~\ref{fig:GP_win24} displays the cluster-specific spike probabilities over time. Solid lines denote posterior point estimates at each time point, and shaded bands represent 95\% point-wise highest-posterior-density (HPD) credible intervals. These trajectories are particularly informative when compared with the detected spikes in Figure~\ref{fig:AA_by_cluster} and the observed calcium traces in Figure~\ref{fig:clusters_and_neurons}. As an illustration, Clusters~2 and~5 both show activation around the middle of the time window (approximately at second 3);  however, neurons assigned to Cluster~5 exhibit a consistently higher spike probability. This pattern is consistent with the calcium traces in Figure~\ref{fig:clusters_and_neurons}, which show a steady and sustained increase for neurons in Cluster~5, whereas the increase is more subtle and gradual for neurons in Cluster~2. Similarly, Figure~\ref{fig:AA_by_cluster} reveals a sharp activation interval for Cluster~5, and a more nuanced pattern for neurons in Cluster~2. Taken together, the latent spike probability trajectories provide an informative and interpretable summary of the latent neuronal ensembles underlying the mouse behavior.

\begin{figure}[t]
    \centering
    \includegraphics[width=.5\linewidth]{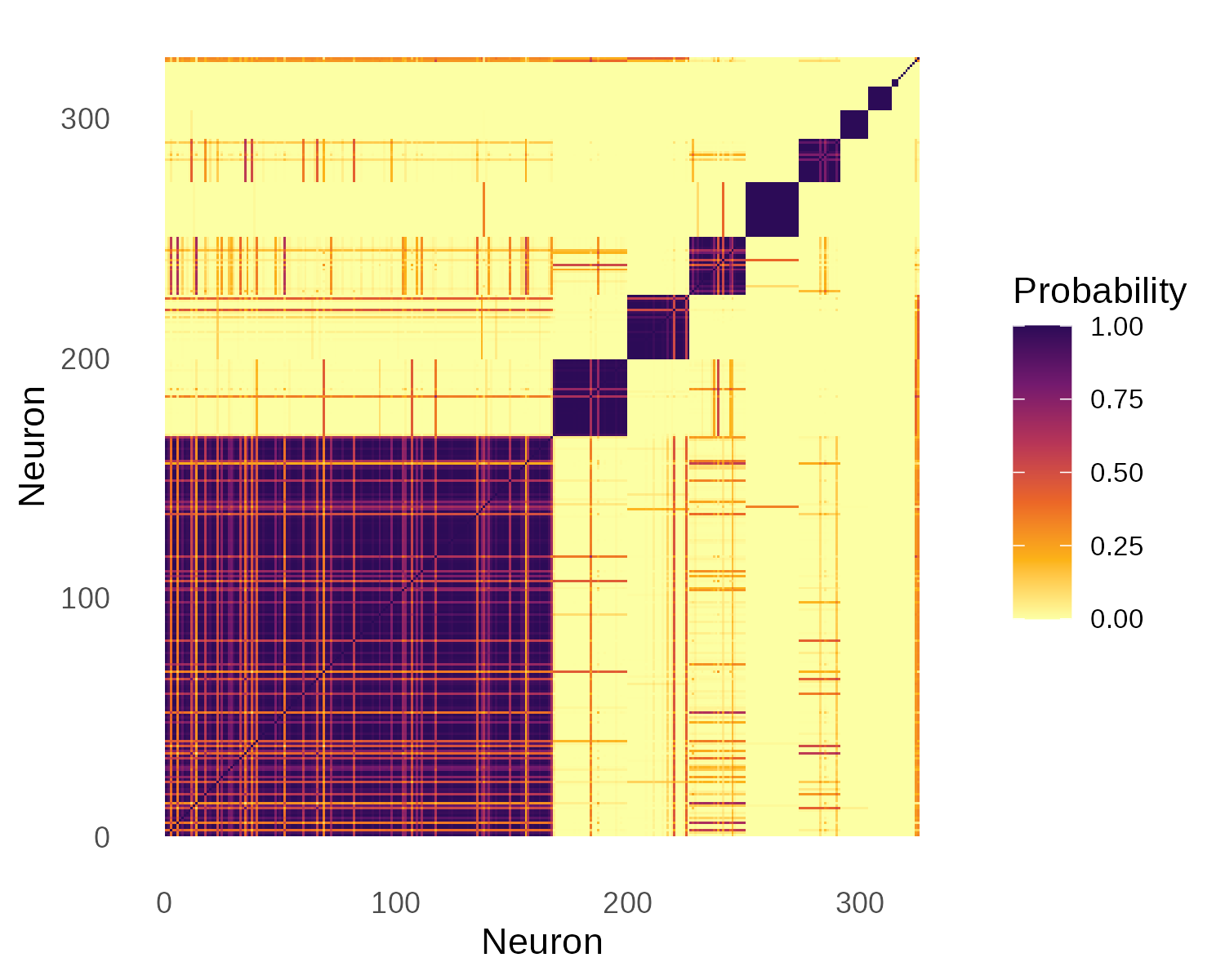}
    \caption{Posterior similarity matrix of the neurons' activity. The color of each cell corresponds to the co-clustering probability of the corresponding pair of neurons.}
    \label{fig:PSM}
\end{figure}

The Bayesian framework provides a direct way to quantify uncertainty in the estimated clustering. Specifically, the partitions sampled during the MCMC yield co-clustering posterior probabilities, computed as the proportion of iterations in which each pair of neurons is assigned to the same cluster. Figure~\ref{fig:PSM} shows the posterior similarity matrix for the analyzed window.  Rows and columns correspond to individual neurons, and the color of each cell is proportional to the co-clustering posterior probability for the corresponding pair. Neurons are ordered according to the estimated cluster allocation, so that each dark block along the diagonal corresponds to a cluster. The clear separation between blocks indicates that the activation clusters are well separated: neurons within the same cluster show relatively homogeneous firing activity, distinct from that of neurons in other clusters.  The first and largest cluster consists of inactive neurons. Most of the allocation uncertainty is concentrated in Clusters~1 and~2. As shown in Figure~\ref{fig:clusters_and_neurons}, trajectories of neurons in Cluster~2 exhibit more scattered spikes and smaller amplitudes,  making some of them similar to those of inactive neurons. This is consistent with the spike-probability trajectory in Figure~\ref{fig:GP_win24}, whose credible intervals during the activation period include zero.

\subsection{Neuronal response to the mouse position}\label{subsec:response_position}
\begin{figure}[t]
    \centering
    \includegraphics[width=\linewidth]{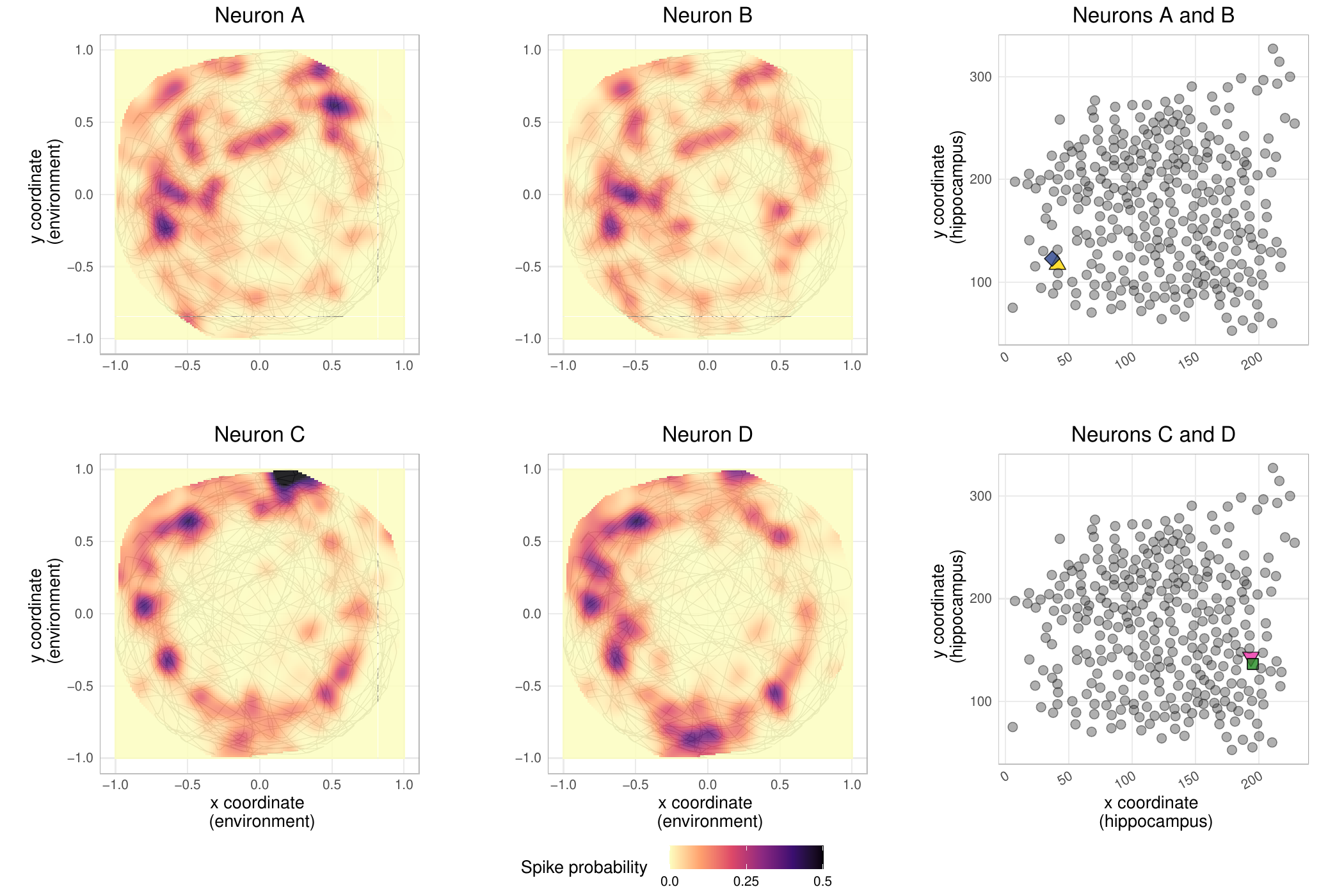}
    \caption{Left and center panels: mouse movements within the arena (grey continuous line) and neuronal activity intensity at that location. Right panels: location of the corresponding neurons in the hippocampus: Neuron A (triangle); Neuron B (diamond); Neuron C (square); Neuron D (down triangle).}
    \label{fig:neuronal_activity}
\end{figure}

Having examined the collective behavior of cell ensembles within a specific time window, we now turn to analyzing how individual neurons respond to the mouse’s position in the environment over the course of the entire experiment. Specifically, we ask whether neurons assigned to the same activation cluster tend to exhibit increased firing rates at similar locations in the arena. To address this question, we first computed the coclustering frequency for each pair of cells, i.e., the proportion of windows in which they are assigned to the same activation ensemble. Remarkably, half of all neuron pairs were clustered together in at least 30\% of the windows, indicating a degree of stability of the activation ensembles over time. We then selected two representative neuron pairs with high coclustering probability and examined their firing rates during the experiment. Specifically, the top row of Figure~\ref{fig:neuronal_activity} shows Neurons A and B, with co-clustering probability 0.84, while the bottom row shows Neurons C and D, with co-clustering probability 0.81.

The left and center panels of the plot show the Monte Carlo posterior estimate of spike probability at each location of the arena for the corresponding neuron. These surfaces are obtained by smoothing the coclustering probability in each location via bivariate multilevel B-splines~\citep{mba}. The similarity within each neuron pair is striking, with both neurons showing increased activity in the same spatial areas. The right panels show the neurons' locations in the hippocampus; notably, they are adjacent in both cases. 
Recent studies suggest that CA1 neurons that become active in similar locations can occur in small nearby groups, although these local groups do not form a smooth large-scale topographic map of environmental position across CA1 \citep{Nakamura2010,WirtshafterDisterhoft2023}. Our results are consistent with this picture, since neurons repeatedly assigned to the same activation ensemble tend to exhibit related spatial responses while also being anatomically proximal. 
In addition, Supplementary Figure~\ref{fig:average_spikes} summarizes the average spike probability of each neuron when the mouse is in the inner circle or in the outer ring of the arena. The figure shows that different local subsets of neurons tend to be more active in the two regions of the arena. 

%
%In addition, Supplementary Figure~SXX (TO BE ADDED) summarizes the average spike probability of each neuron when the mouse is in the inner circle or in the outer ring of the arena, providing a coarse summary of how neural activity varies with the animal's position.

\begin{figure}[t]
    \centering
    \includegraphics[width=.4\linewidth]{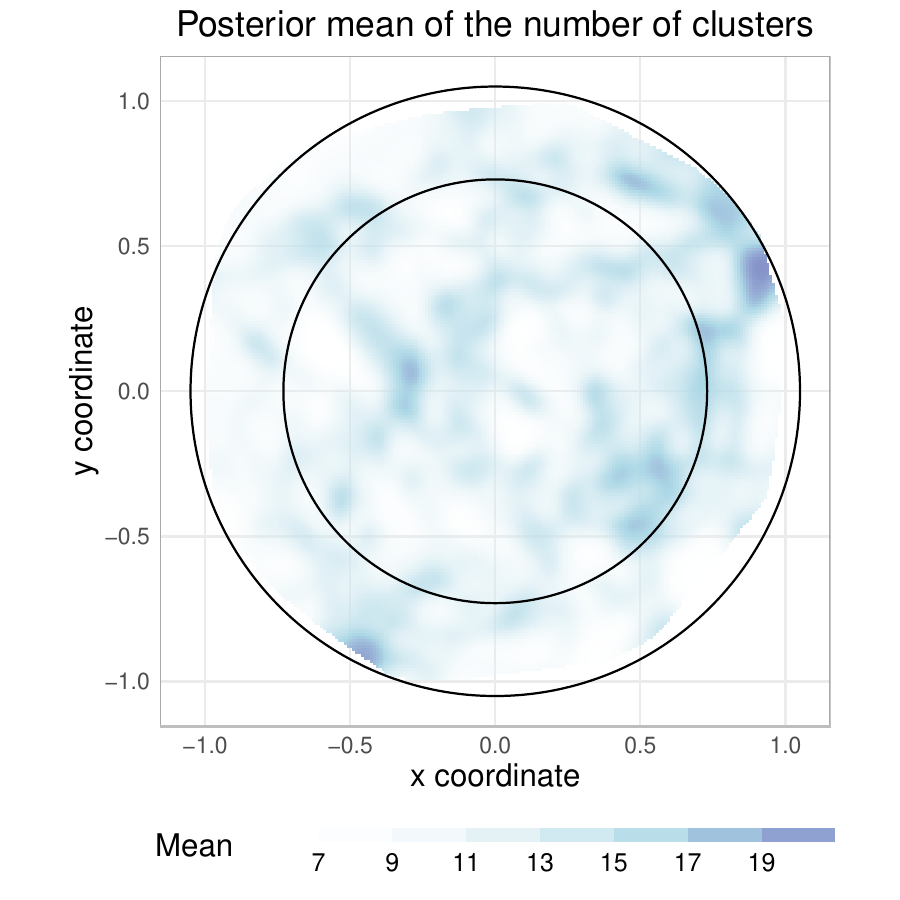}
    \includegraphics[width=.4\linewidth]{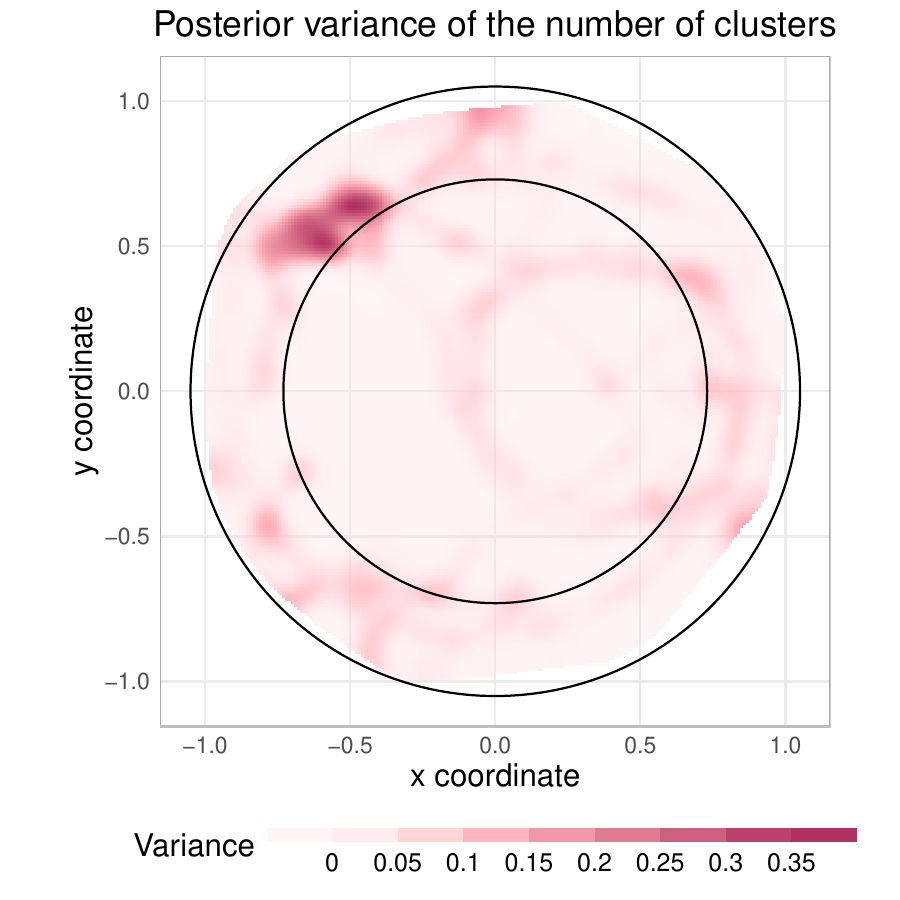}
    \caption{
        Heatmaps showing the spatial distribution of the clustering complexity and variability within the environment. Each point of the mouse trajectory in the arena is reweighted by the mean (left panel) and variance (right panel) of the posterior distribution of the number of clusters. }
    \label{fig:mode_var}
\end{figure}

The second goal is to assess whether the partition's complexity, as measured by the number of clusters, and its uncertainty are related to the mouse's position. For each time window, we examined the posterior mean and variance of the number of clusters. To enhance interpretability, we created spatially smoothed versions of these quantities, shown in Figure~\ref{fig:mode_var}. The left panel displays a heatmap of the posterior mean number of clusters in relation to the mouse's position. In contrast, the right panel shows the corresponding posterior variance. Interestingly, the clustering complexity and variability appear to be larger when the mouse is close to the arena boundaries, suggesting a spatial dependence of the functional clustering. This is in line with previous findings in the literature~\citep[e.g., see][]{Stewart2014}, suggesting varying levels of complexity and neuronal activity depending on proximity to boundaries. While we observe a general relationship between these quantities, certain spatial regions exhibit both a low posterior number of clusters and high uncertainty, for example, near coordinates $(-0.5, 0.5)$.

Additional results obtained fitting Bayesian GPFA \citep{Jensen2021} are reported in the Supplementary Material and yield qualitatively consistent evidence of position-dependent neural dynamics. 
 
Finally, simulation studies investigating the model's sensitivity to prior specification and computational cost are available in Section~\ref{sec:simulations} of the Supplementary Material.

\subsection{Comparison with the two-step deconvolution and clustering}\label{sec:realdata_comparison}

To benchmark our approach, we implemented a two-stage procedure that first performs spike detection by deconvolving the calcium traces and then applies cluster analysis to the resulting spike trains. 
Specifically, for the first step, we used the $\ell_0$ optimization algorithm of \cite{jewell2019}, openly available in the R package \texttt{FastLZeroSpikeInference} and widely used for spike detection in calcium imaging data. This approach is based on the same biophysical model of calcium dynamics described in Equation~\eqref{eq:calcium_dyn}, thereby ensuring a fair comparison. Unlike our Bayesian approach, which learns from the spike-amplitude distribution which fluctuations should be treated as spikes, this optimization method requires setting a fixed threshold parameter $\lambda$. To calibrate this threshold, we first estimated the noise level of each trace and then chose $\lambda$ as the smallest value such that no detected spike amplitudes lay below one standard deviation of the estimated noise. The algorithm returns both the estimated spike times and the denoised calcium signal. 
In the second stage, to identify neuronal ensembles, we followed \cite{chen2023anatomical} and applied consensus $K$-means clustering to the deconvolved traces using the R package \texttt{coca} \citep{coca}. The $K$-means algorithm was applied to a randomly selected subset comprising 80\% of the neurons. The procedure was repeated over 200 replications to account for variability due to subsampling and initialization. The optimal number of clusters $K$ was determined by maximizing the average silhouette score across replications.

To provide a comparison, we ran this two-step method on the representative window analyzed in Section~\ref{sec:representative_win}. Results for additional windows are reported in the Supplementary Material. Overall, this method consistently produced fewer clusters than our model. Although a coarser partition may be easier to interpret, it can also reduce within-cluster homogeneity, making activation patterns harder to distinguish.
\begin{figure}[t]
    \centering
    \includegraphics[width=.75\linewidth]{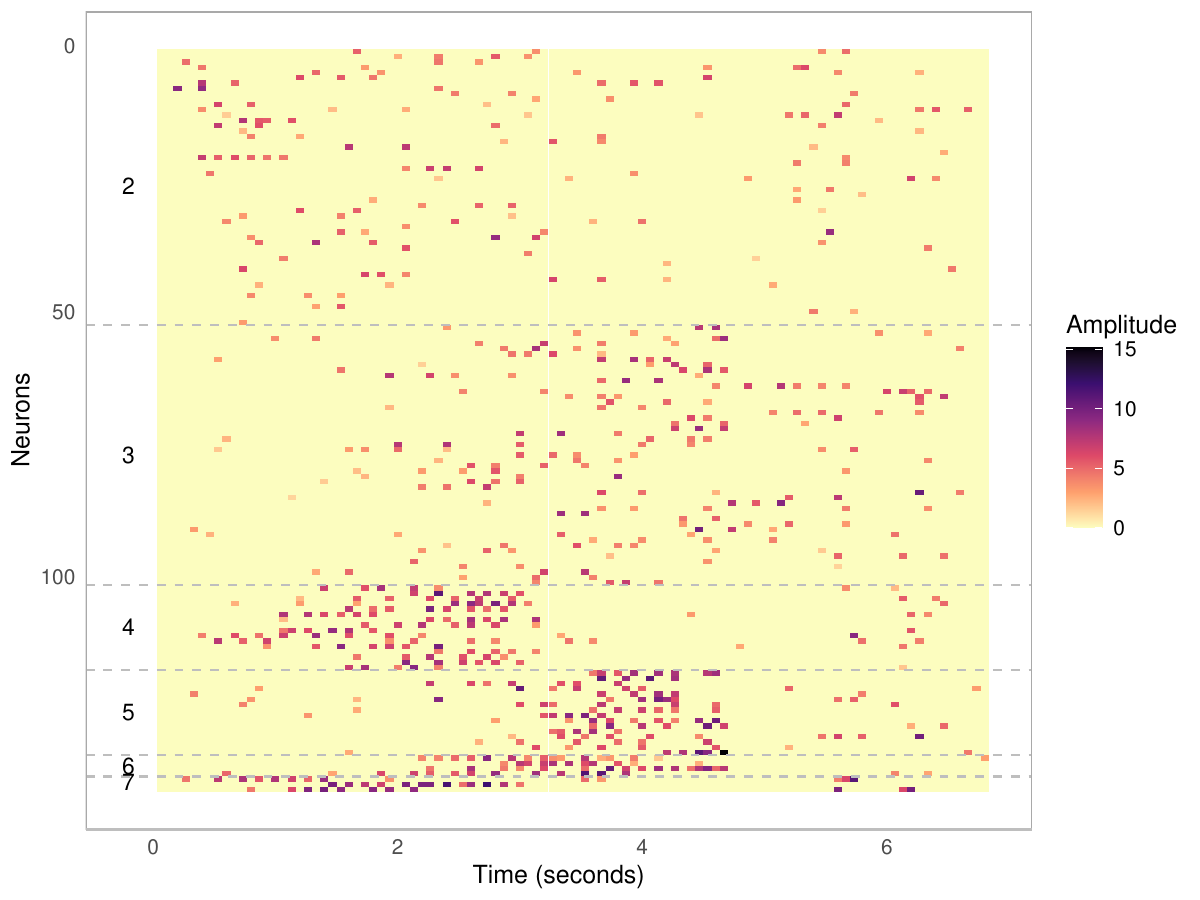}
    \caption{Estimated spike amplitudes over time for active neurons using the two-step procedure. Rows are sorted by the neurons' cluster allocation (Clusters 2-6, left labels).}
    \label{fig:heatmap_two-step_24}
\end{figure}
Figure~\ref{fig:heatmap_two-step_24} shows the heatmap of the estimated spike amplitudes over the selected time window for the active neurons, sorted by their cluster assignment under the two-step procedure. A first notable feature is that this approach detects fewer spikes overall. This likely reflects the use of the fixed threshold parameter $\lambda$ in the spike-detection step, which may be less flexible in accommodating heterogeneity in spike amplitudes. Additional insight into the inferred partition is provided by Figure~\ref{fig:clusters_and_neurons_two-step} in the Supplementary Material, which shows the observed calcium traces colored and sorted by cluster allocation. Comparison with the corresponding results under the proposed model reveals partial agreement between the two procedures, as both recover some groups with similar activity patterns. At the same time, the two-step approach tends to produce broader and less homogeneous groups, especially among neurons with weaker or more scattered activity, making the resulting patterns less clearly defined. By contrast, the proposed method yields a finer partition, leading to a cleaner segmentation of neuronal activity and improved detection of outlying patterns.

\section{Discussion}\label{sec:Discussion}

In this work, we introduced a Bayesian semiparametric model for the joint deconvolution and clustering of calcium imaging traces, combining a latent continuous process for estimating spike probabilities with a spatially-informed clustering prior. The framework enables the identification of functionally coherent neuronal clusters while quantifying uncertainty and allowing comparisons of clustering structure across time. Applying the model to hippocampal CA1 data, we examined how inferred co-activation patterns relate to the mouse's spatial behavior. We found that neurons frequently co-assigned to the same cluster across time windows also tend to exhibit elevated firing rates in overlapping regions of the arena, indicating a link between functional clustering and spatial coding. These co-clustered neurons were often anatomically proximal, suggesting that spatially organized activity patterns may reflect underlying structural organization.  

Conventional unsupervised approaches like GPFA and other latent variable models may also reveal this anatomical organization in the data. However, the detected structure ultimately depends on the assumed latent representation and is therefore not directly comparable across methods. 
In addition, we observed that both the complexity of the inferred partitions and its uncertainty varied across the environment: clustering was more fragmented and uncertain when the mouse was near boundaries, and more stable in central regions of the arena.  
Overall, these findings support the interpretation that the inferred groups are related to local functional ensembles within CA1. This interpretation should nevertheless be viewed as qualitative, since the two-dimensional imaging-plane coordinates in our dataset are not registered to histological or laminar landmarks and therefore do not allow direct assignment of the inferred groups to specific anatomical subregions. 
%Taken together, these findings illustrate how the model captures not only latent neuronal dynamics but also how these dynamics adapt to behavioral context and spatial features of the environment. 

Looking ahead, our framework opens several promising avenues for future research. Methodological extensions could include the incorporation of hierarchical structure across multiple brain regions, explicit modeling of behavioral covariates, or the integration of complementary data modalities such as electrophysiology. From a neuroscience perspective, applying this approach to diverse behavioral contexts, developmental stages, or disease models could yield new insights into the principles governing neural population coding and plasticity. In addition, although the current dataset did not allow this level of granularity, incorporating more detailed representations of the animal's movement, e.g.,  conditioning directly on spatial trajectories or segmenting the arena into functionally distinct zones, could offer a more precise understanding of how ensemble activity varies across the environment.

Finally, substantial computational challenges would need to be addressed for extensions of the model or more complex datasets. 
As the number of neurons, time points, and covariates increases, posterior inference becomes more demanding and may no longer be tractable using standard MCMC techniques.
The time scale, in particular, places substantial limits on the applicability of the proposed algorithm. Long experiments would require adopting strategies to improve the efficiency of the estimation of latent GPs. 
To address this, future work could explore variational inference methods, low-rank approximations, or amortized inference approaches that balance computational efficiency with modeling flexibility.

\section*{Data Availability}
Data are openly available at this \href{https://data.mendeley.com/datasets/tnbwfw2pg2/3}{Mendeley repository}, doi: \texttt{10.17632/tnbwfw2pg2.3}

\section*{Acknowledgments} LD acknowledges support of MUR - Prin 2022 - Grant no. 2022CLTYP4, funded by the European Union – Next Generation EU. FD acknowledges support by the STARS@UNIPD 2025 program, with a contribution from the Fondazione Cassa di Risparmio di Padova e Rovigo. AC acknowledges support of the Royal Society Wolfson Visiting Fellowship. 

\bibliographystyle{agsm5}
\bibliography{references}

@article{Jimenez2020,
  title={{Contextual Fear Memory Retrieval by Correlated Ensembles of Ventral CA1 Neurons}},
  author={Jimenez, Jessica C and Berry, Jack E and Lim, Sean C and Ong, Samantha K and Kheirbek, Mazen A and Hen, Rene},
  journal={Nature communications},
  volume={11},
  number={1},
  pages={3492},
  year={2020},
  publisher={Nature Publishing Group UK London}
}

@article{Modi2014,
	author = {Modi, Mehrab N. and Dhawale, Ashesh K. and Bhalla, Upinder S.},
	doi = {10.7554/eLife.01982},
	journal = {eLife},
	pages = {e01982},
	title = {{CA1 Cell Activity Sequences Emerge After Reorganization of Network Correlation Structure During Associative Learning}},
	volume = {3},
	year = {2014}}

@article{Taxidis2020,
	author = {Taxidis, Jiannis and Pnevmatikakis, Eftychios A. and Dorian, Conor C. and Mylavarapu, Apoorva L. and Arora, Jagmeet S. and Samadian, Kian D. and Hoffberg, Emily A. and Golshani, Peyman},
	doi = {10.1016/j.neuron.2020.08.028},
	journal = {Neuron},
	number = {5},
	pages = {984--998.e9},
	title = {{Differential Emergence and Stability of Sensory and Temporal Representations in Context-Specific Hippocampal Sequences}},
	volume = {108},
	year = {2020}}

@article{Slettmoen2026,
	author = {Slettmoen, Torstein and de Jong, Nienke L. and Eneqvist, Hanna and Skyt{\o}en, Emilie R. and Zong, Weijian and Moser, May-Britt and Moser, Edvard I.},
	doi = {10.1073/pnas.2528601123},
	journal = {Proceedings of the National Academy of Sciences of the United States of America},
	number = {8},
	pages = {e2528601123},
	title = {{Place Cells in {CA1} Lack Topographical Organization of Firing Locations}},
	volume = {123},
	year = {2026}}

@article{WilsonMcNaughton1993,
  author  = {Wilson, Matthew A. and McNaughton, Bruce L.},
  title   = {{Dynamics of the Hippocampal Ensemble Code for Space}},
  journal = {Science},
  year    = {1993},
  volume  = {261},
  number  = {5124},
  pages   = {1055--1058},
  doi     = {10.1126/science.8351520}
}

@article{HarrisEtAl2003,
  author  = {Harris, Kenneth D. and Csicsvari, Jozsef and Hirase, Hajime and Dragoi, George and Buzs{\'a}ki, Gy{\"o}rgy},
  title   = {{Organization of Cell Assemblies in the Hippocampus}},
  journal = {Nature},
  year    = {2003},
  volume  = {424},
  number  = {6948},
  pages   = {552--556},
  doi     = {10.1038/nature01834}
}

@article{Nakamura2010,
  author  = {Nakamura, Nozomu H. and Fukunaga, Masaki and Akama, Keith T. and Soga, Tomoko and Ogawa, Sonoko and Pavlides, Constantine},
  title   = {{Hippocampal Cells Encode Places by Forming Small Anatomical Clusters}},
  journal = {Neuroscience},
  year    = {2010},
  volume  = {166},
  number  = {3},
  pages   = {994--1007},
  doi     = {10.1016/j.neuroscience.2009.12.069}
}

@article{WirtshafterDisterhoft2023,
  author  = {Wirtshafter, Hannah S. and Disterhoft, John F.},
  title   = {{Place Cells are Nonrandomly Clustered by Field Location in {CA1} Hippocampus}},
  journal = {Hippocampus},
  year    = {2023},
  volume  = {33},
  number  = {2},
  pages   = {65--84},
  doi     = {10.1002/hipo.23489}
}

@inproceedings{Yu2008,
 author = {Yu, Byron M and Cunningham, John P and Santhanam, Gopal and Ryu, Stephen and Shenoy, Krishna V and Sahani, Maneesh},
 booktitle = {Advances in Neural Information Processing Systems},
 pages = {1--8},
 publisher = {Curran Associates, Inc.},
 title = {{Gaussian-Process Factor Analysis for Low-Dimensional Single-Trial Analysis of Neural Population Activity}},
 volume = {21},
 year = {2008}
}

@inproceedings{Jensen2021,
 author = {Jensen, Kristopher and Kao, Ta-Chu and Stone, Jasmine and Hennequin, Guillaume},
 booktitle = {Advances in Neural Information Processing Systems},
 pages = {10613--10626},
 publisher = {Curran Associates, Inc.},
 title = {{Scalable Bayesian GPFA with Automatic Relevance Determination and Discrete Noise Models}},
 volume = {34},
 year = {2021}
}

@inproceedings{Duncker2018,
 author = {Duncker, Lea and Sahani, Maneesh},
 booktitle = {Advances in Neural Information Processing Systems},
 pages = {1--11},
 publisher = {Curran Associates, Inc.},
 title = {{Temporal Alignment and Latent Gaussian Process Factor Inference in Population Spike Trains}},
 volume = {31},
 year = {2018}
}

@article {Levi9807,
	author = {Levi, Rafael and Varona, Pablo and Arshavsky, Yuri I. and Rabinovich, Mikhail I. and Selverston, Allen I.},
	title = {{The Role of Sensory Network Dynamics in Generating a Motor Program}},
	volume = {25},
	number = {42},
	pages = {9807--9815},
	year = {2005},
	doi = {10.1523/JNEUROSCI.2249-05.2005},
	publisher = {Society for Neuroscience},
	eprint = {https://www.jneurosci.org/content/25/42/9807.full.pdf},
	journal = {Journal of Neuroscience}
}

@article{STOPFER2003991,
title = {{Intensity Versus Identity Coding in an Olfactory System}},
journal = {Neuron},
volume = {39},
number = {6},
pages = {991-1004},
year = {2003},
issn = {0896-6273},
doi = {https://doi.org/10.1016/j.neuron.2003.08.011},
author = {Mark Stopfer and Vivek Jayaraman and Gilles Laurent}
}

@article{BROOME2006467,
title = {{Encoding and Decoding of Overlapping Odor Sequences}},
journal = {Neuron},
volume = {51},
number = {4},
pages = {467-482},
year = {2006},
issn = {0896-6273},
doi = {https://doi.org/10.1016/j.neuron.2006.07.018},
author = {Bede M. Broome and Vivek Jayaraman and Gilles Laurent}
}

@article{LFADS,
  title={{Inferring Single-Trial Neural Population Dynamics Using Sequential Auto-Encoders}},
  author={Pandarinath, Chethan and O’Shea, Daniel J and Collins, Jasmine and Jozefowicz, Rafal and Stavisky, Sergey D and Kao, Jonathan C and Trautmann, Eric M and Kaufman, Matthew T and Ryu, Stephen I and Hochberg, Leigh R and others},
  journal={Nature Methods},
  volume={15},
  number={10},
  pages={805--815},
  year={2018},
  publisher={Nature Publishing Group US New York}
}

@article{Radical,
  title={{A Deep Learning Framework for Inference of Single-Trial Neural Population Dynamics From Calcium Imaging With Subframe Temporal Resolution}},
  author={Zhu, Feng and Grier, Harrison A and Tandon, Raghav and Cai, Changjia and Agarwal, Anjali and Giovannucci, Andrea and Kaufman, Matthew T and Pandarinath, Chethan},
  journal={Nature Neuroscience},
  volume={25},
  number={12},
  pages={1724--1734},
  year={2022},
  publisher={Nature Publishing Group US New York}
}

@article{ganmor2016,
      title={{Direct Estimation of Firing Rates from Calcium Imaging Data}}, 
      author={Elad Ganmor and Michael Krumin and Luigi F. Rossi and Matteo Carandini and Eero P. Simoncelli},
      year={2016},
      eprint={1601.00364},
      journal={arXiv:1601.00364},
      primaryClass={q-bio.NC}
}

@article {Rupasinghe,
article_type = {journal},
title = {{Direct Extraction of Signal and Noise Correlations From Two-Photon Calcium Imaging of Ensemble Neuronal Activity}},
author = {Rupasinghe, Anuththara and Francis, Nikolas and Liu, Ji and Bowen, Zac and Kanold, Patrick O and Babadi, Behtash},
editor = {Bathellier, Brice and Shinn-Cunningham, Barbara G and Bathellier, Brice},
volume = 10,
year = 2021,
pages = {e68046},
citation = {eLife 2021;10:e68046},
doi = {10.7554/eLife.68046},
journal = {eLife}
}

@article{Keeley2026, 
    title={{Improved Inference of Latent Neural States From Calcium Imaging Data}}, 
    DOI={10.7554/elife.109405.1}, 
    journal = {eLife},
    publisher={eLife Sciences Publications, Ltd}, 
    author={Keeley, Stephen and Zoltowski, David and Charles, Adam and Pillow, Jonathan}, 
    year={2026},
    number = {10},
    volume = {e68046}
}

@article {Prince2021,
	author = {Prince, Luke Y. and Bakhtiari, Shahab and Gillon, Colleen J. and Richards, Blake A.},
	title = {{Parallel Inference of Hierarchical Latent Dynamics in Two-Photon Calcium Imaging of Neuronal Populations}},
	elocation-id = {2021.03.05.434105},
	year = {2021},
	doi = {10.1101/2021.03.05.434105},
	journal = {bioRxiv: 10.1101/2021.03.05.434105}
}

@article{Robinson2020,
	author = {Robinson, Nicholas T. M. and Descamps, Lauranne A. L. and Russell, Luke E. and Buchholz, Max O. and Bicknell, Brendan A. and Antonov, Georgi K. and Lau, Janice Y. N. and Nutbrown, Rachel and Schmidt-Hieber, Christoph and H{\"a}usser, Michael},
	doi = {10.1016/j.cell.2020.09.061},
	journal = {Cell},
	number = {6},
	pages = {1586--1599.e10    },
	title = {{Targeted Activation of Hippocampal Place Cells Drives Memory-Guided Spatial Behavior}},
	volume = {183},
	year = {2020}}

@article{ChiuENEURO2023,
	author = {Chiu, YuHung and Dong, Can and Krishnan, Seetha and Sheffield, Mark E. J.},
	doi = {10.1523/ENEURO.0261-23.2023},
	elocation-id = {ENEURO.0261-23.2023},
	journal = {eNeuro},
	pages = {1-10},
        number = {12},
	publisher = {Society for Neuroscience},
	title = {{The Precision of Place Fields Governs Their Fate Across Epochs of Experience}},
	volume = {10},
	year = {2023},
}

@article{Gauthier2022,
	author = {Gauthier, Jeffrey L. and Koay, Sue Ann and Nieh, Edward H. and Tank, David W. and Pillow, Jonathan W. and Charles, Adam S.},
	id = {Gauthier2022},
	isbn = {1548-7105},
	journal = {Nature Methods},
	number = {4},
	pages = {470--478},
	title = {{Detecting and Correcting False Transients in Calcium Imaging}},
	volume = {19},
	year = {2022}}

@article{Shibue2020,
	author = {Shibue, Ryohei and Komaki, Fumiyasu},
	journal = {PLOS Computational Biology},
	month = {03},
	number = {3},
	pages = {1-25},
	publisher = {Public Library of Science},
	title = {{Deconvolution of Calcium Imaging Data Using Marked Point Processes}},
	volume = {16},
	year = {2020}}

@article{Terashima2022,
	author = {Terashima, Hiroshi and Minatohara, Keiichiro and Maruoka, Hisato and Okabe, Shigeo},
	journal = {Microscopy},
	month = {02},
	number = {1},
	pages = {i81-i99},
	title = {{Imaging Neural Circuit Pathology of Autism Spectrum Disorders: Autism-Associated Genes, Animal Models and the Application of in Vivo Two-Photon Imaging}},
	volume = {71},
	year = {2022}}

@article{Chen2021,
        author = {Chen, Min and Tian, Hongjun and Huang, Guoyong and Fang, Tao and Lin, Xiaodong and Shan, Jianmin and Cai, Ziyao and Chen, Gaungdong and Chen, Suling and Chen, Ce and Ping, Jing and Cheng, Langlang and Chen, Chunmian and Zhu, Jingjing and Zhao, Feifei and Jiang, Deguo and Liu, Chuanxin and Huang, Guangchuan and Lin, Chongguang and Zhuo, Chuanjun},
	journal = {Translational Psychiatry},
	number = {1},
	pages = {619},
	title = {{Calcium Imaging Reveals Depressive- and Manic-Phase-Specific Brain Neural Activity Patterns in a Murine Model of Bipolar Disorder: a Pilot Study}},
	volume = {11},
	year = {2021}}

@article{Beacher2021,
	author = {Beacher, Nicholas J. and Washington, Kayden A. and Werner, Craig T. and Zhang, Yan and Barbera, Giovanni and Li, Yun and Lin, Da-Ting},
	journal = {Frontiers in Neural Circuits},
	title = {{Circuit Investigation of Social Interaction and Substance Use Disorder Using Miniscopes}},
	number = {762441},
	volume = {15},
        pages = {1--9},
	year = {2021}}

@article{Grienberger2022,
	author = {Grienberger, Christine and Magee, Jeffrey C.},
	isbn = {1476-4687},
	journal = {Nature},
	number = {7936},
	pages = {554--562},
	title = {{Entorhinal Cortex Directs Learning-Related Changes in CA1 Representations}},
	volume = {611},
	year = {2022}}

@article{Siciliano2019,
	author = {Cody A. Siciliano and Kay M. Tye},
	journal = {Alcohol},
	pages = {47-63},
	title = {{Leveraging Calcium Imaging to Illuminate Circuit Dysfunction in Addiction}},
	volume = {74},
	year = {2019}}

@article{Aharoni2019,
	author = {Aharoni, Daniel and Khakh, Baljit S. and Silva, Alcino J. and Golshani, Peyman},
	journal = {Nature Methods},
	number = {1},
	pages = {11--13},
	title = {{All the Light That We Can See: A New Era in Miniaturized Microscopy}},
	volume = {16},
	year = {2019}}

@article{chen2023anatomical,
	author = {Chen, Lujia and Lin, Xiaoxiao and Ye, Qiao and Nenadic, Zoran and Holmes, Todd C and Nitz, Douglas A and Xu, Xiangmin},
	journal = {iScience},
	number = {5},
	pages = {1--24},
	publisher = {Elsevier},
	title = {{Anatomical Organization of Temporally Correlated Neural Calcium Activity in the Hippocampal {CA1} Region}},
	volume = {26},
	year = {2023}}

@article{oneill,
	author = {Joseph O'Neill and Timothy Senior and Jozsef Csicsvari},
	doi = {https://doi.org/10.1016/j.neuron.2005.10.037},
	issn = {0896-6273},
	journal = {Neuron},
	number = {1},
	pages = {143-155},
	title = {{Place-Selective Firing of {CA1} Pyramidal Cells During Sharp Wave/Ripple Network Patterns in Exploratory Behavior}},
	volume = {49},
	year = {2006}}

@article{Ainge,
	author = {Ainge, James A. and van der Meer, Matthijs A.A. and Langston, Rosamund F. and Wood, Emma R.},
	doi = {https://doi.org/10.1002/hipo.20301},
	journal = {Hippocampus},
	number = {10},
	pages = {988-1002},
	title = {{Exploring the Role of Context-Dependent Hippocampal Activity in Spatial Alternation Behavior}},
	volume = {17},
	year = {2007}}

@article{bird2008hippocampus,
	author = {Bird, Chris M and Burgess, Neil},
	journal = {Nature Reviews Neuroscience},
	number = {3},
	pages = {182--194},
	publisher = {Nature Publishing Group UK London},
	title = {{The Hippocampus and Memory: Insights From Spatial Processing}},
	volume = {9},
	year = {2008}}

@article{OKeefe1971hippomap,
	author = {J. O'Keefe and J. Dostrovsky},
	doi = {https://doi.org/10.1016/0006-8993(71)90358-1},
	issn = {0006-8993},
	journal = {Brain Research},
	number = {1},
	pages = {171-175},
	title = {{The Hippocampus as a Spatial Map. Preliminary Evidence From Unit Activity in the Freely-Moving Rat}},
	volume = {34},
	year = {1971}}

@article{Krupic2018,
	author = {Julija Krupic and Marius Bauza and Stephen Burton and John O'Keefe},
	doi = {10.1126/science.aao4960},
	journal = {Science},
	number = {6380},
	pages = {1143-1146},
	title = {{Local Transformations of the Hippocampal Cognitive Map}},
	volume = {359},
	year = {2018}}

@article{moser2014grid,
	author = {Moser, Edvard I and Roudi, Yasser and Witter, Menno P and Kentros, Clifford and Bonhoeffer, Tobias and Moser, May-Britt},
	journal = {Nature Reviews Neuroscience},
	number = {7},
	pages = {466--481},
	publisher = {Nature Publishing Group UK London},
	title = {{Grid Cells and Cortical Representation}},
	volume = {15},
	year = {2014}}

@book{bookHippoc,
	address = {Oxford},
	author = {O'Keefe, J. and Nadel, L.},
	publisher = {Clarendon},
	title = {{The Hippocampus as a Cognitive Map}},
	year = {1978}}

@article{calcium_review2,
	author = {Christine Grienberger and Arthur Konnerth},
	issn = {0896-6273},
	journal = {Neuron},
	number = {5},
	pages = {862-885},
	title = {{Imaging Calcium in Neurons}},
	volume = {73},
	year = {2012}}

@article{Stewart2014,
	author = {Stewart, Sarah and Jeewajee, Ali and Wills, Thomas J. and Burgess, Neil and Lever, Colin},
	doi = {10.1098/rstb.2012.0514},
	journal = {Philosophical Transactions of the Royal Society B: Biological Sciences},
	number = {1635},
	pages = {20120514},
	publisher = {The Royal Society},
	title = {{Boundary Coding in the Rat Subiculum}},
	volume = {369},
	year = {2014}}

@article{churchland2012neural,
author = {Churchland, M. M. and Cunningham, J. P. and Kaufman, M. T. and Foster, J. D. and Nuyujukian, P. and Ryu, S. I. and Shenoy, K. V. },
	journal = {Nature},
	number = {7405},
	pages = {51--56},
	publisher = {Nature Publishing Group UK London},
	title = {{Neural Population Dynamics During Reaching}},
	volume = {487},
	year = {2012}}

@article{adler2012temporal,
	author = {Adler, Avital and Katabi, Shiran and Finkes, Inna and Israel, Zvi and Prut, Yifat and Bergman, Hagai},
	journal = {Journal of Neuroscience},
	number = {7},
	pages = {2473--2484},
	publisher = {Society for Neuroscience},
	title = {{Temporal Convergence of Dynamic Cell Assemblies in the Striato-Pallidal Network}},
	volume = {32},
	year = {2012}}

@article{dombeck2010,
	author = {Dombeck, D. A. and Harvey, C. D. and Tian, L. and Looger, L. L. and Tank, D. W.},
	journal = {Nature Neuroscience},
	pages = {1433--1440},
	title = {{Functional Imaging of Hippocampal Place Cells at Cellular Resolution During Virtual Navigation}},
	volume = {13},
	year = {2010}}

@article{shen2022,
	author = {Tong Shen and Gyorgy Lur and Xiangmin Xu and Zhaoxia Yu},
	doi = {https://doi.org/10.1016/j.jneumeth.2021.109431},
	issn = {0165-0270},
	journal = {Journal of Neuroscience Methods},
	pages = {109431},
	title = {{To Deconvolve, or Not to Deconvolve: Inferences of Neuronal Activities Using Calcium Imaging Data}},
	volume = {366},
	year = {2022}}

@article{pnevmatikakis2016,
	author = {Pnevmatikakis, Eftychios A and Soudry, Daniel and Gao, Yuanjun and Machado, Timothy A and Merel, Josh and Pfau, David and Reardon, Thomas and Mu, Yu and Lacefield, Clay and Yang, Weijian and others},
	journal = {Neuron},
	number = {2},
	pages = {285--299},
	publisher = {Elsevier},
	title = {{Simultaneous Denoising, Deconvolution, and Demixing of Calcium Imaging Data}},
	volume = {89},
	year = {2016}}

@article{deneux2016,
	author = {Deneux, Thomas and Kaszas, Attila and Szalay, Gergely and Katona, Gergely and Lakner, Tam{\'a}s and Grinvald, Amiram and R{\'o}zsa, Bal{\'a}zs and Vanzetta, Ivo},
	journal = {Nature communications},
	number = {1},
	pages = {12190},
	publisher = {Nature Publishing Group UK London},
	title = {{Accurate Spike Estimation From Noisy Calcium Signals for Ultrafast Three-Dimensional Imaging of Large Neuronal Populations in Vivo}},
	volume = {7},
	year = {2016}}

@incollection{paiva2010inner,
	author = {Paiva, Antonio RC and Park, Il and Principe, Jose C},
	booktitle = {Statistical signal processing for neuroscience and neurotechnology},
	pages = {265--309},
	publisher = {Elsevier},
	title = {{Inner Products for Representation and Learning in the Spike Train Domain}},
	year = {2010}}

@article{humphries2011spike,
	author = {Humphries, Mark D},
	journal = {Journal of Neuroscience},
	number = {6},
	pages = {2321--2336},
	publisher = {Society for Neuroscience},
	title = {{Spike-Train Communities: Finding Groups of Similar Spike Trains}},
	volume = {31},
	year = {2011}}

@article{diana2019bayesian,
	author = {Diana, Giovanni and Sainsbury, Thomas TJ and Meyer, Martin P},
	journal = {PLoS computational biology},
	number = {10},
	pages = {e1007481},
	publisher = {Public Library of Science San Francisco, CA USA},
	title = {{Bayesian Inference of Neuronal Assemblies}},
	volume = {15},
	year = {2019}}

@article{shen2024time,
	author = {Shen, Tong and Du, Mingyu and Johnston, Kevin and Grieco, Steven F and Crary, Rachel and Guzowski, John F and Lur, Gyorgy and Xu, Xiangmin and Ombao, Hernando and Guindani, Michele and others},
	journal = {Data Science in Science},
	number = {1},
	pages = {2407770},
	publisher = {Taylor \& Francis},
	title = {{Time-Varying $\ell_0$ Optimization for Spike Inference from Multi-Trial Calcium Recordings}},
	volume = {3},
	year = {2024}}

@article{vogelstein2010fast,
	author = {Vogelstein, Joshua T and Packer, Adam M and Machado, Timothy A and Sippy, Tanya and Babadi, Baktash and Yuste, Rafael and Paninski, Liam},
	journal = {Journal of Neurophysiology},
	number = {6},
	pages = {3691--3704},
	title = {{Fast Nonnegative Deconvolution for Spike Train Inference From Population Calcium Imaging}},
	volume = {104},
	year = {2010}}

@article{jewell2018,
	author = {Sean Jewell and Daniela Witten},
	journal = {The Annals of Applied Statistics},
	number = {4},
	pages = {2457 - 2482},
	title = {{Exact Spike Train Inference via {$\ell_0$} Optimization}},
	volume = {12},
	year = {2018}}

@article{jewell2019,
	author = {Jewell, Sean W. and Hocking, Toby Dylan and Fearnhead, Paul and Witten, Daniela M.},
	journal = {Biostatistics},
	month = {02},
	number = {4},
	pages = {709-726},
	title = {{Fast Nonconvex Deconvolution of Calcium Imaging Data}},
	volume = {21},
	year = {2020}}

@article{dangeloetal2023biometrics,
	author = {D'Angelo, Laura and Canale, Antonio and Yu, Zhaoxia and Guindani, Michele},
	doi = {10.1111/biom.13626},
	journal = {Biometrics},
	number = {2},
	pages = {1370--1382},
	title = {{Bayesian Nonparametric Analysis for the Detection of Spikes in Noisy Calcium Imaging Data}},
	volume = {79},
	year = {2023}}

@article{rodriguez2011,
	author = {Rodr\'iguez, Abel and Dunson, David},
	journal = {Bayesian Analysis},
	month = {03},
	pages = {145-178},
	title = {{Nonparametric {Bayesian} Models Through Probit Stick-Breaking Processes}},
	volume = {6},
	year = {2011}}

@article{ishwaran2001,
	author = {Hemant Ishwaran and Lancelot F James},
	doi = {10.1198/016214501750332758},
	journal = {Journal of the American Statistical Association},
	number = {453},
	pages = {161-173},
	publisher = {Taylor & Francis},
	title = {{Gibbs Sampling Methods for Stick-Breaking Priors}},
	volume = {96},
	year = {2001}}

@article{albertchib,
	author = {James H. Albert and Siddhartha Chib},
	journal = {Journal of the American Statistical Association},
	number = {422},
	pages = {669-679},
	publisher = {Taylor & Francis},
	title = {{Bayesian Analysis of Binary and Polychotomous Response Data}},
	volume = {88},
	year = {1993}}

@article{coca,
	author = {Alessandra Cabassi and Paul DW Kirk},
	journal = {Bioinformatics},
	number = {18},
	pages = {4789 --4796},
	title = {{Multiple Kernel Learning for Integrative Consensus Clustering of Genomic Datasets}},
	volume = {36},
	year = {2020}}

@article{canale2017pitman,
	author = {Canale, Antonio and Lijoi, Antonio and Nipoti, Bernardo and Pr{\"u}nster, Igor},
	journal = {Biometrika},
	number = {3},
	pages = {681--697},
	publisher = {Oxford University Press},
	title = {{On the Pitman--Yor Process With Spike and Slab Base Measure}},
	volume = {104},
	year = {2017}}

@article{wade2018,
	author = {Sara Wade and Zoubin Ghahramani},
	doi = {10.1214/17-BA1073},
	journal = {Bayesian Analysis},
	number = {2},
	pages = {559 -- 626},
	title = {{{Bayesian} Cluster Analysis: Point Estimation and Credible Balls (with Discussion)}},
	volume = {13},
	year = {2018}}

@article{salso,
	author = {David B. Dahl and Devin J. Johnson and Peter M\"uller},
	note = {\texttt{R} package, version 0.3.35},
	title = {{\texttt{salso}: Search Algorithms and Loss Functions for Bayesian Clustering}},
	year = {2023}
}

@article{Dahl2022,
	author = {David B. Dahl and Devin J. Johnson and Peter M\"uller},
	number = {4},
	journal = {Journal of Computational and Graphical Statistics},
	month = {10},
	pages = {1189--1201},
	title = {{Search Algorithms and Loss Functions for Bayesian Clustering}},
	volume = {31},
	year = {2022}}

@article{Mishchenko2011,
    author = {Mishchenko, Y. and Vogelstein, J. T. and Paninski, L.},
    title = {{A Bayesian Approach for Inferring Neuronal Connectivity from Calcium Fluorescent Imaging Data}},
    journal = {{The Annals of Applied Statistics}},
    year = {2011},
    volume = {5},
    number = {2B},
    pages = {1229--1261},
    doi = {10.1214/09-AOAS303},
    publisher = {Institute of Mathematical Statistics}
}

@article{Quintana2022,
  title = {{The Dependent Dirichlet Process and Related Models}},
  volume = {37},
  ISSN = {0883-4237},
  DOI = {10.1214/20-sts819},
  number = {1},
  journal = {Statistical Science},
  publisher = {Institute of Mathematical Statistics},
  author = {Quintana,  Fernando A. and M\"{u}ller,  Peter and Jara,  Alejandro and MacEachern,  Steven N.},
  year = {2022},
  month = feb 
}

@Manual{mba,
    title = {MBA: Multilevel B-Spline Approximation},
    author = {Finley, A. and Banerjee, S. and Hjelle, {\O}.},
    year = {2024},
    note = {R package version 0.1-2},
    doi = {10.32614/CRAN.package.MBA},
  }

\clearpage
\appendix

\setcounter{figure}{0}
\setcounter{table}{0}
\setcounter{equation}{0}
\renewcommand{\thetable}{\thesection\arabic{table}}
\renewcommand{\thefigure}{\thesection\arabic{figure}}
\renewcommand{\shorttitle}{Supplementary Material}

\thispagestyle{empty}

\begin{center}
   {\Large\textsc{ Supplementary Material for: \\
   \vspace{.3cm}
``Decoding Neuronal Ensembles from Spatially-Referenced Calcium Traces: \\ A Bayesian Semiparametric Approach''} \par} 
\end{center}

 \vspace{.5cm}

\section{Additional details on posterior inference }\label{Asec:posterior_inference}

In this section, we report the graphical representation of the proposed model, which highlights the dependencies among random variables. 
Moreover, we provide the pseudo-code of the Gibbs sampler, outlined in Algorithm~\ref{algo::gibbs}, and the details on the full conditional distributions of Steps 1 and 6 of the algorithm. 

\begin{figure}[ht]
\centering
\resizebox{.85\textwidth}{!}{%
\begin{circuitikz}
\tikzstyle{every node}=[font=\Large]
\draw  (6.75,14.25) circle (0cm);
\draw  (9,14.75) circle (0cm);
\draw  (21.25,11.75) circle (0cm);
\draw  (21.25,12) circle (0cm);
\draw  (21.25,12) circle (0cm);
\draw  (4.5,15.25) circle (0.75cm) node {\Large $y_{i,t-1}$} ;
\draw  (9.25,15.25) circle (0.75cm) node {\Large $y_{i,t}$} ;
\draw  (13.5,15.25) circle (0.75cm) node {\Large $y_{i,t+1}$} ;
\draw  (4.5,12.5) circle (0.75cm) node {\Large $c_{i,t-1}$} ;
\draw  (9.25,12.5) circle (0.75cm) node {\Large $c_{i,t}$} ;
\draw  (13.5,12.5) circle (0.75cm) node {\Large $c_{i,t+1}$} ;
\draw  (9.25,18.75) circle (0.75cm) node {\Large $b_i$} ;
\draw [->, >=Stealth] (5.25,12.5) -- (8.5,12.5);
\draw [->, >=Stealth] (10,12.5) -- (12.75,12.5);
\draw [->, >=Stealth] (4.5,13.25) -- (4.5,14.5);
\draw [->, >=Stealth] (9.25,13.25) -- (9.25,14.5);
\draw [->, >=Stealth] (13.5,13.25) -- (13.5,14.5);
\draw [->, >=Stealth] (14.25,12.5) -- (16.5,12.5);
\draw [->, >=Stealth] (1.25,12.5) -- (3.75,12.5);
\node [font=\Large] at (0.75,12.5) {$\dots$};
\node [font=\Large] at (17,12.5) {$\dots$};
\draw [->, >=Stealth] (9,18) -- (5,15.75);
\draw [->, >=Stealth] (9.25,18) -- (9.25,16);
\draw [->, >=Stealth] (9.5,18) -- (13,15.75);
\draw [ color={rgb,255:red,143; green,143; blue,143} ] (2.25,14.25) circle (0.75cm) node {\Large $\epsilon_{i,t-1}$} ;
\draw [ color={rgb,255:red,143; green,143; blue,143} ] (7,14.25) circle (0.75cm) node {\Large $\epsilon_{i,t}$} ;
\draw [ color={rgb,255:red,143; green,143; blue,143} ] (11.25,14.25) circle (0.75cm) node {\Large $\epsilon_{i,t+1}$} ;
\draw [->, >=Stealth] (7.75,14.5) -- (8.5,15);
\draw [->, >=Stealth] (12,14.5) -- (12.75,15);
\draw [->, >=Stealth] (3,14.5) -- (3.75,15);
\draw [ color={rgb,255:red,143; green,143; blue,143} ] (2.5,11.25) circle (0.75cm) node {\Large $\eta_{i,t-1}$} ;
\draw [ color={rgb,255:red,143; green,143; blue,143} ] (7.25,11.25) circle (0.75cm) node {\Large $\eta_{i,t}$} ;
\draw [ color={rgb,255:red,143; green,143; blue,143} ] (15.75,11.25) circle (0.75cm) node {\Large $\eta_{i,t+1}$} ;
\draw [->, >=Stealth] (3.25,11.5) -- (4,12);
\node [font=\Large] at (8.75,14.25) {};
\node [font=\Large] at (8.75,14.25) {};
\draw [->, >=Stealth] (8,11.5) -- (8.75,12);
\node [font=\Large] at (9,11.25) {};
\node [font=\Large] at (9,11.25) {};
\draw [ color={rgb,255:red,0; green,123; blue,255} ] (11.75,5) circle (0.75cm) node {\Large $a^*_{h}$} ;
\draw [ color={rgb,255:red,0; green,123; blue,255} ] (14.5,5) circle (0.75cm) node {\Large $\xi_{i,t}$} ;
\draw [ color={rgb,255:red,255; green,89; blue,0} ] (17.25,5) circle (0.75cm) node {\Large $s_{i,t}$} ;
\draw  (9.25,7.5) circle (0.75cm) node {\Large $\gamma$} ;
\draw  (16.25,10.5) circle (0cm);
\draw [->, >=Stealth] (9.25,8.25) -- (9.25,11.75);
\draw [->, >=Stealth] (9.5,8.25) -- (13.25,11.75);
\draw [->, >=Stealth] (9,8.25) -- (5,12);
\draw [->, >=Stealth] (15,11.5) -- (14,12);
\draw [ color={rgb,255:red,255; green,89; blue,0}, ->, >=Stealth] (16.75,5.5) -- (10,12.25);
\draw [ color={rgb,255:red,0; green,123; blue,255}, ->, >=Stealth] (14.25,5.75) -- (10,12);
\draw [ color={rgb,255:red,0; green,123; blue,255}, ->, >=Stealth] (11.75,5.75) -- (9.75,12);
\draw [ color={rgb,255:red,255; green,89; blue,0} ] (16,1.25) circle (0.75cm) node {\Large $\tilde{s}^*_k$} ;
\draw [ color={rgb,255:red,0; green,123; blue,255} ] (13.5,1.25) circle (0.75cm) node {\Large $v_h$} ;
\draw [ color={rgb,255:red,0; green,123; blue,255}, ->, >=Stealth] (15.75,2) -- (14.75,4.25);
\draw [ color={rgb,255:red,0; green,123; blue,255}, ->, >=Stealth] (13.75,2) -- (14.5,4.25);
\draw [ color={rgb,255:red,255; green,89; blue,0} ] (19.25,1.25) circle (0.75cm) node {\Large $\zeta_{i}$} ;
\draw [ color={rgb,255:red,255; green,89; blue,0}, ->, >=Stealth] (19,2) -- (17.5,4.25);
\draw [ color={rgb,255:red,255; green,89; blue,0}, ->, >=Stealth] (16.25,2) -- (17,4.25);
\node [font=\Large, color={rgb,255:red,0; green,123; blue,255}] at (15.75,2.75) {};
\node [font=\Large, color={rgb,255:red,0; green,123; blue,255}] at (15.75,2.75) {};
\draw [ color={rgb,255:red,0; green,123; blue,255}, ->, >=Stealth] (18.75,1.75) -- (15,4.5);
\draw [ color={rgb,255:red,255; green,89; blue,0} ] (22.25,1.25) circle (0.75cm) node {\Large $\pi_k(\ell_i)$} ;
\draw [ color={rgb,255:red,255; green,89; blue,0}, ->, >=Stealth] (21.5,1.25) -- (20,1.25);
\end{circuitikz}
}%
\caption{Graphical representation of the model for the calcium dynamics (black) and of the hierarchical prior on an individual $c_{i,t}$. Error terms are colored in gray. The prior on the spike amplitudes is colored in blue, while the prior on the latent signal is colored in orange. For simplicity and ease of interpretation, we are omitting the hyperparameters.}
\label{fig:graphical_model}
\end{figure}

\begin{algorithm}[ht!]
 \caption{Gibbs sampler algorithm}\label{algo::gibbs}
\setstretch{1.35}
\SetAlgoLined
{\footnotesize
\vspace{.2cm} 
Define the number of MCMC iterations $H$.\\
\For{$h=1,\dots,H$}{    
    \begin{enumerate}
    \item Sample the unobserved calcium concentration $\bm{c}_i$, independently for each $i=1,\dots,n$, using a forward filtering backward sampling scheme.
    \item Update the baseline parameters $b_i$, for $i=1,\dots,n$, sampling each of them from a normal distribution with mean
$\tilde{b}_i = \frac{\sigma^2 B_0}{\sigma^2 + T B_0} \left( \frac{\sum_{t=1}^{T}(y_{i,t} - c_{i,t})}{\sigma^2} + \frac{b_0}{B_0} \right)$ and variance $\tilde{B}_i=\frac{\sigma^2 B_0}{\sigma^2 + T B_0}$.
\item Update the variance of the output and state equations, sampling $1/\sigma^2$ from a gamma with shape and rate parameters $\tilde{\alpha}_{\sigma} = \alpha_{\sigma} + \frac{nT}{2}$ and $\tilde{\beta}_{\sigma} = \beta_{\sigma} + \frac{1}{2}\sum_{i=1}^n\sum_{t=1}^T(y_{i,t} - b_i - c_{i,t})^2$, respectively; and $1/\tau^2$ from a gamma with parameters $\tilde{\alpha}_{\tau} =\alpha_{\tau} + \frac{nT}{2}$ and $\tilde{\beta}_{\tau}  = \beta_{\tau} + \frac{1}{2}\sum_{i=1}^n\sum_{t=1}^T(c_{i,t} - \gamma c_{i,t-1} - s_{i,t}a_{i,t})^2$.
\item Perform a Metropolis-Hastings step to sample the decay $\gamma$ from a distribution with density
\vspace{-.4cm}
    \begin{equation*}
p(\gamma\mid \{\bm{c}_i,\bm{s}_i,\bm{a}_i\}_{i=1,\dots,n},\tau^2 ) \propto \exp\left\{ -\frac{1}{2\tau^2}  \sum_{i=1}^n \sum_{t=1}^T (c_{i,t} - \gamma c_{i,t-1}-a_{i,t})^2
    \right\} \gamma^{(\alpha_{\gamma}-1)} (1-\gamma)^{(\beta_{\gamma}-1)}.
    \end{equation*} \vspace{-.8cm}
\item Perform the spike detection and update the positive amplitudes. Introduce the cluster allocation variables $\xi_{i,t}$, for $i=1,\dots,n$ and $t=1,\dots,T$, with values in $\{0,1,2,\dots\}$, where 0 represents the absence of a spike, and $\{a_{i,t}\mid \xi_{i,t}=j\} = a^*_j$.
\vspace{-.2cm}
\begin{itemize}
    \item[a)] Sample the cluster assignments from a categorical distribution
    \vspace{-.4cm}
    \begin{equation*}
        \mathbb{P}(\xi_{i,t} = j \mid c_{i,t}, \zeta_i = k, \tilde{s}^*_k, \gamma,\tau^2) \propto \begin{cases}
            \Phi\big( -\tilde{s}^*_{k,t} \big)  \, \phi(c_{i,t}  \mid \gamma c_{i,t-1}, \: \tau^2) \quad \text{if }\, j=0,\\
            \Phi\big( \tilde{s}^*_{k,t} \big)\, \omega_j \,  \phi(c_{i,t}  \mid \gamma c_{i,t-1} + a^*_j, \: \tau^2) \quad \text{if }\, j\geq 1.
        \end{cases}
    \end{equation*} \vspace{-.7cm}
    \item[b)] Sample the unique amplitude values $a^*_j$ (atoms of the DP), for $j\geq 1$, using a Metropolis-Hastings step from a distribution with density 
        \vspace{-.4cm}
    \begin{equation*}
    p(a^*_j \mid \gamma, \tau^2, \{c_{i,t}, \xi_{i,t}\})\propto \left\{ (a^*_j)^{(\alpha_a - 1)} e^{-\beta_a a^*_j}  \right\} \exp\left\{ -\frac{1}{2(\tau^2)} \sum_{(i,t): \xi_{i,t} = j} (c_{i,t} - \gamma c_{i,t-1} - a^*_{j})^2 \right\}.
    \end{equation*} \vspace{-.7cm}
    \item[c)] For $j\geq 1$, let $n_j = \sum_{i=1}^n \sum_{t=1}^T \mathbb{I}_{(\xi_{i,t} = j)}$ and sample the variables $v_j$ from a
    $\mathrm{Beta}\left(1 + n_j, \alpha + \sum_{r>j}n_r \right)$ distribution. 
    Use these variables to construct the weights of the DP through the stick-breaking construction.
\end{itemize}
\item Update the clustering of neurons and the realizations of the latent GPs. 
\vspace{-.2cm}
\begin{itemize}
    \item[a)] For $i=1,\dots,n$, sample the cluster allocation variables $\zeta_{i}$ with values in $\{1,2,\dots\}$, from the categorical distribution
    \vspace{-.5cm}
    \begin{equation*}
    \mathbb{P}(\zeta_{i} = k \mid \bm{s}_i, \tilde{s}^*_k )
    \propto \pi_k(\bm{\ell}_i ) \prod_{t=1}^T  \Phi\big( \tilde{s}^*_{k,t} \big)^{s_{i,t}} \bigg[ 1- \Phi\big( \tilde{s}^*_{k,t} \big) \bigg]^{(1-s_{i,t})}.
    \end{equation*} \vspace{-1cm}
    \item[b)] Update the probabilities $\pi_k(\bm{\ell}_i)$, $i=1,\dots,n$, $k\geq 1$, using the probit stick-breaking process. 
    \item[c)] Sample the latent GPs using the data augmentation scheme of~\citet{albertchib}.
\end{itemize}
 \end{enumerate} }}
 
\end{algorithm}

\clearpage

\paragraph{STEP 1: Calcium concentration.} Update the unobserved calcium concentration $\bm{c}_i$, independently for each $i=1,\dots,n$, using a forward filtering backward sampling scheme:
\begin{itemize}
    \item[a)] Apply the Kalman filter: set $\bar{q}_0 = \bar{m}_0 = 0$, $\bar{R}_0 = C_0 = \mathrm{var}(c_{i,0})$. For $t = 1,\dots,T$ let 
    \begin{gather*}
        \bar{q}_t = \gamma \, \bar{m}_{t-1} + s_{i,t}a_{i,t},\\
        \bar{R}_t = \gamma^2 \, \bar{C}_{t-1} + \tau^2.
    \end{gather*} 
    Compute the filtering distribution's parameters, $\bar{m}_t$ and $\bar{C}_{t}$, for $t = 1,\dots,T$, where
    \begin{gather*}
        \bar{m}_t = \bar{q}_t + \bar{R}_t\, (\bar{R}_t + \sigma^2)^{-1} \, (y_{i,t} - b_i - \bar{q}_t),\\
        \bar{C}_t = \bar{R}_t -  \bar{R}_t^2 \, (\bar{R}_t + \sigma^2)^{-1}.
    \end{gather*}            
    \item[b)] Draw $c_{i,T} \sim \mathrm{N}(\bar{m}_T, \bar{C}_T)$;
    \item[c)] For $t = T-1, \dots, 0$, draw $c_{i,t} \sim \mathrm{N}(\bar{h}_t, \bar{H}_t)$, with 
    \begin{gather*}
        \bar{h}_t = \bar{m}_t + \gamma \, \bar{C}_t \, \bar{R}_{t+1}^{-1} (c_{t+1} - \bar{q}_{t+1}), \quad\quad
        \bar{H}_t = \bar{C}_t - \gamma^2 \, \bar{C}_t^2 \, \bar{R}_{t+1}^{-1}.
    \end{gather*}   
\end{itemize}

\paragraph{STEP 6: Cluster of the neurons.} Here, the quantities to update are, for $i=1,\dots,n$
\begin{itemize}[noitemsep]
\item[a)] the cluster allocation variables $\zeta_{i}$ with values in $\{1,2,\dots\}$;
    \item[b)] the mixing probabilities, using the probit stick-breaking process. These probabilities depend on the location of neurons in the hippocampus, $\bm{\ell}_i$, encoded through the ``proximity'' matrix $\Sigma = \Sigma(\bm{\ell}_i, \bm{\ell}_{i'})$.
    \item[c)] the unique values of the latent GP (atoms of the mixture).
\end{itemize}
First of all, define the mixing weights, for $k\geq 1$, as
$$\pi_k(\bm{\ell}_i ) = \Phi(\alpha_k(\bm{\ell}_i )) \prod_{r<k} \big[1-\Phi(\alpha_r(\bm{\ell}_i )) \big].$$
Then, update:
\begin{itemize}
 \item[a)] The cluster assignment: for $i=1,\dots,n$
    \begin{align*}
    \mathbb{P}(\zeta_{i} = k \mid \bm{s}_i, \tilde{s}^*_k )& \propto \pi_k(\bm{\ell}_i ) p(\bm{s}_i \mid \tilde{s}^*_k) 
    \propto \pi_k(\bm{\ell}_i ) \prod_{t=1}^T  \Phi\big( \tilde{s}^*_{k,t} \big)^{s_{i,t}} \bigg[ 1- \Phi\big( \tilde{s}^*_{k,t} \big) \bigg]^{(1-s_{i,t})}.
    \end{align*}
    \item[b)] The mixing probabilities, which are based on the variables $\{\alpha_k(\bm{\ell}_1 ),\dots,\alpha_k(\bm{\ell}_n )\}_{k\geq 1}$. This update is based on the data augmentation scheme with truncated normals outlined in~\citet{rodriguez2011}: 
    \begin{itemize}
        \item[b1)] conditionally on the past values of $\{\alpha_k(\bm{\ell}_1 ),\dots,\alpha_k(\bm{\ell}_n )\}_{k\geq 1}$ and of the cluster assignment $\zeta_{i}$, for $i=1,\dots,n$ and $k\geq 1$, sample
        \begin{equation*}
            z_{k}(\bm{\ell}_i) \mid  \alpha_k(\bm{\ell}_i) , \zeta_i \sim \begin{cases}
                \mathrm{N}( \alpha_k(\bm{\ell}_i), \,1 )^- \:\:\text{if} \:\: k < \zeta_i,\\
                \mathrm{N}( \alpha_k(\bm{\ell}_i), \,1 )^+ \:\:\text{if} \:\: k = \zeta_i.
            \end{cases}
        \end{equation*}
        where $N(\mu,1)^+$ denotes the density of a Gaussian random variable of mean $\mu$ and variance 1, truncated to $(0,-\infty)$ (and, similarly, $N(\mu,1)^-$ when it is truncated to $(-\infty,0)$).
        \item[b2)] sample the latent location-dependent variables. For $k\geq 1$,
        \begin{equation*}
            [\alpha_k(\bm{\ell}_1 ),\dots,\alpha_k(\bm{\ell}_n )]^T \sim \mathrm{N}_n \bigg( (\Sigma^{-1}+ \frac{1}{\sigma^2_{\alpha}} I)^{-1} [\mu_{\alpha}\Sigma^{-1} \bm{1}_n+ \frac{1}{\sigma^2_{\alpha}}\bm{z}_{k}], (\Sigma^{-1}+\frac{1}{\sigma^2_{\alpha}}I)^{-1}  \bigg),
        \end{equation*}
        where $\bm{1}_n$ is a unit column vector of length $n$, and $I$ is the identity matrix.
    \end{itemize}
    \item[c)] The unique realizations of the GP (atoms of the probit stick-breaking DP). For $k\geq 1$, consider the neurons allocated to cluster $k$, i.e., all the observations $i=1\ldots,n$ such that $\zeta_i = k$. We need to update the GP $[\tilde{s}^*_{k,1}.\dots,\tilde{s}^*_{k,T}]$ given the binary series $\{[s_{i,1},\dots,s_{i,T}]\}_{i:\zeta_i = k}$.

    We resort to the data augmentation scheme of~\citet{albertchib} to relate the latent $\tilde{s}^*_{k}$ with the Bernoulli variables: we introduce independent Gaussian random variables $\tilde{z}_{i,t}\mid\tilde{s}^*_{k,t} \sim N(\tilde{s}^*_{k,t},1)$, for $i=1,\dots,n$ and $t=1,\dots,T$. This is equivalent to writing
\begin{equation*}
    s_{i,t}\mid \tilde{z}_{i,t}= \begin{cases}
    1 \quad \text{if } \tilde{z}_{i,t}>0, \\
    0 \quad\text{if } \tilde{z}_{i,t}\leq 0.
    \end{cases}
\end{equation*}

With this in mind, perform the following steps.

\begin{itemize}
    \item[c1)] The full conditional distribution of $\tilde{z}_{i,t}$ is a truncated normal distribution. Specifically, independently for $i=1,\dots,n$ and $t=1,\dots,T$, sample  
\begin{equation*}
    \tilde{z}_{it}\mid \tilde{s}^*_{k,t},s_{i,t} \sim \begin{cases}
    N(\tilde{s}^*_{k,t},1)^+ \quad \text{if } s_{i,t} = 1, \\
    N(\tilde{s}^*_{k,t},1)^- \quad\text{if } s_{i,t} = 0.
    \end{cases}
\end{equation*}
\item[c2)] For $k\geq 1$, define $\bar{\bm{z}}_k = 1/(\sum_{i=1}^n \mathbb{I}(\zeta_i=k)) \sum_{i: \zeta_i=k} \tilde{\bm{z}}_i$ and update $(\tilde{{s}}^*_{k,t}\mid \{\tilde{\bm{z}}_{i}\}_{i:\zeta_i=k})$ for $t=1,\dots,T$ from a normal with mean 
$\tilde{\mu}_{\tilde{s}_t} = \bm{k}^T \tilde{\Omega}^{-1} \bar{z}_{k,t}$
and variance 
$\tilde{\Omega}_{\tilde{s}} = (\sigma^2_{\Omega} + \eta^2) - \bm{k}^T \tilde{\Omega}^{-1} \bm{k}$, where $\tilde{\Omega} = \Omega + \frac{1}{n_k \eta^2}I_T$, $\eta^2$ is the noise variance of the GP, and $\bm{k}$ is a vector of the covariances $\Omega(t,t')$ for $t'=1,\dots,T$. 
\end{itemize}
\end{itemize}

\section{Additional details on the calcium imaging data analysis}\label{Asec:read}

\subsection{Sensitivity analysis}\label{Asec:read_sensitivity}
\paragraph{Influence of $\bar{a}$}
To assess the robustness of our results with respect to the choice of $\bar{a}$, we conducted a sensitivity analysis on the representative window used in the main paper (Window \#24). Posterior estimates of the model parameters remain stable across different values of $\bar{a}\in\{0,0.5,1\}$, indicating consistent inference. Specifically, under the three specifications, the posterior means of the variances $\sigma^2$ and $\tau^2$ are $\{ 0.8013, 0.8010, 0.7981 \}$ and $\{ 0.3715, 0.3790, 0.3840 \}$, respectively. The posterior estimates of the decay parameter $\gamma$ are stable as well, being $\{ 0.9577, 0.9587, 0.9587 \}$. The estimated firing rates (i.e., proportion of detected spikes over the total number of observations) are $\{0.0372, 0.0382, 0.0389\}$.

Assessing the coherence of the inferred neuronal partitions is more nuanced. The Adjusted Rand Indices (ARI) between pairs of estimated partitions are $0.7456$ for $\bar{a} = 0$ and $0.5$, $0.6457$ for $\bar{a} = 0.5$ and $1$, and $0.6819$ for $\bar{a} = 0$ and $1$, suggesting a reasonably high degree of similarity across the different configurations.

\paragraph{Influence of the PSB prior}\label{sec:sensitivity_PSB}
We then assess the sensitivity of the inferred neuronal partition to the degree of spatial borrowing induced by the prior. Recall that the location-dependent probit stick-breaking process incorporates neuronal
proximity through the matrix
$\Sigma(\bm{\ell}_i,\bm{\ell}_{i'})
= \exp\{-\theta\lVert\bm{\ell}_i-\bm{\ell}_{i'}\rVert^2\}$, for $\theta>0$ and $i, i' =1,\dots,n$.
Smaller values of $\theta$ induce stronger dependence among the prior
allocation probabilities of nearby neurons, whereas larger values
attenuate this dependence. We conducted the sensitivity analysis for
Window~\#24 varying $\theta\in\{2,10,100\}$.
The results outlined in the main analysis correspond to $\theta=2$. We first evaluate the robustness of the estimates of the model's parameters: here, inference should not be affected by the prior specification. This is consistent with our findings: the posterior means of the variances $\sigma^2$ and $\tau^2$ are $\{ 0.8002, 0.8016, 0.8047\}$ and $\{ 0.3744, 0.3831, 0.3723 \}$, respectively. The posterior estimates of the decay parameter $\gamma$ are $\{ 0.9581, 0.9590, 0.9589 \}$; and the estimated firing rates are $\{0.0379, 0.0377, 0.0390\}$.

\begin{figure}[t]
    \centering
    \includegraphics[width=.9\linewidth]{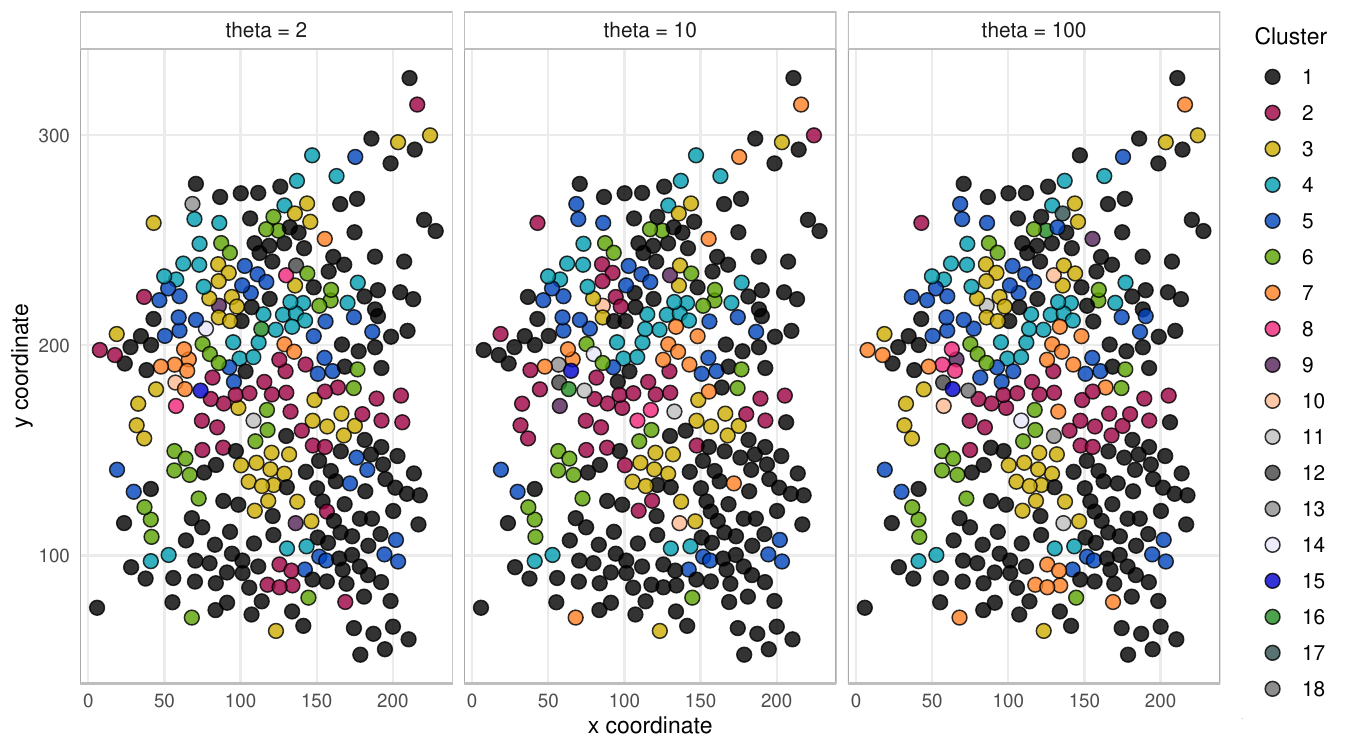}
    \caption{Neurons' locations in the hippocampus colored by cluster allocation, estimated varying the PSB parameter $\theta\in\{2,10,100\}$.}
    \label{fig:PSB_sensitivity}
\end{figure}
We then assess the stability of the estimated clustering of neurons by computing the ARI between pairs of estimated partitions. The ARI between the partitions obtained assuming $\theta=2$ and $\theta=10$ is equal to $0.7363$, when $\theta=10$ and $\theta=100$ is equal to $0.8079$, and between $\theta=2$ and $\theta=100$ is $0.7212$. Hence, for all specifications, the estimated clusterings are quite stable. Figure~\ref{fig:PSB_sensitivity} shows how the estimated partitions are spatially organized in the hippocampus. It is notable how the largest clusters are quite stable and preserve the spatial organization under all specifications, and most discrepancies occur for the smaller clusters. However, it is reasonable for results to slightly differ, especially for neurons that do not clearly belong to a specific activation cluster. For these neurons, a smaller $\theta$ would suggest clustering them in the spatially closest similar activation cluster, while a large $\theta$ would allocate them randomly between the similar clusters.

\subsection{Additional figures on the neuronal response to the mouse position}
\begin{figure}[ht]
    \centering
    \includegraphics[width=.85\linewidth]{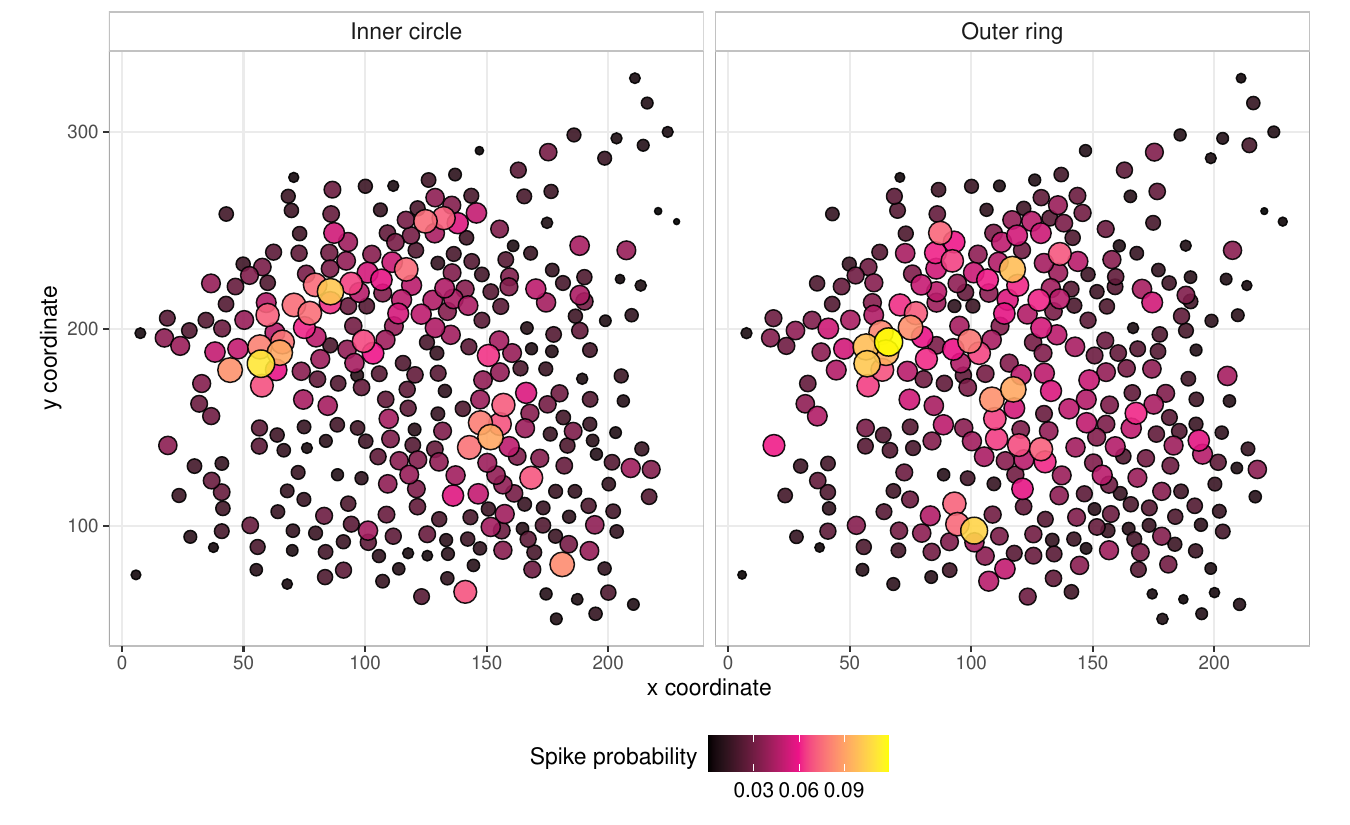}
    \caption{Neurons' locations in the hippocampus. The color and size of each point is proportional to the average spike probability of each neuron when the mouse is in the inner circle (left) or in the outer ring (right) of the arena.}
    \label{fig:average_spikes}
\end{figure}

Figure~\ref{fig:average_spikes} summarizes the average spike probability of each neuron when the mouse is in the inner circle or in the outer ring of the arena. 
The average spike probability was computed as the mean number of spikes over time for each neuron, averaged over all time windows in which the mouse was in the same region of the arena. The figure provides a coarse summary of how neural activity varies with the animal's position.

\subsection{Analysis of additional windows}\label{Asec:additional_windows}
\begin{figure}[hb!]
    \centering
    \includegraphics[width=.9\linewidth]{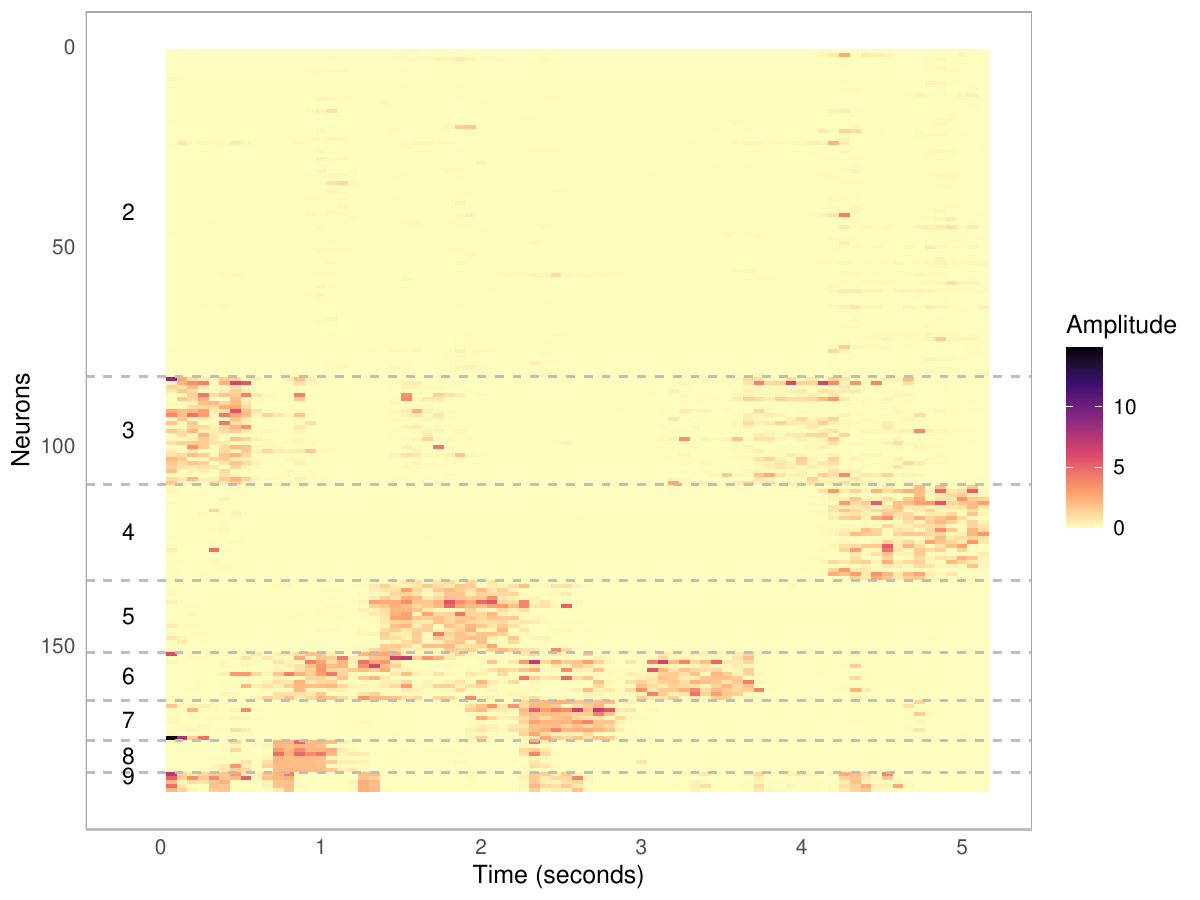}
    \caption{Window \#15: estimated spike amplitudes. The traces are sorted according to the neuron's cluster allocations (Clusters 2 to 9, left labels).}
    \label{fig:AA_by_cluster4}
\end{figure}

\begin{figure}[ht]
    \centering
    \includegraphics[width=\linewidth]{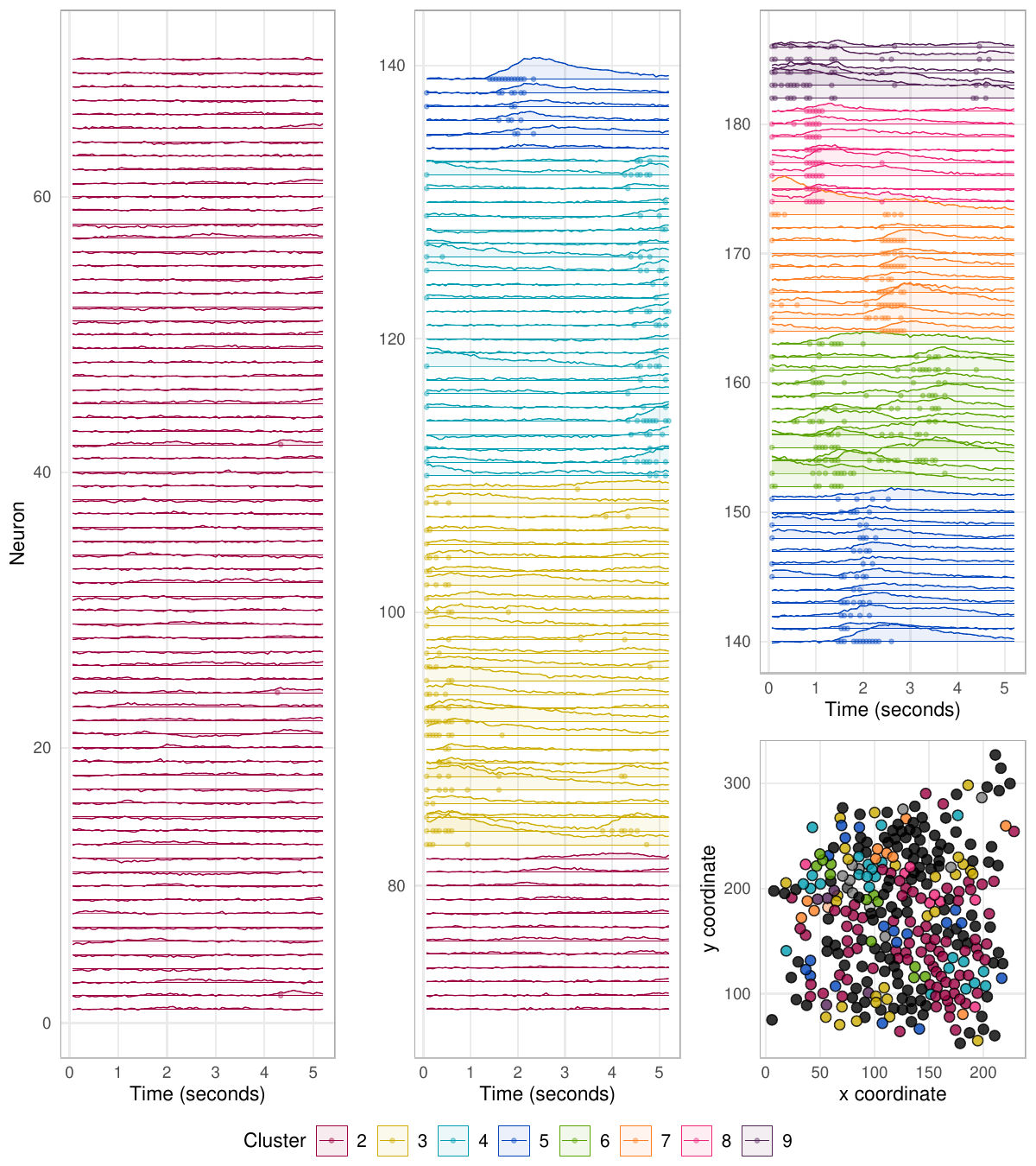}
    \caption{Window \#15. Left, center, and top-right panels: observed calcium traces sorted (and colored) by estimated cluster allocation (only active neurons). Points correspond to the times of the detected firing events. Bottom-right panel: neurons' location in the hippocampus, colored according to the estimated cluster allocations.}
    \label{fig:clusters_and_neurons4}
\end{figure}

\begin{figure}[ht]
    \centering
    \includegraphics[width=\linewidth]{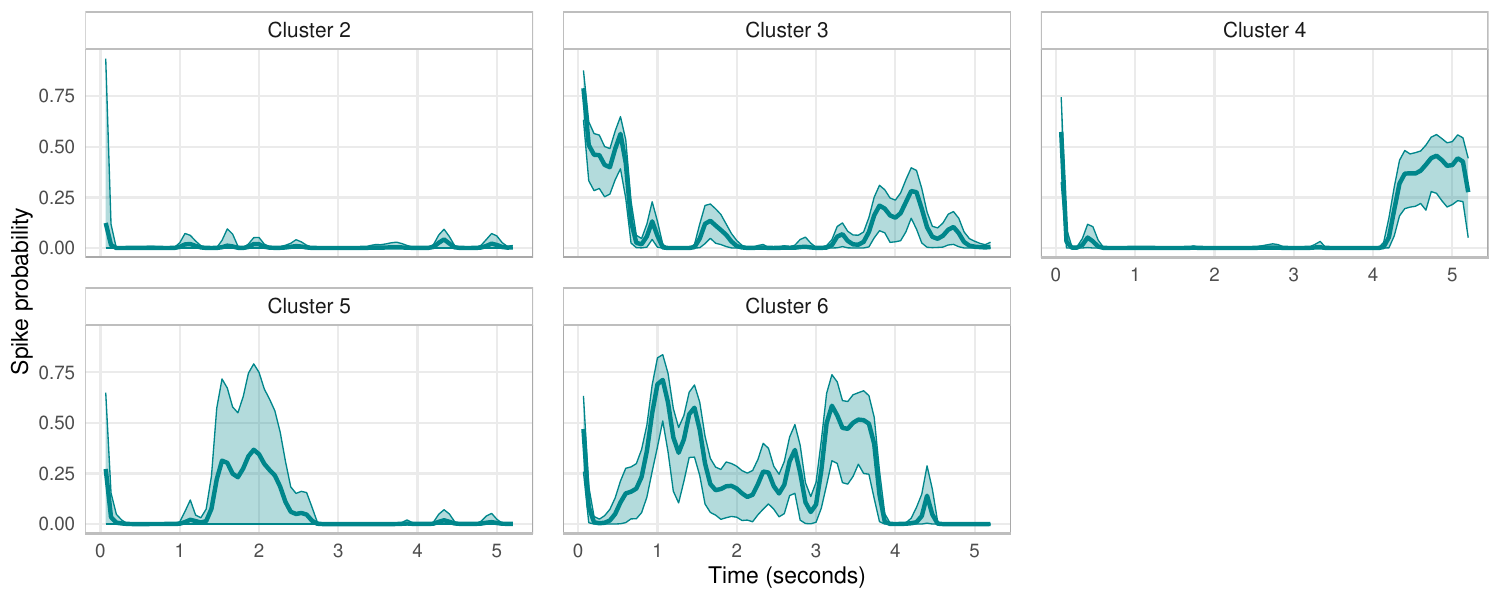}
    \caption{Window \#15. Posterior estimates of the underlying spike probability trajectories for the activation clusters with at least ten neurons. Shaded bands correspond to the point-wise 95\% HPD credible intervals.}
    \label{fig:GP4}
\end{figure}

\begin{figure}[hb!]
    \centering
    \includegraphics[width=\linewidth]{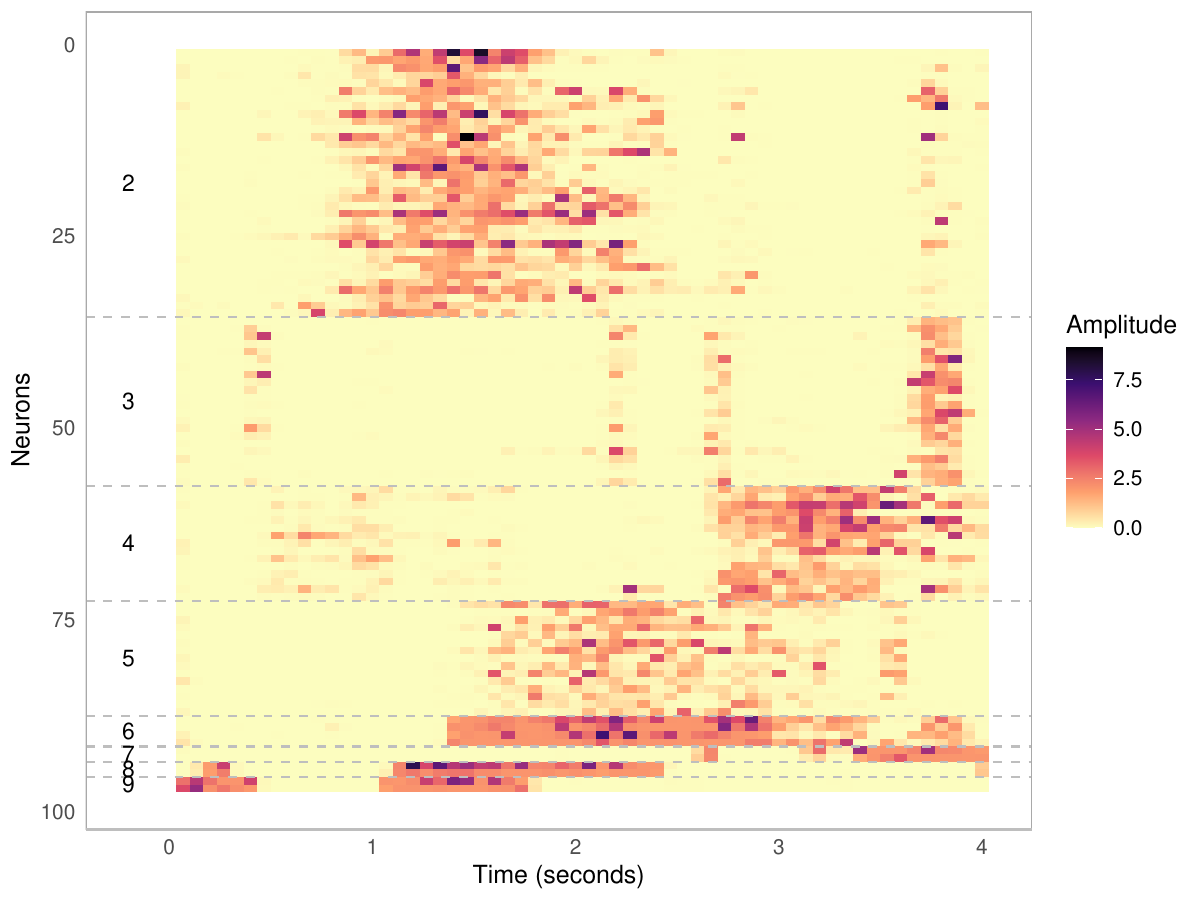}
    \caption{Window \#18: estimated spike amplitudes. The traces are sorted according to the neuron's cluster allocations (Clusters 2 to 9, left labels).}
    \label{fig:AA_by_cluster36}
\vspace{1.5cm}
    \includegraphics[width=\linewidth]{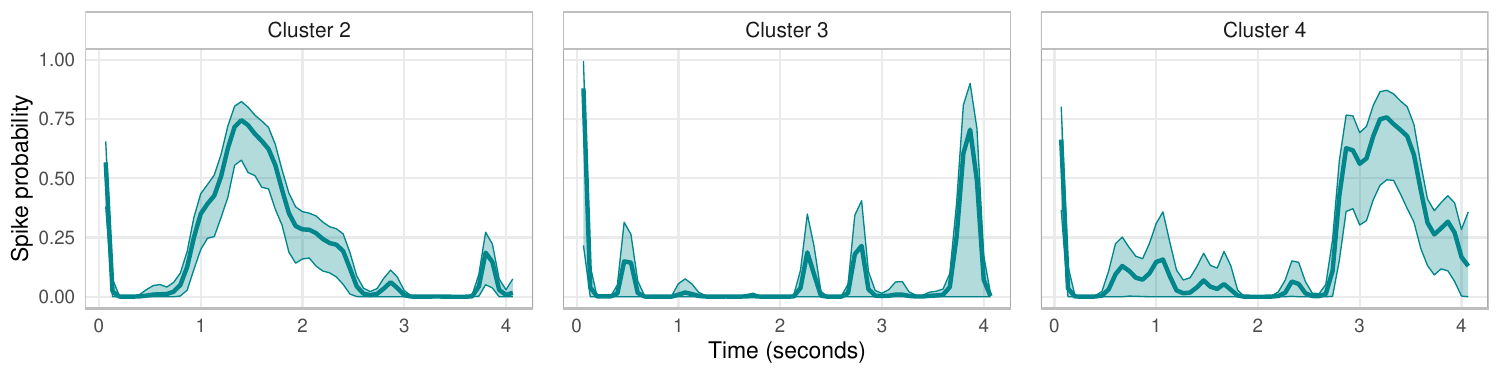}
    \caption{Window \#18. Posterior estimates of the underlying spike probability trajectories for the activation clusters with at least ten neurons. Shaded bands correspond to the point-wise 95\% HPD credible intervals.}
    \label{fig:GP36}
\end{figure}

\begin{figure}[ht]
    \centering
    \includegraphics[width=\linewidth]{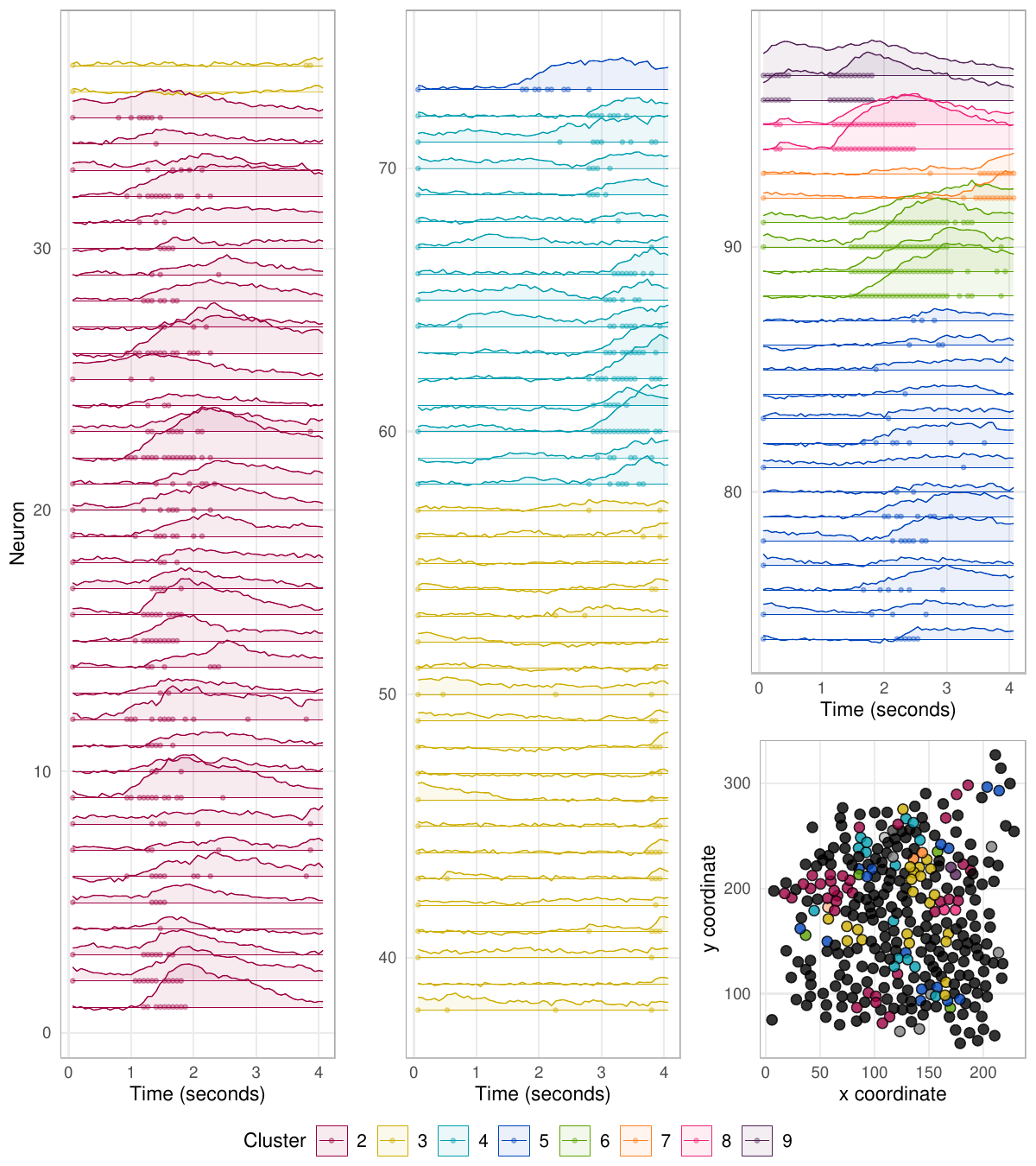}
    \caption{Window \#18. Left, center, and top-right panels: observed calcium traces sorted (and colored) by estimated cluster allocation (only active neurons). Points correspond to the times of the detected firing events. Bottom-right panel: neurons' location in the hippocampus, colored according to the estimated cluster allocations.}
    \label{fig:clusters_and_neurons36}
\end{figure}

\begin{figure}[hb!]
    \centering
    \includegraphics[width=\linewidth]{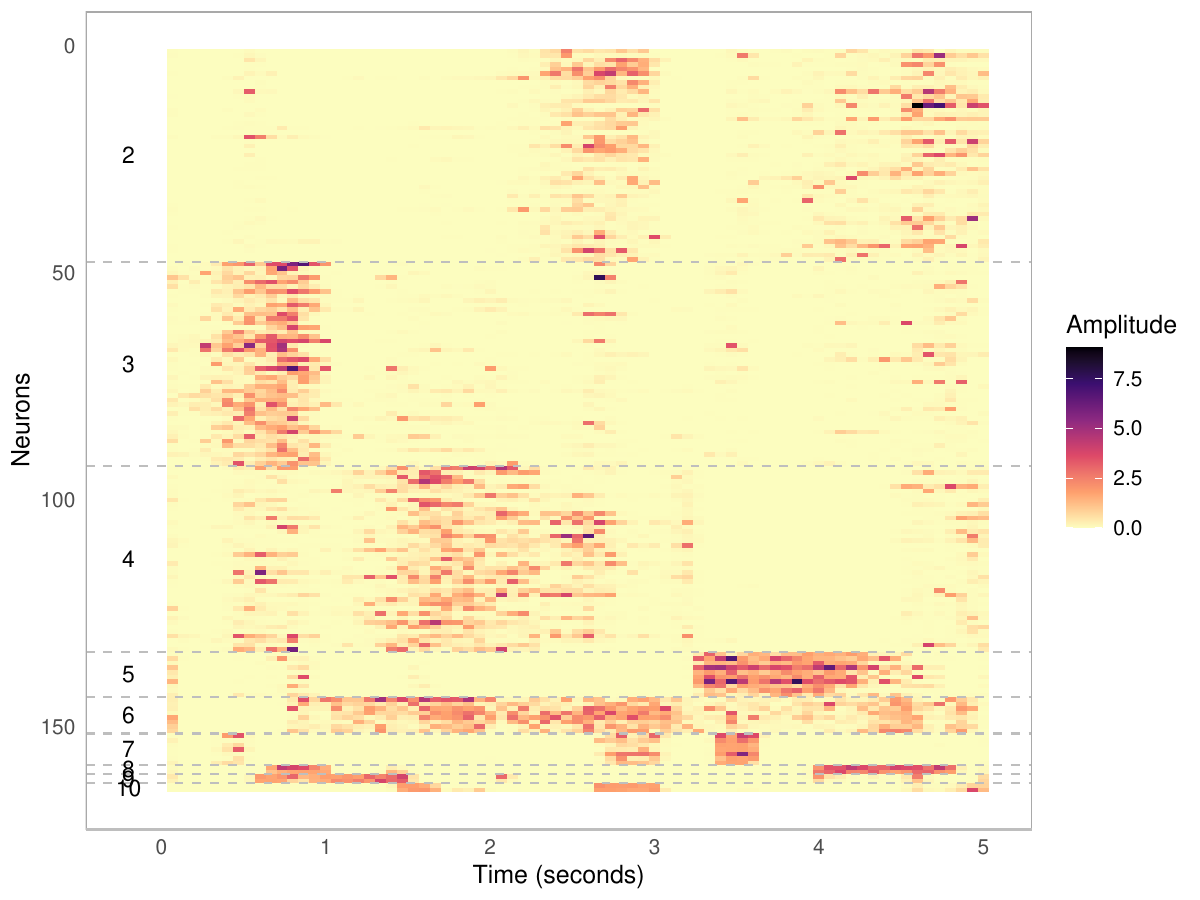}
    \caption{Window \#85: estimated spike amplitudes. The traces are sorted according to the neuron's cluster allocations (Clusters 2 to 10, left labels).}
    \label{fig:AA_by_cluster77}
\vspace{1.5cm}
    \includegraphics[width=\linewidth]{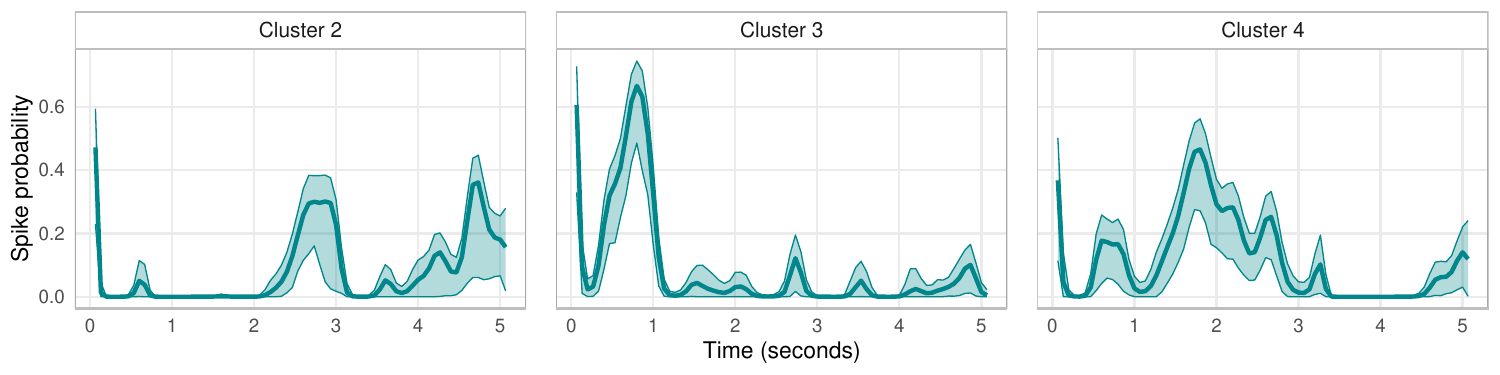}
    \caption{Window \#85. Posterior estimates of the underlying spike probability trajectories for the activation clusters with at least ten neurons. Shaded bands correspond to the point-wise 95\% HPD credible intervals.}
    \label{fig:GP77}
\end{figure}

\begin{figure}[ht]
    \centering
    \includegraphics[width=\linewidth]{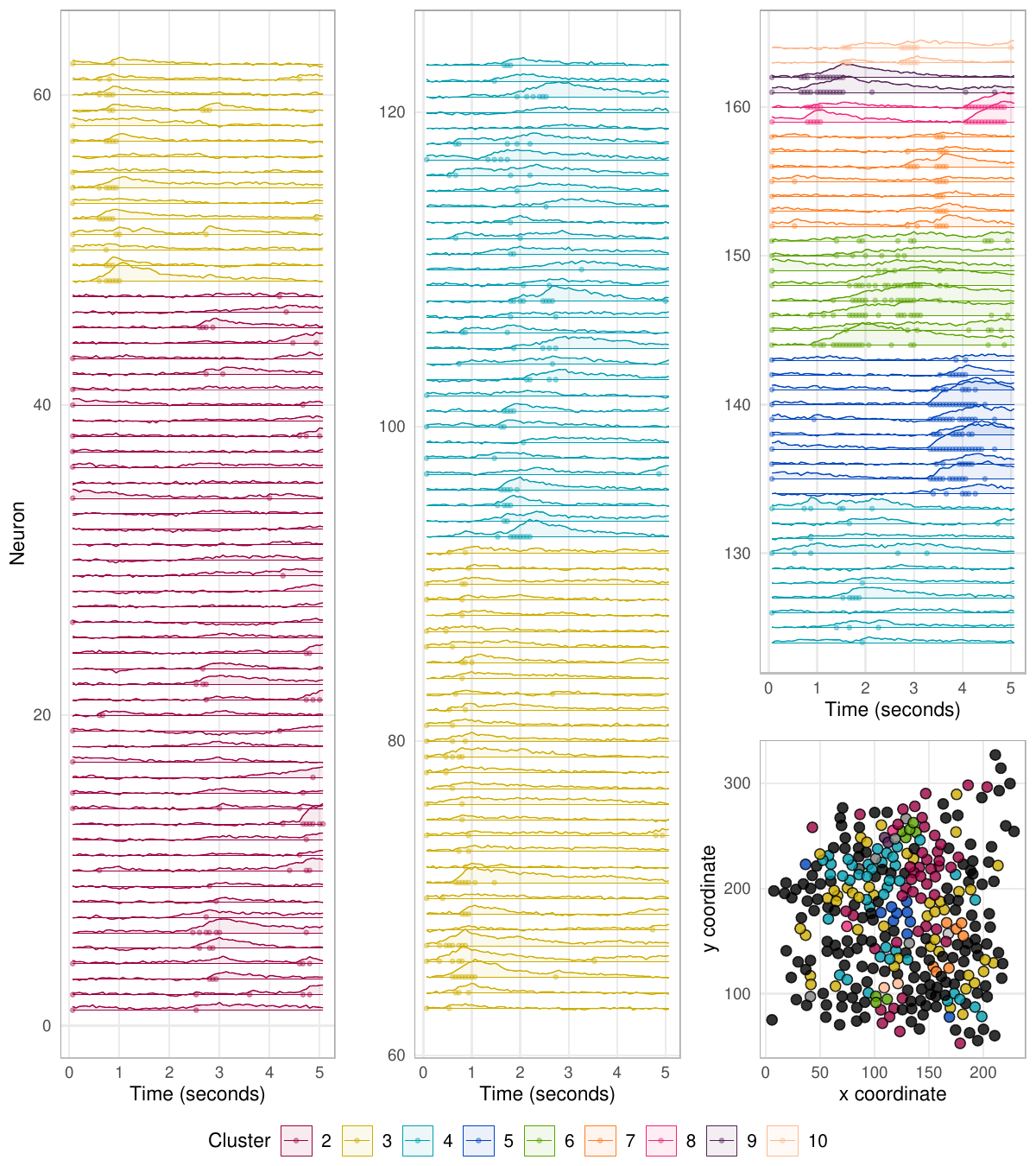}
    \caption{Window \#85. Left, center, and top-right panels: observed calcium traces sorted (and colored) by estimated cluster allocation (only active neurons). Points correspond to the times of the detected firing events. Bottom-right panel: neurons' location in the hippocampus, colored according to the estimated cluster allocations.}
    \label{fig:clusters_and_neurons77}
\end{figure}

Here, we report analogous plots to the ones in the main paper to analyze the results for additional windows. Specifically, we report the heatmap of activation, the calcium traces sorted by cluster allocation, and the cluster-specific spike probability trajectories for three additional time windows, namely, Window~\#15, Window~\#18, and Window~\#85.
Figures~\ref{fig:AA_by_cluster4},~\ref{fig:clusters_and_neurons4}, and~\ref{fig:GP4} correspond to Window~\#15; Figures~\ref{fig:AA_by_cluster36},~\ref{fig:clusters_and_neurons36}, and~\ref{fig:GP36} to Window~\#18; and Figures~\ref{fig:AA_by_cluster77},~\ref{fig:clusters_and_neurons77}, and~\ref{fig:GP77} to Window~\#85.

%%%% window 15
\subsubsection*{Window \#15}

%%%% window 18
\clearpage
\subsubsection*{Window \#18}

%%%% window 85
\clearpage
\subsubsection*{Window \#85}

\begin{figure}[ht]
    \centering
    \includegraphics[width=\linewidth]{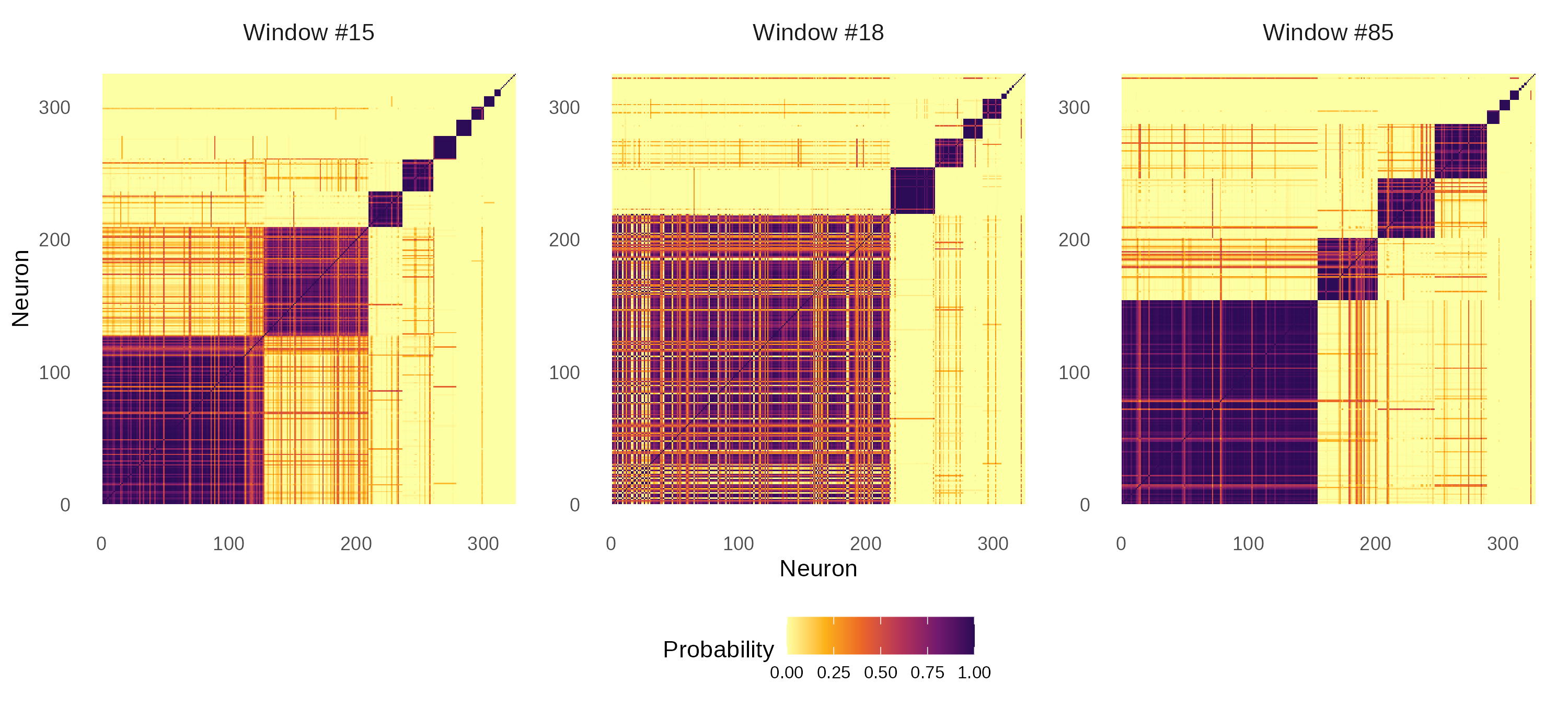}
    \caption{Posterior similarity matrices for the considered time windows. The color of each cell corresponds to the co-clustering probability of the corresponding pair of neurons.}
    \label{fig:PSM2}
\end{figure}

\clearpage
Finally, Figure~\ref{fig:PSM2} reports the posterior similarity matrices for the additional windows analyzed here.

\section{Additional analyses with state-of-the-art methods}\label{Asec:comparisons}
\subsection{Gaussian process factor analysis}\label{Asec:GPFA}
\begin{figure}[ht!]
    \centering
    \includegraphics[width=.8\linewidth]{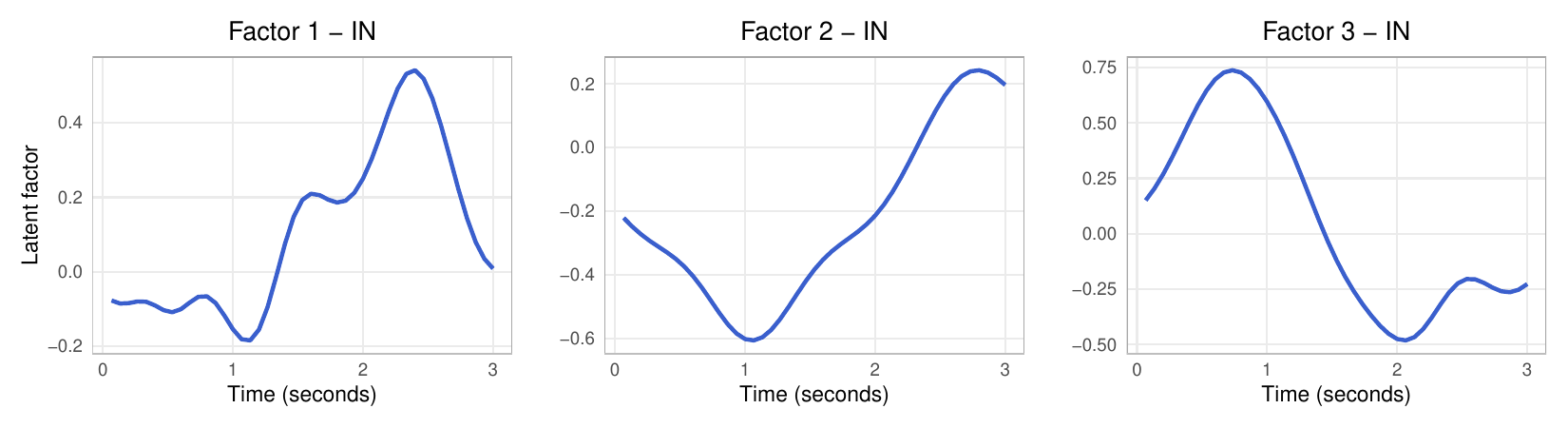}
    \includegraphics[width=.8\linewidth/3]{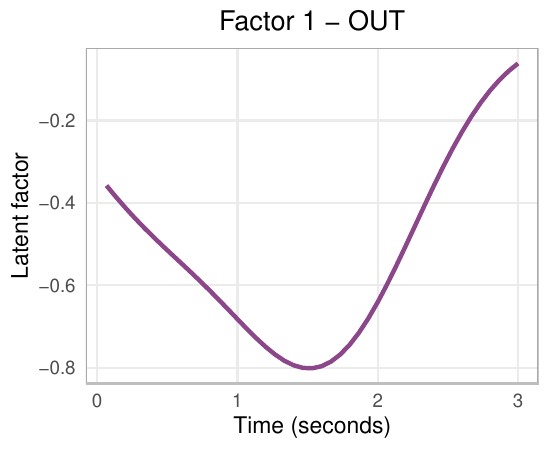}
    \caption{Latent factors estimated with bGPFA on the deconvolved signal. Top: factors when the mouse is in the inner circle; bottom: factors when the mouse is on the border. }
    \label{fig:GPFA_factors}
\end{figure}

In this section we report the results of the Bayesian Gaussian process factor analysis (bGPFA) of \cite{Jensen2021}, whose implementation is available in the Github repository \url{github.com/tachukao/mgplvm-pytorch}. To investigate whether the neuronal activity is different in the center and outer ring of the environment, we set up a multi-trial formulation. We consider the previously defined time windows and then construct two 3-dimensional arrays for the two experimental conditions. Each slice of the array contains the neuronal activity recordings for the observed neurons, while slices correspond to the various time windows in which the mouse is in the same region of the arena. To make the temporal dimension coherent among windows, we consider the first three seconds of each chunk of the experiment. bGPFA assumes a negative binomial likelihood for the neuronal activity. Thus, it works best when the data consists of spike counts. To obtain such data, we first deconvolved the traces using the method of \cite{jewell2019}, and, for each time point, we counted how many unitary spikes were observed (the spike amplitude was set as described in Section~\ref{sec:realdata_comparison} of the main paper). We then ran the algorithm separately for the inner and outer trials, setting the maximum number of factors to 25. The algorithm allows inferring the number of factors from the data by discarding the unnecessary ones through an approximate marginal likelihood approach. On our data, the model identifies three informative factors in the inner circle, and only one factor in the outer ring. This is an indication of a different level of complexity of the neural states depending on the mouse's position in the arena, and it is consistent with our findings in Section~\ref{subsec:response_position}.
Figure~\ref{fig:GPFA_factors} shows the posterior estimates of the relevant factor as functions of time for the two experimental conditions. However, these dynamics are not directly comparable with those obtained with our method, since the latent processes describe fundamentally different aspects of the neural state.  

\subsection{Two-step deconvolution and clustering}

\label{sec:realdata_comparison_suppl}
\begin{figure}[ht!]
    \centering
    \includegraphics[width=.7\linewidth]{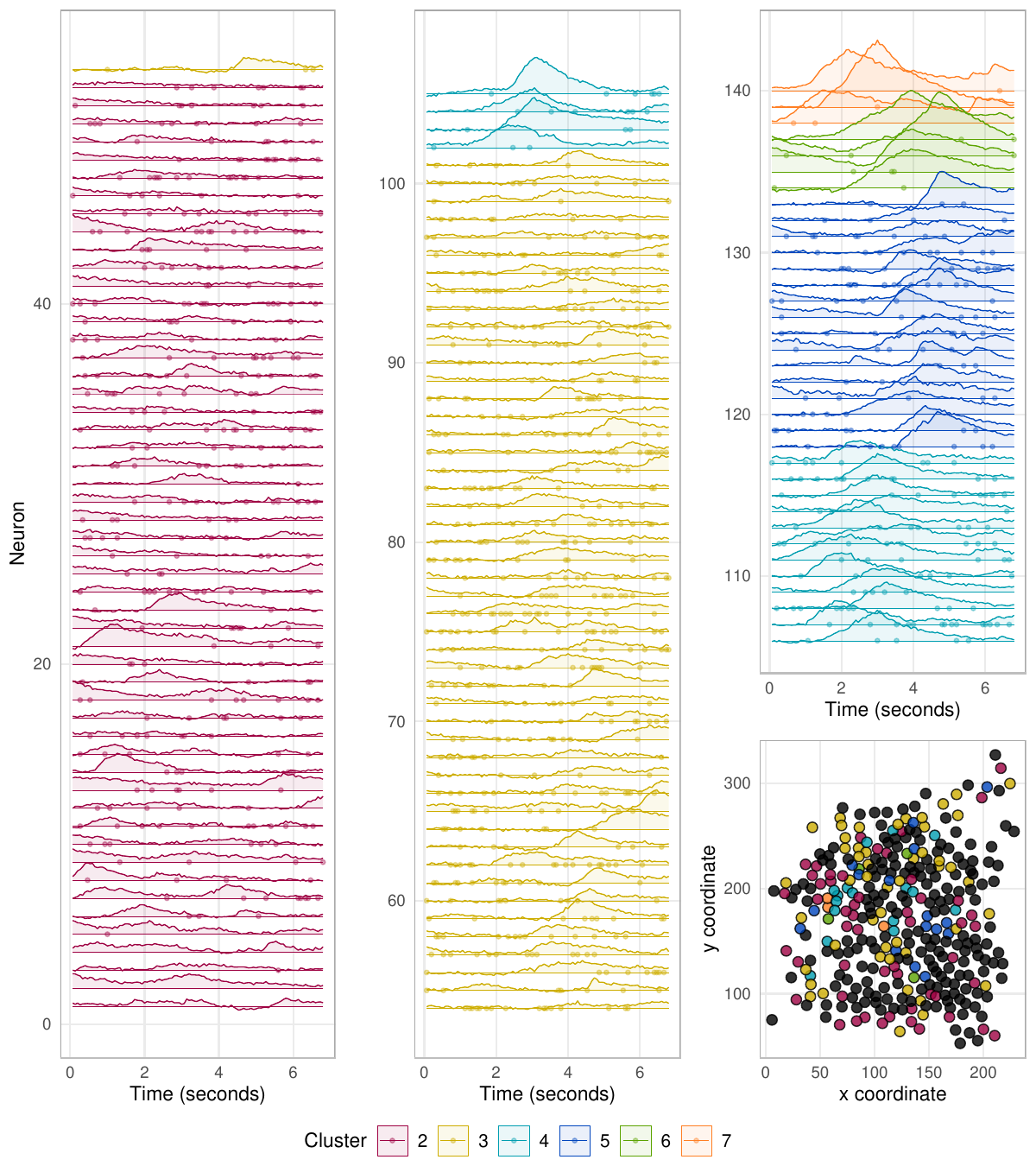}
    \caption{Left, center, and top-right panels: observed calcium traces sorted (and colored) by estimated cluster allocation (only active neurons), using the two-step procedure. Points correspond to the times of the detected firing events. Bottom-right panel: neurons' location in the hippocampus, colored according to the estimated cluster allocation. \label{fig:clusters_and_neurons_two-step}}
\end{figure}

\begin{figure}[hb!]
    \centering
    \includegraphics[width=\linewidth]{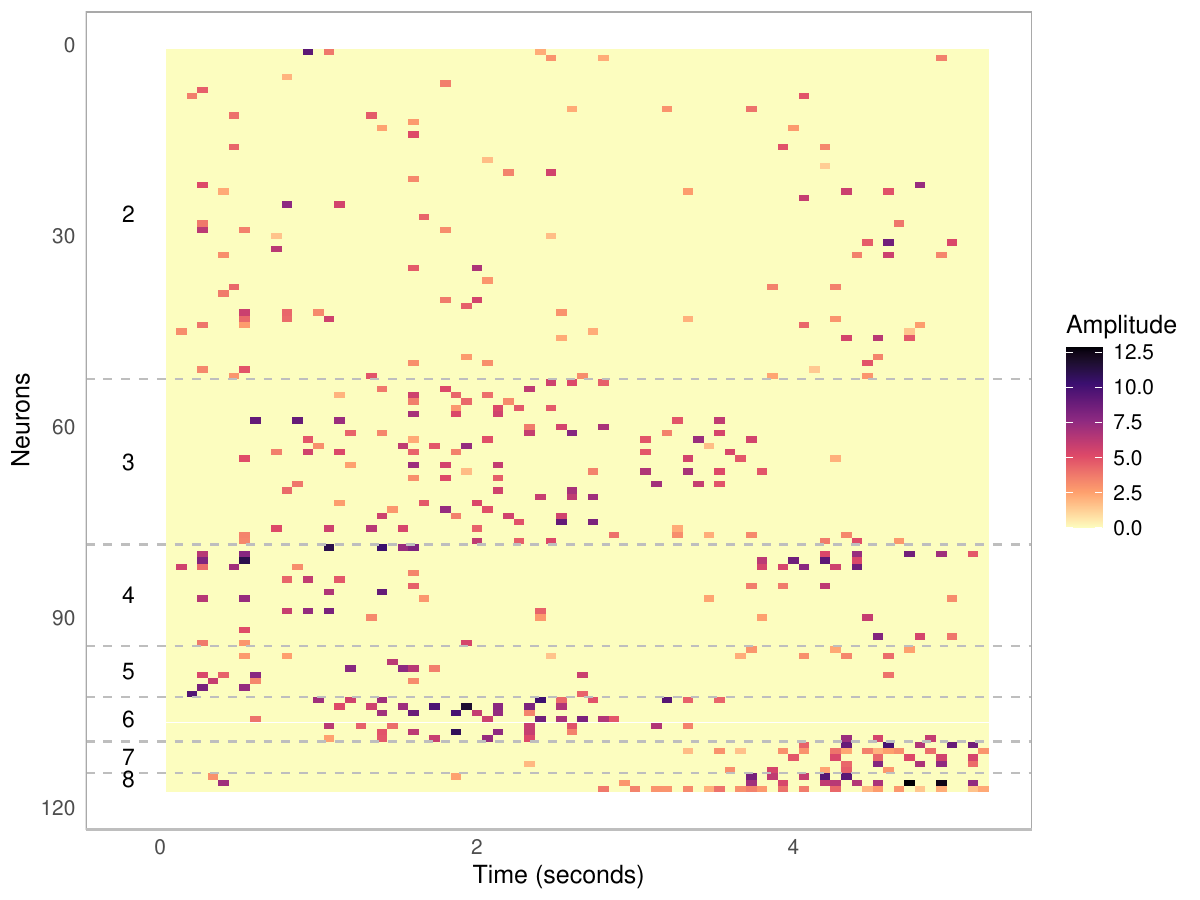}
    \caption{Window \#15: estimated spike amplitudes using the two-step approach. The traces are sorted according to the neuron's cluster allocations (Clusters 2 to 4, left labels).}
    \label{fig:AA_by_cluster4_two-step}
\end{figure}

\begin{figure}[ht]
    \centering
    \includegraphics[width=\linewidth]{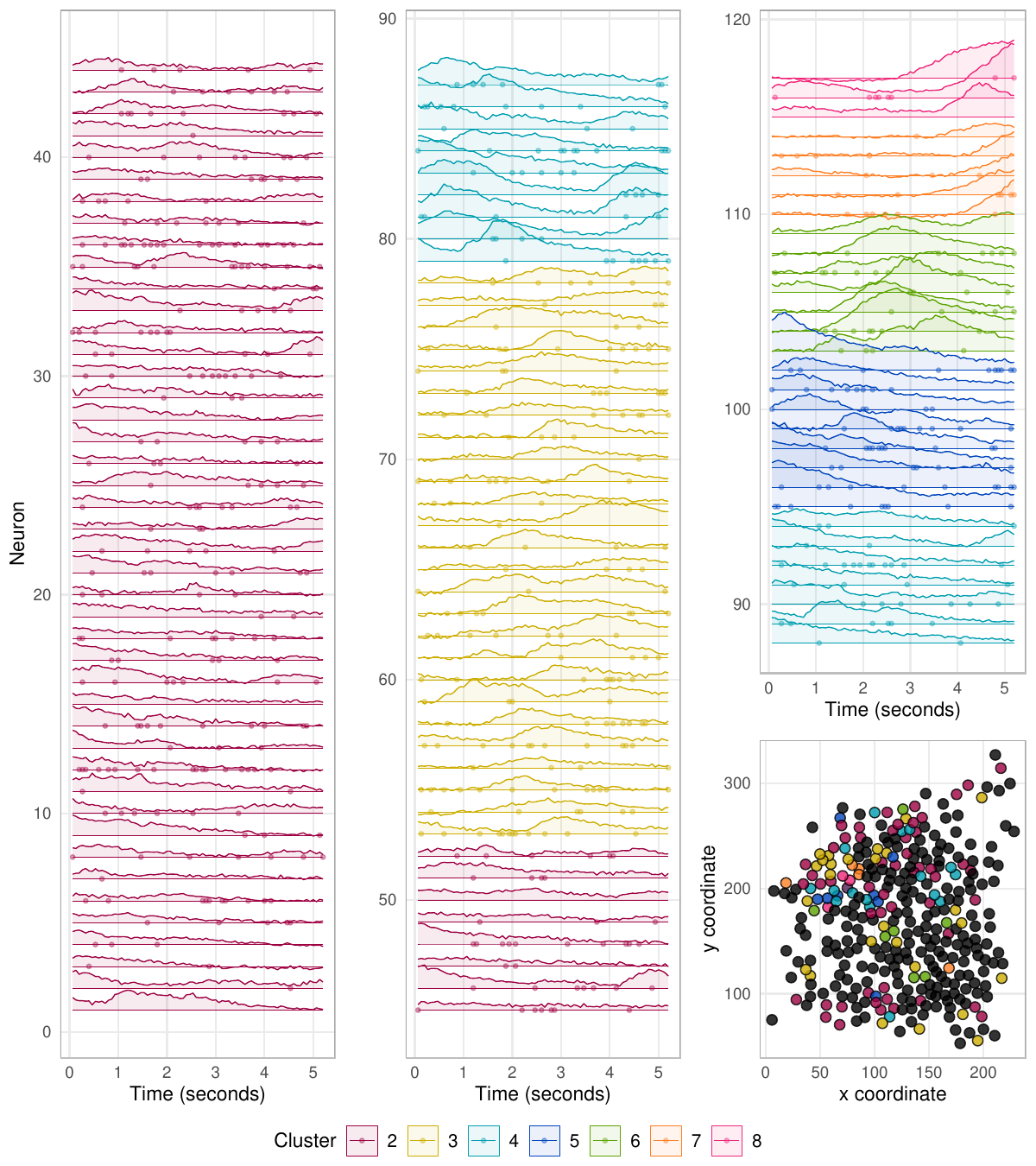}
    \caption{Window \#15. Left, center, and top-right panels: observed calcium traces sorted (and colored) by estimated cluster allocation (only active neurons) using the two-step approach. Points correspond to the times of the detected firing events. Bottom-right panel: neurons' location in the hippocampus, colored according to the estimated cluster allocations.}
    \label{fig:clusters_and_neurons4_two-step}
\end{figure}

\begin{figure}[hb!]
    \centering
    \includegraphics[width=\linewidth]{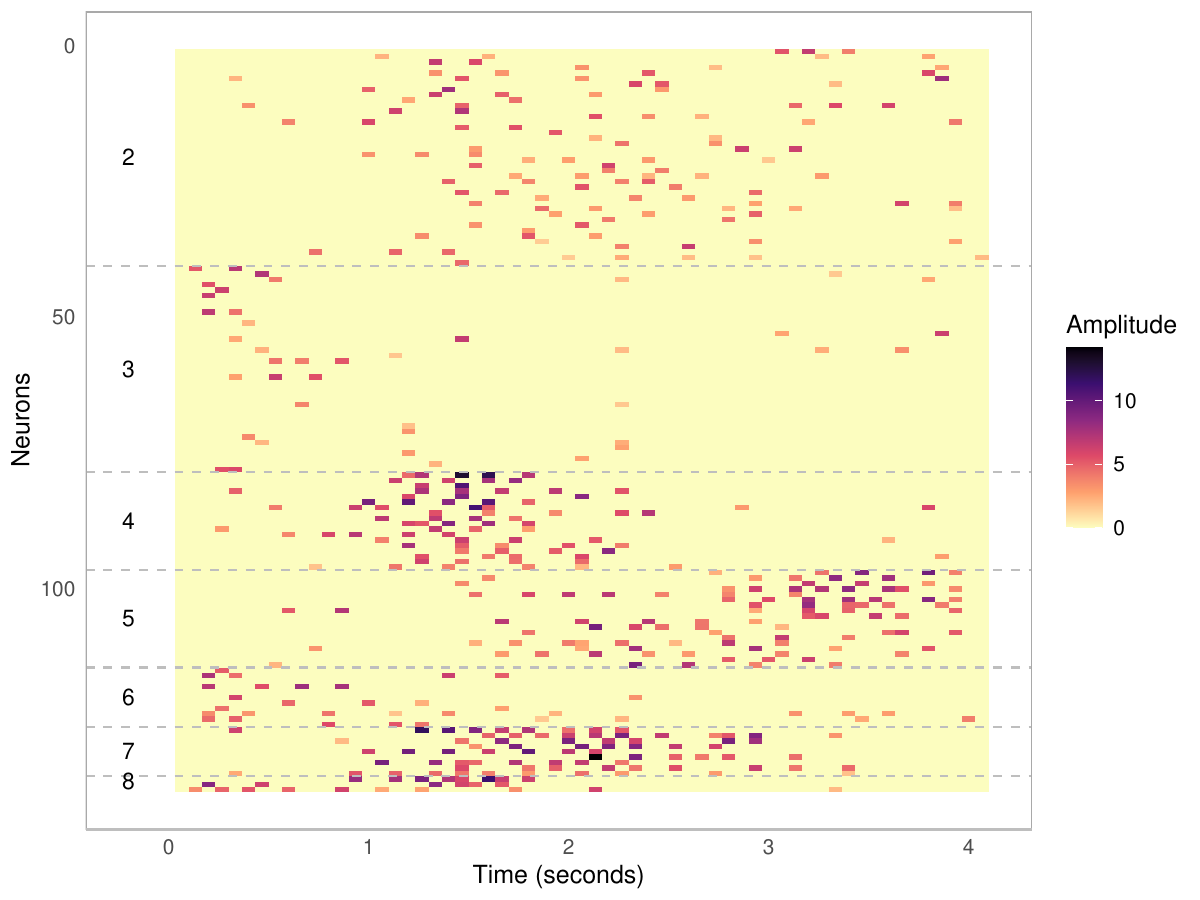}
    \caption{Window \#18: estimated spike amplitudes using the two-step approach. The traces are sorted according to the neuron's cluster allocations (Clusters 2 to 7, left labels).}
    \label{fig:AA_by_cluster36_two-step}
\end{figure}

\begin{figure}[ht]
    \centering
    \includegraphics[width=\linewidth]{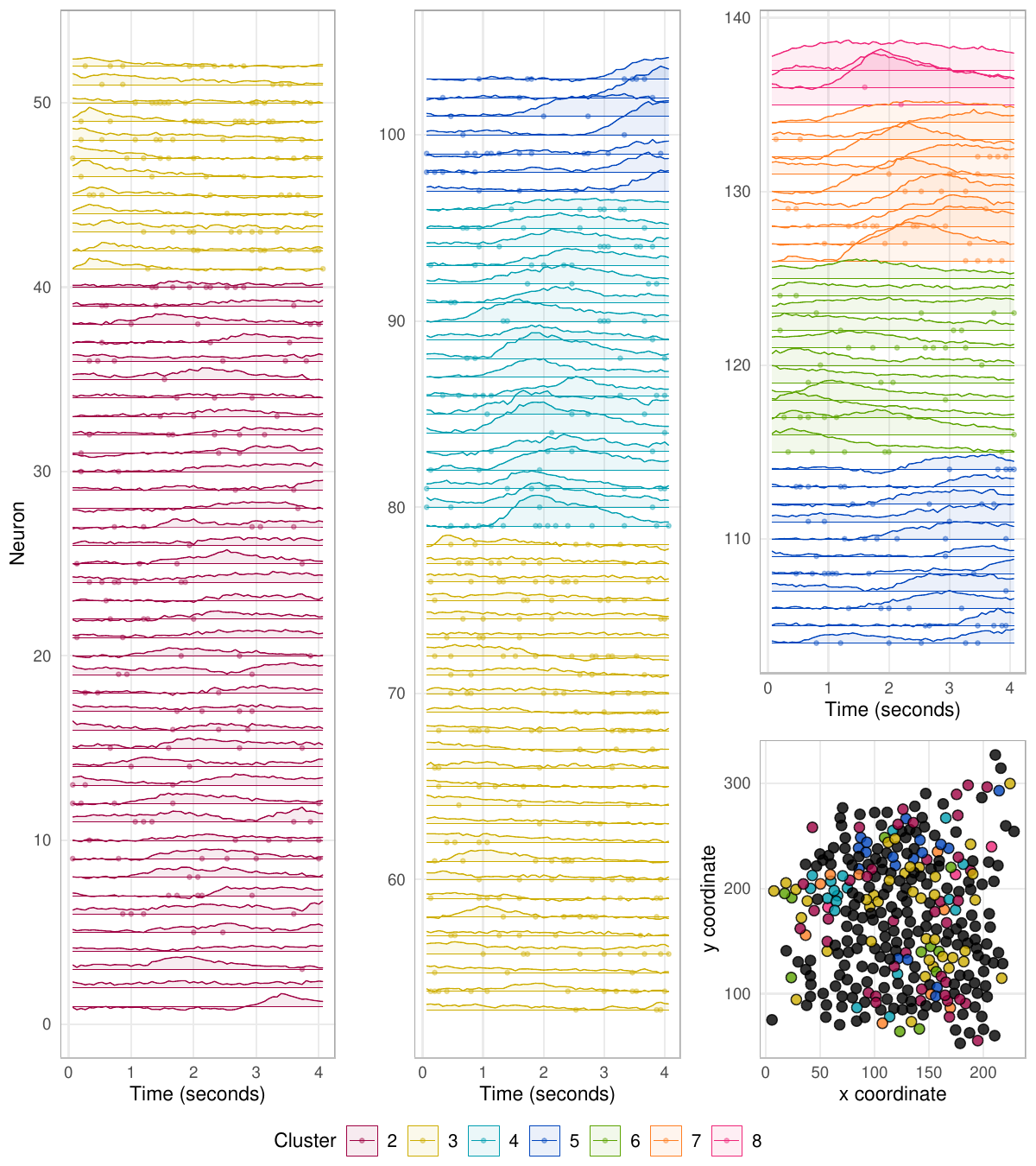}
    \caption{Window \#18. Left, center, and top-right panels: observed calcium traces sorted (and colored) by estimated cluster allocation (only active neurons) using the two-step approach. Points correspond to the times of the detected firing events. Bottom-right panel: neurons' location in the hippocampus, colored according to the estimated cluster allocations.}
    \label{fig:clusters_and_neurons36_two-step}
\end{figure}

\begin{figure}[hb!]
    \centering
    \includegraphics[width=\linewidth]{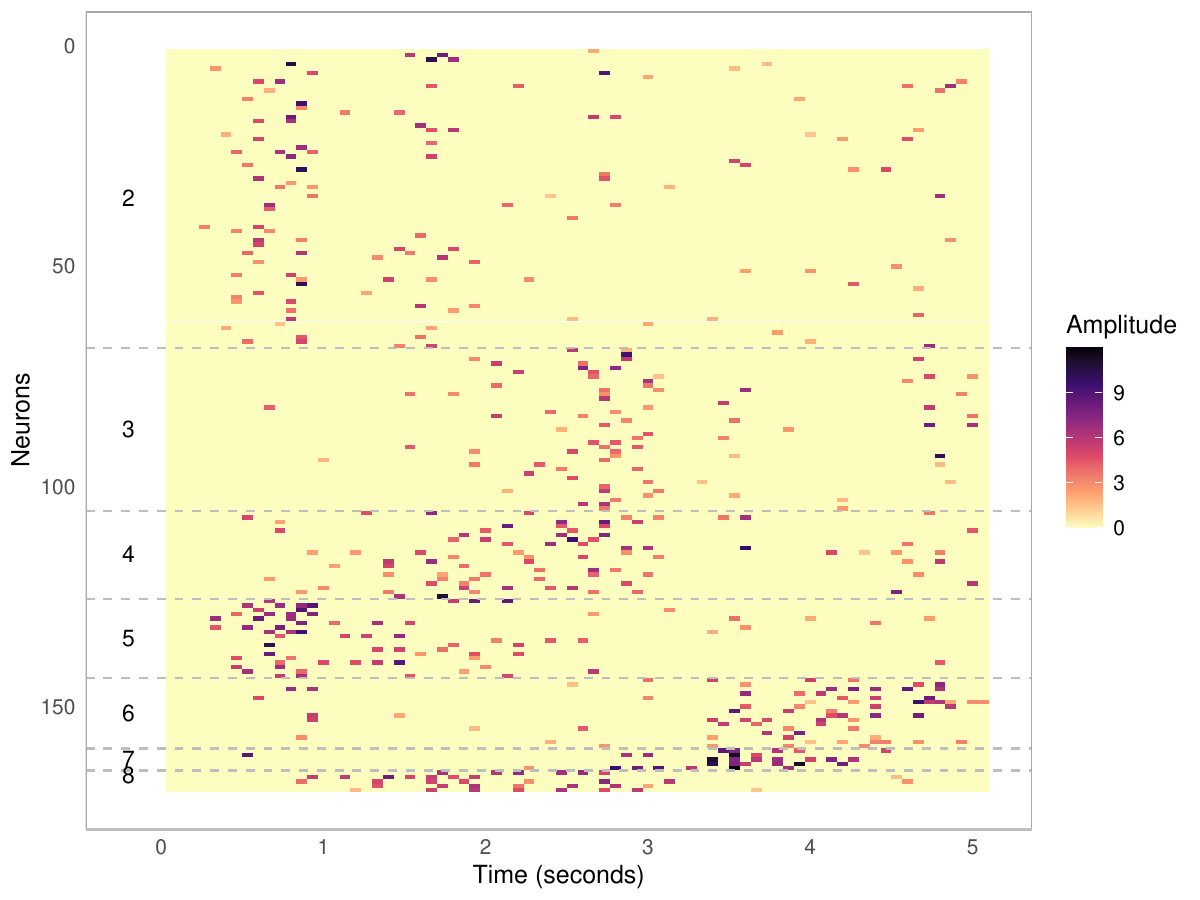}
    \caption{Window \#85: estimated spike amplitudes using the two-step approach. The traces are sorted according to the neuron's cluster allocations (Clusters 2 to 5, left labels).}
    \label{fig:AA_by_cluster77_two-step}
\end{figure}

\begin{figure}[ht]
    \centering
    \includegraphics[width=\linewidth]{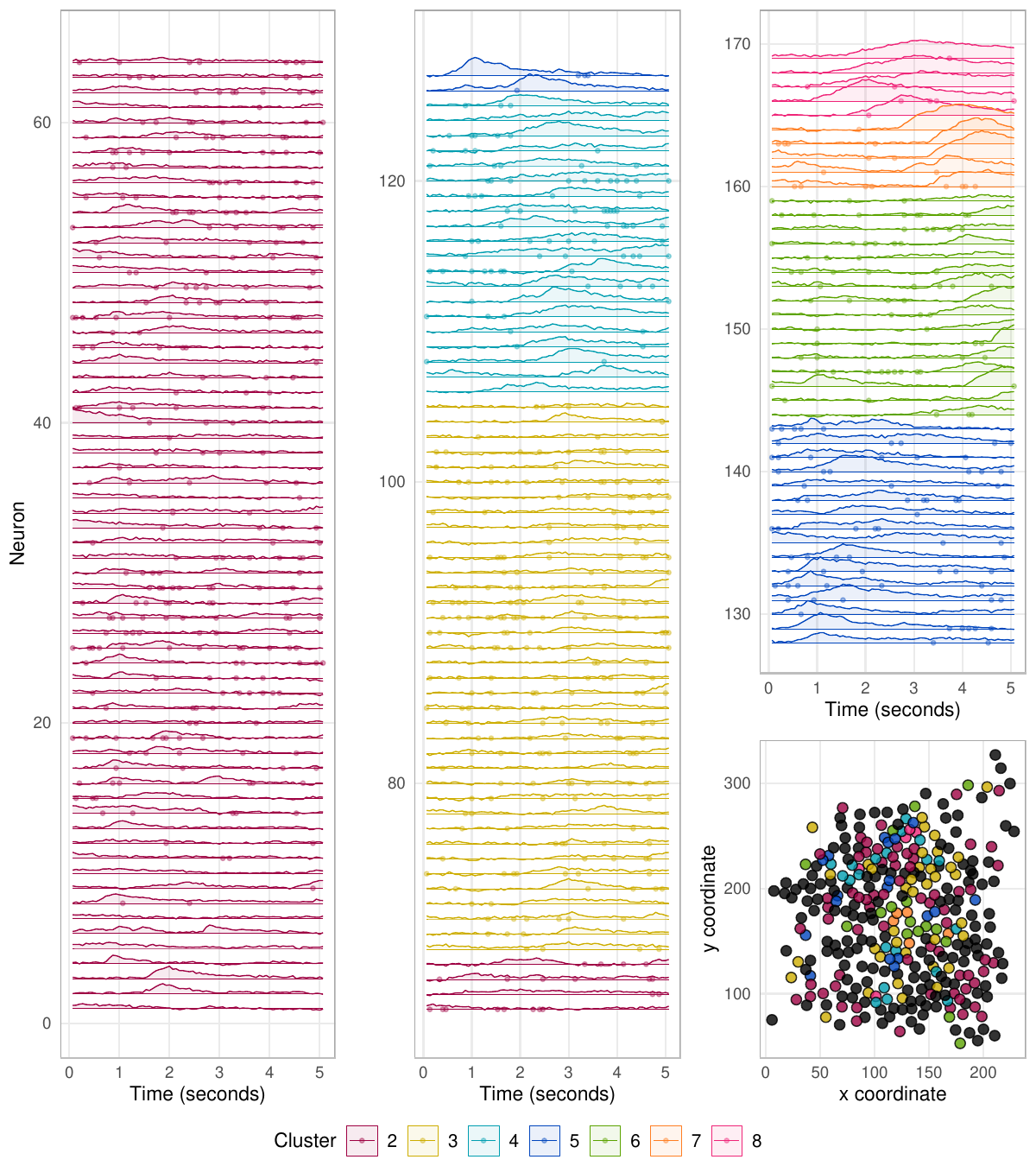}
    \caption{Window \#85. Left, center, and top-right panels: observed calcium traces sorted (and colored) by estimated cluster allocation (only active neurons) using the two-step approach. Points correspond to the times of the detected firing events. Bottom-right panel: neurons' location in the hippocampus, colored according to the estimated cluster allocations.}
    \label{fig:clusters_and_neurons77_two-step}
\end{figure}

In this section, we report analogous plots to the ones in the main paper to analyze the results using the two-step approach. 
Specifically, in Figure~\ref{fig:clusters_and_neurons_two-step} we report the analogous of Figure~\ref{fig:clusters_and_neurons} for the data of Section~\ref{sec:representative_win}. We then report the heatmap of activation and the calcium traces sorted by cluster allocation for three additional time windows, namely, window~\#15, window~\#18, and window~\#85. These plots are displayed in  
Figures~\ref{fig:AA_by_cluster4_two-step} and~\ref{fig:clusters_and_neurons4_two-step} (window~\#15); Figures~\ref{fig:AA_by_cluster36_two-step} and ~\ref{fig:clusters_and_neurons36_two-step} (window~\#18); and Figures~\ref{fig:AA_by_cluster77_two-step} and ~\ref{fig:clusters_and_neurons77_two-step} (window~\#85).

\clearpage

%%%% window 4

\subsubsection*{Window \#15}

%%%% window 36
\clearpage
\subsubsection*{Window \#18}

%%%% window 77
\clearpage
\subsubsection*{Window \#85}

\clearpage
\section{Model evaluation and comparison on synthetic data}\label{sec:simulations}

In this section, we test and validate the performance of the proposed approach on simulated data. In particular, we focus on evaluating the accuracy of the spike detection and the clustering of neurons. For our method, we ran the algorithm of Section~\ref{sec::posterior_inference} for 15,000 iterations and discarded the first 10,000 as burn-in.

\subsection{Data-generating mechanism}
We generated synthetic data to closely replicate the observed characteristics of calcium traces. The dataset consists of artificial fluorescence traces spanning $T=80$ time points for $n=100$ neurons. 
Neurons were assigned to $K = 3$ activation ensembles with spatial proximity guiding cluster membership.
In particular, we resorted to a finite mixture model with largely overlapping components to reflect the idea that similar activation patterns should be spatially related but not clearly distinct in real data. We replicated the experiment on 50 independent datasets.

\paragraph{Activation patterns.} 
For each cluster, we generated a time-varying function to model the evolution of spike probabilities.
Specifically, we began by constructing piecewise constant functions to delineate periods of activity and rest, and smoothed them using a polynomial regression. These functions served as the mean trajectories of Gaussian processes, which were then sampled to introduce irregularity and variability across simulation replicates. The resulting realizations were transformed to the $[0,1]$ interval via a probit link and interpreted as time-dependent spike probabilities.

\begin{figure}
    \centering
    \includegraphics[width=0.5\linewidth]{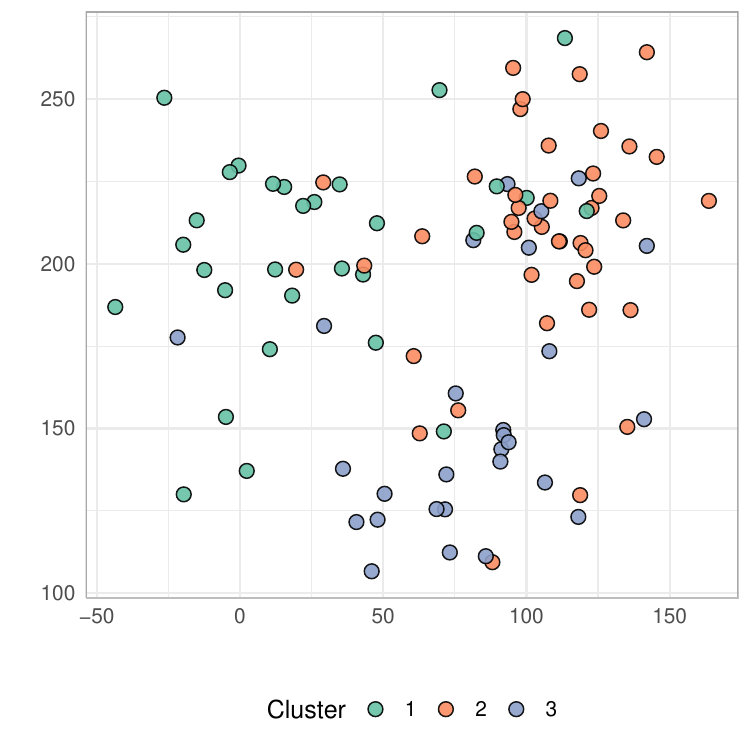}
    \caption{Example of the generated locations of neurons and their cluster allocation.}
    \label{fig:simulated_location_neurons}
\end{figure}

\paragraph{Spatially related neuronal clusters.} We generated the spatial locations of the neurons along with their cluster assignments $(\zeta_1, \dots, \zeta_n)$ from three clusters. Recall that neurons belonging to the same cluster share the same activation pattern, and clusters are assumed to exhibit spatial coherence. 
Hence, we generated the locations and cluster labels jointly from a mixture of three components with weights $\{0.25, 0.33, 0.42\}$. Each component is a bivariate normal whose realizations encode the spatial coordinates of neurons. Specifically, the first component is a $\mathrm{N}_2([0, 200]^T, 1000 \cdot I)$, the second component is a $\mathrm{N}_2([110, 220]^T, 500 \cdot I)$, while the third one is a $\mathrm{N}_2([100, 150]^T, 700 \cdot I)$. 
This setup yields spatially contiguous yet partially overlapping clusters, as illustrated in Figure~\ref{fig:simulated_location_neurons}.
Consistent with our model, all neurons within the same cluster share the same baseline spike probabilities. Hence, given the cluster allocation variables, we generated each spike train $\bm{s}_i$ so that, when $\zeta_i = k$, the spike probabilities were governed by the $k$-th cluster-specific function. This construction ensures that neurons within the same cluster exhibit coordinated activation patterns, while allowing for variability in the precise timing of spiking events across individual neurons—reflecting the natural heterogeneity of co-active neuronal ensembles.
The actual spike trains were generated with Bernoulli random variables using the simulated dynamic probabilities.
Each non-zero spike indicator was then associated with a corresponding spike amplitude. These amplitudes were sampled from five possible distinct values, $a^*_j \in \{3, 6, 10, 14, 25\}$, with probabilities $\{0.1, 0.3, 0.4, 0.1, 0.1\}$. These values are coherent with the amplitudes estimated on the true traces from a preliminary analysis. 
These spike trains were then used as input signals in the calcium dynamics model described in Equation~\eqref{eq:calcium_dyn}, to simulate the autoregressive structure typically observed in real fluorescence data. The calcium decay parameter was set to $\gamma = 0.9$, while the variances were fixed at $\tau^2 = 1$ and $\sigma^2 = 1.5$.\\

\subsection{Sensitivity study}\label{Asec:sensit}
We investigated the sensitivity of the model to the prior specification on spike amplitudes by evaluating its impact on both spike detection (in terms of misclassification error) and neuronal clustering. Recall that spike amplitudes are modeled using a shifted gamma distribution, $a^*_j - \bar{a} \sim \text{Gamma}(\alpha_a, \beta_a)$. The choice of this prior is crucial: a distribution concentrated near zero may lead the model to misinterpret random fluctuations in the calcium signal as spikes, while a prior with negligible mass near zero may cause the model to overlook small-amplitude spikes. At the same time, the prior must be sufficiently flexible to allow for the detection of unusually large amplitudes. Hence, to assess robustness to prior elicitation, we conducted a sensitivity analysis.

\begin{figure}
\centering
    \includegraphics[width=.9\linewidth]{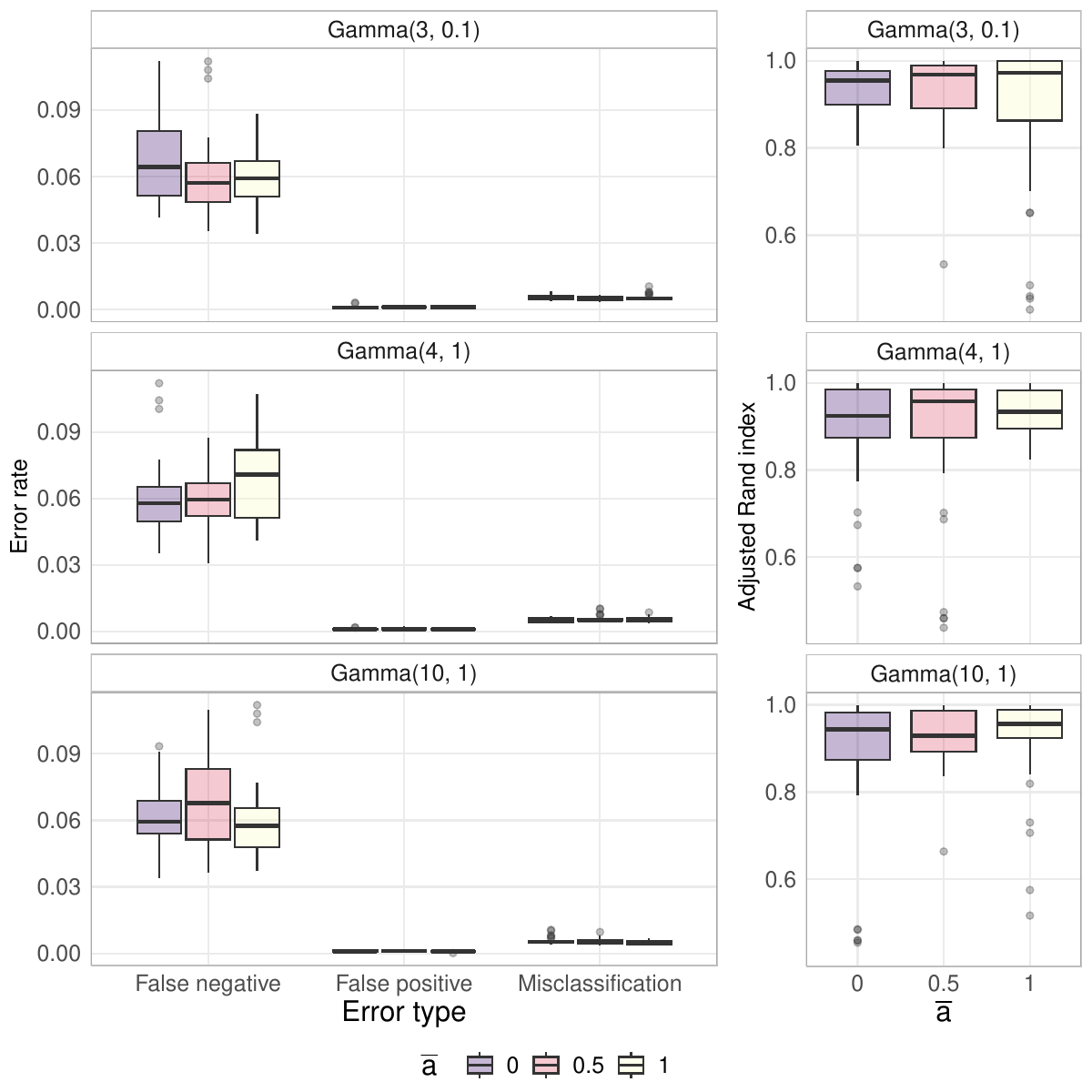}
    \caption{Sensitivity analysis on $(\alpha_a,\beta_a)$ and $\bar{a}$. Rows correspond to different gamma hyperparameters, and colors correspond to different shift parameters. Left panel: boxplots of the false negative, false positive, and misclassification error rates of the spike detection for different combinations of hyperparameters. Right panel: boxplots of the Adjusted Rand Index between the true and estimated partition of neurons.}
    \label{fig:sim_pars_error}
\end{figure}

We considered three different hyperparameter configurations of the gamma density and three shift parameters, studying all possible combinations, for a total of nine configurations. Specifically, the gamma hyperparameters were fixed to $(\alpha_a, \beta_a) \in \{(3, 0.1), (4, 1), (10, 1)\}$. The first specification induces a high-variance prior with substantial mass near zero, allowing for a broad range of amplitudes but at the risk of detecting spurious spikes. In contrast, the latter two priors assign negligible probability mass near zero, reducing the risk of false positives. However, their lower variance may limit the model’s ability to capture spikes with atypical amplitudes, especially under the $(4,1)$ specification. The shift parameter was varied in $\bar{a} \in \{0, 0.5, 1\}$.

The left panels of Figure~\ref{fig:sim_pars_error} report the distribution of spike detection error rates across 50 simulated datasets. All nine prior specifications yield comparable and satisfactory performance, suggesting robustness of the spike detection to prior specification. 
The right panels of the same figure display the distribution of the ARI comparing the estimated and true neuron partitions. 
Also in this case, the results are stable under all prior specifications.
Overall, the prior with hyperparameters $(10,1)$ and $\bar{a}=1$ achieves a slightly lower misclassification error, higher ARI, and smaller variability, supporting it as the most effective among those considered.

\subsection{Computational cost} \label{sec:comput_cost}
\begin{figure}
\centering
\includegraphics[width=0.85\linewidth]{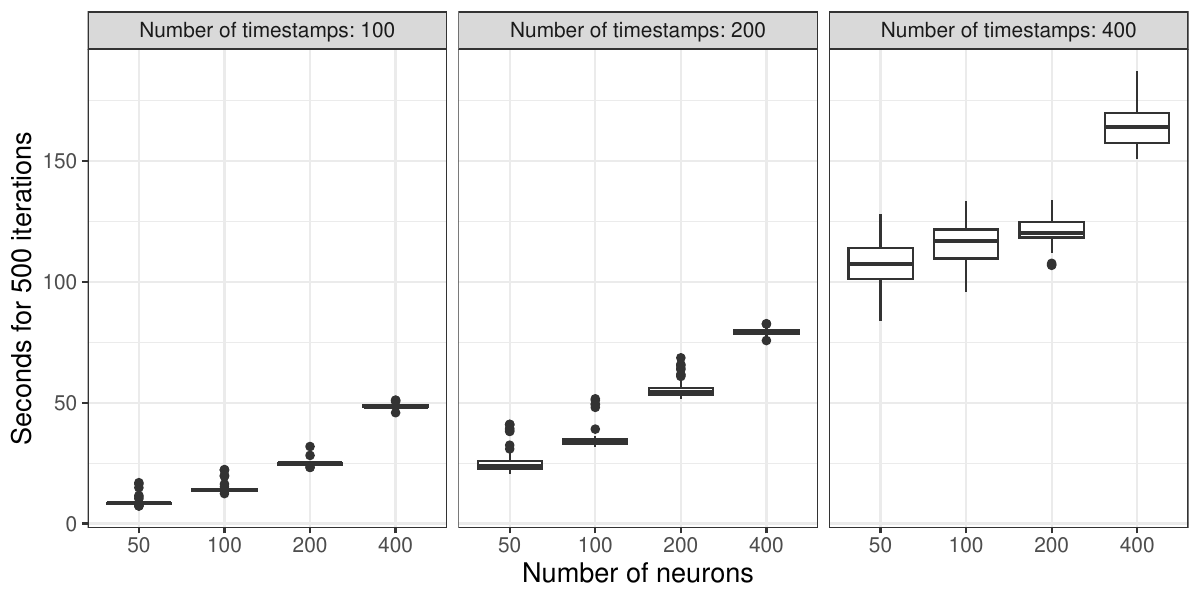} 
\caption{Computational time (in seconds) to perform 500 iterations of the Gibbs sampler on simulated data for varying $N$ and $T$.}\label{fig:comput_cost}
\end{figure}

We conducted a simulation study to study how the proposed algorithm scales with the number of neurons ($N$) and the duration of the experiment ($T$). We generated 50 replicates of synthetic datasets with varying $N\in\{50,100,200,400\}$ and $T\in\{ 100,200,400 \}$, and ran the proposed Gibbs sampler for 500 iterations. 
Figure~\ref{fig:comput_cost} shows the distribution of the computational time (in seconds) to perform 500 iterations of the Gibbs sampler on simulated data with varying numbers of neurons and recording duration, for the 50 replicates. As expected, the computing time increases with $N$ and $T$. While the population size ($N$, reported on the $x$-axis) is not problematic, the true bottleneck lies in the experiment duration ($T$, shown across the three panels). This is primarily driven by the need to sample the Gaussian processes that model the temporal evolution of spike probabilities.  
These results confirm that the proposed model is best suited for the in-depth analysis of the dynamics of even large neuronal populations over short time intervals, effectively capturing fine-scale dynamics.

\subsection{Comparison with state-of-the-art methods}
\begin{figure}
\centering
    \includegraphics[width=\linewidth]{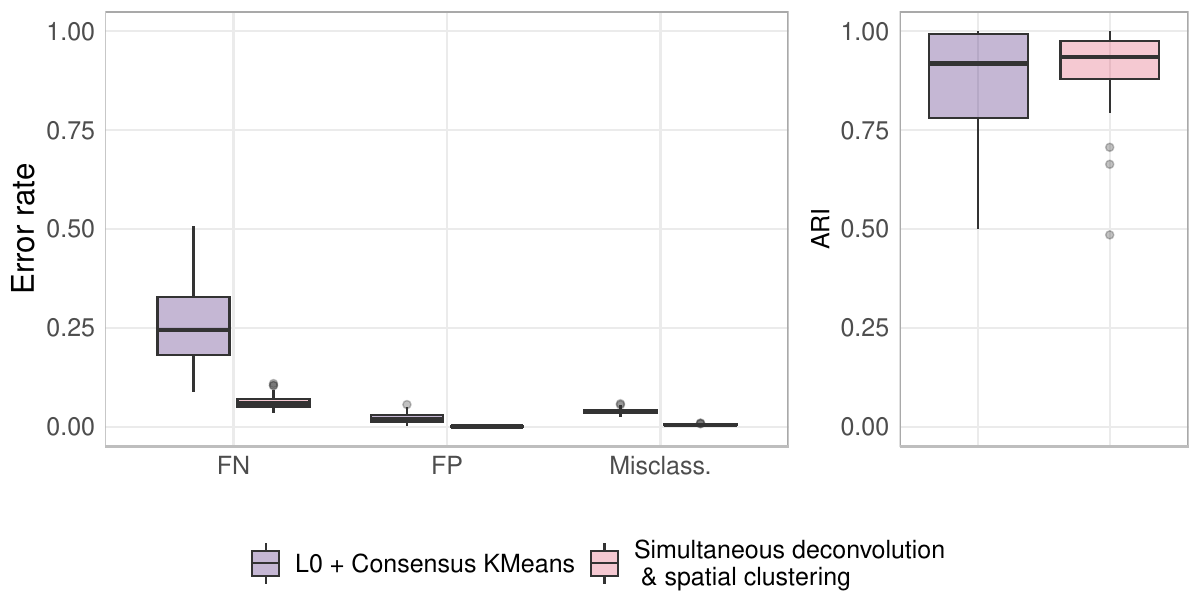}
    \caption{Left panel: boxplot of the false negative (FN), false positive (FP), and misclassification error rates of the spike detection obtained with the $\ell_0$ method and the proposed model. Right panel: boxplot of the Adjusted Rand Index between the true and estimated partition of neurons obtained with the consensus $K$-means and the proposed model. }
    \label{fig:sim_comparison}
\end{figure}

We applied the two competing methods discussed in the main paper to synthetic data to evaluate and compare their results under controlled conditions. 
To assess the spike detection performance, we examine the false positive, false negative, and global misclassification error rates. 
The left panel of Figure~\ref{fig:sim_comparison} shows the error rates obtained with the two competing models. The $\ell_0$ optimization method yields slightly worse results than the proposed approach. While the false positive rates are comparable, the competitor exhibits a considerably higher false negative rate. Lowering the threshold $\lambda$ would reduce this error. However, it would increase the false positive rate, hindering a reduction of the overall error.

We now assess and compare the clustering performance of both methods using the ARI. The right panel of Figure~\ref{fig:sim_comparison} displays the ARI distribution across 50 replicated datasets. Our unified modeling framework demonstrates superior performance in both spike detection accuracy and latent cluster recovery compared to the two-stage approach. Estimating all model parameters in a single framework improves accuracy by using all available information and controls the propagation of the error typical of a multi-stage analysis.

\end{document}